\documentclass{sigchi}

\toappear{\scriptsize Permission to make digital or hard copies of all or part of this work for personal or classroom use is granted without fee provided that copies are not made or distributed for profit or commercial advantage and that copies bear this notice and the full citation on the first page. Copyrights for components of this work owned by others than ACM must be honored. Abstracting with credit is permitted. To copy otherwise, or republish, to post on servers or to redistribute to lists, requires prior specific permission and/or a fee. Request permissions from permissions@acm.org. \\
{\emph{CHI '20, April 25--30, 2020, Honolulu, HI, USA.} } \\
\textcopyright~2020 Association of Computing Machinery. \\
ACM ISBN 978-1-4503-6708-0/20/04\ ...\$15.00. \\
https://dx.doi.org/10.1145/10.1145/3313831.3376873}

\usepackage{balance}       %
\usepackage{graphics}      %
\usepackage[T1]{fontenc}   %
\usepackage{txfonts}
\usepackage{mathptmx}
\usepackage[pdflang={en-US},pdftex]{hyperref}
\usepackage{color}
\usepackage{booktabs}
\usepackage{textcomp}
\usepackage{siunitx}
\usepackage{xspace}
\usepackage{enumitem}
\usepackage{bbold}
\usepackage{float}

\usepackage{microtype}        %
\usepackage{ccicons}          %
\usepackage{todonotes}

\newif{\ifhidecomments}
\hidecommentsfalse
\ifhidecomments
    \newcommand{\chenhao}[1]{}
\else
    \newcommand{\chenhao}[1]{\textcolor{blue}{[#1 ---\textsc{ct}]}}
\fi

\def\plaintitle{%
``Why is `Chicago' deceptive?''\\
Towards Building Model-Driven Tutorials for Humans
}
\def\plainauthor{Vivian Lai, Han Liu and Chenhao Tan
}
\def\plainkeywords{explanations; interpretable machine learning; tutorials; deception detection}

\makeatletter
\def\url@leostyle{%
  \@ifundefined{selectfont}{
    \def\UrlFont{\sf}
  }{
    \def\UrlFont{\small\bf\ttfamily}
  }}
\makeatother
\def\pprw{8.5in}
\def\pprh{11in}

\definecolor{linkColor}{RGB}{6,125,233}
\hypersetup{%
  pdftitle={\plaintitle},
  pdfauthor={\plainauthor},
  pdfkeywords={\plainkeywords},
  pdfdisplaydoctitle=true, %
  bookmarksnumbered,
  pdfstartview={FitH},
  colorlinks,
  citecolor=black,
  filecolor=black,
  linkcolor=black,
  urlcolor=linkColor,
  breaklinks=true,
  hypertexnames=false
}

\DeclareMathOperator*{\argmax}{argmax}

\newcommand{\para}[1]{\noindent{\bf #1}}

\newcommand{\figref}[1]{Fig.~\ref{#1}}

\newcommand{\themethodssection}{the Methods section\xspace}
\newcommand{\thediscussionsection}{the Discussion section\xspace}
\newcommand{\limeglobal}{SP-LIME\xspace}

\newcommand{\citep}[1]{\cite{#1}}
\graphicspath{{figs/}}
\begin{document}

\title{\plaintitle}

\numberofauthors{3}
\author{%
  \alignauthor{Vivian Lai\\
    \affaddr{University of Colorado}\\
    \email{vivian.lai@colorado.edu}}\\
  \alignauthor{Han Liu\\
    \affaddr{University of Colorado}\\
    \email{han.liu@colorado.edu}}\\
  \alignauthor{Chenhao Tan\\
    \affaddr{University of Colorado}\\
    \email{chenhao@chenhaot.com}}\\
}

\maketitle

\begin{abstract}

To support human decision making with machine learning models, we often need to elucidate patterns embedded in the models that are unsalient, unknown, or counterintuitive to humans. While existing approaches focus on explaining machine predictions with real-time assistance, we explore model-driven tutorials to help humans understand these patterns in a training phase. We consider both tutorials with guidelines from scientific papers, analogous to current practices of science communication, and automatically selected examples from training data with explanations. We use deceptive review detection as a testbed and conduct large-scale, randomized human-subject experiments to examine the effectiveness of such tutorials. We find that tutorials indeed improve human performance, with and without real-time assistance. In particular, although deep learning provides superior predictive performance than simple models, tutorials and explanations from simple models are more useful {\em to humans}. Our work suggests future directions for human-centered tutorials and explanations towards a synergy between humans and AI.

\end{abstract}

\section{Introduction}

Interpretable machine learning (ML) has attracted significant interest as %
ML models are used to support human decision making in societally critical domains such as justice systems and healthcare \cite{doshi2017towards,guidotti2019survey,lipton2016mythos}.
In these domains, full automation is often not desired and humans are the final decision makers 
for legal and ethical reasons.
In fact, the Wisconsin Supreme Court ruled that
``a COMPAS risk assessment should
not be used to determine the severity of a sentence or whether
an offender is incarcerated'', but does not eliminate the use of ML models if ``judges be made aware of the limitations of risk assessment tools''~\citep{nytimes,wicourts}.
Therefore, it is crucial to {\em enhance} human performance with the assistance of machine learning models, e.g., by explaining the recommended decisions.

However, recent human-subject studies 
tend to show limited effectiveness of explanations in improving human performance~\citep{bussone2015role,horne2019rating,lai+tan:19,weerts2019human}. 
For instance,
Lai and Tan \citep{lai+tan:19} 
show that explanations alone only slightly improve human performance in deceptive review detection;
Weerts et al. \citep{weerts2019human} similarly find that explanations do not improve human performance in predicting whether one's income exceeds 50,000 in the Adult dataset.
These studies
explain a machine prediction by revealing model internals, e.g., via
attributing importance weights to features and then visualizing feature importance.
We refer to such assistance as real-time assistance because they are provided as humans make individual decisions.

To understand such limited effectiveness, we argue that it is useful to distinguish
two {\em distinct} modes in which 
ML models are being used: {\em emulating} and {\em discovering}.
In tasks such as object recognition \citep{deng2009imagenet,he2015delving}, %
datasets are crowdsourced because humans are considered the gold standard, and ML models are designed to emulate human intelligence.\footnote{As a corollary, it is usually considered overfitting the dataset when machine learning models outperform humans in these tasks.}
In contrast,
in the discovering mode, datasets are usually collected from observing social processes, e.g., whether a person commits crime on bail for bail decisions \citep{kleinberg2017human} and what the writer intention is for deceptive review detection~\citep{abouelenien2014deception,ott2011finding}.
ML models can thus often identify patterns that are unsalient, unknown, and even counterintuitive to humans,
and may even outperform humans in {\bf constrained} datasets \citep{kleinberg2017human,ott2011finding,tan2014effect}.
Notably, many critical policy decisions such as bail decisions resemble the discovering mode more than the emulating mode because policy decisions are usually challenging (to humans) in nature \citep{kleinberg2015prediction}.

Studies on how explanations affect human performance tend to employ these challenging tasks for humans (the {\em discovering} mode for ML models) because humans need {\em little} assistance to perform tasks in the emulating mode (except for scalability).
This observation highlights different roles of explanations in these two modes.
In the emulating mode, explanations can help debug and identify biases and robustness issues in the models for future {\em automation}.
In the discovering mode, if the patterns embedded in ML models can be elucidated for humans, they may enhance human knowledge and improve human decision making.\footnote{It is worth noting that these two modes represent two ends of a continuum, e.g., emulating experts lead to discoveries for novices.}
Moreover, it might help humans identify spurious patterns in ML models and account for potential mistakes to generalize beyond a {\bf constrained} dataset.

\begin{figure*}[t]
  \centering
  \includegraphics[width=0.98\textwidth]{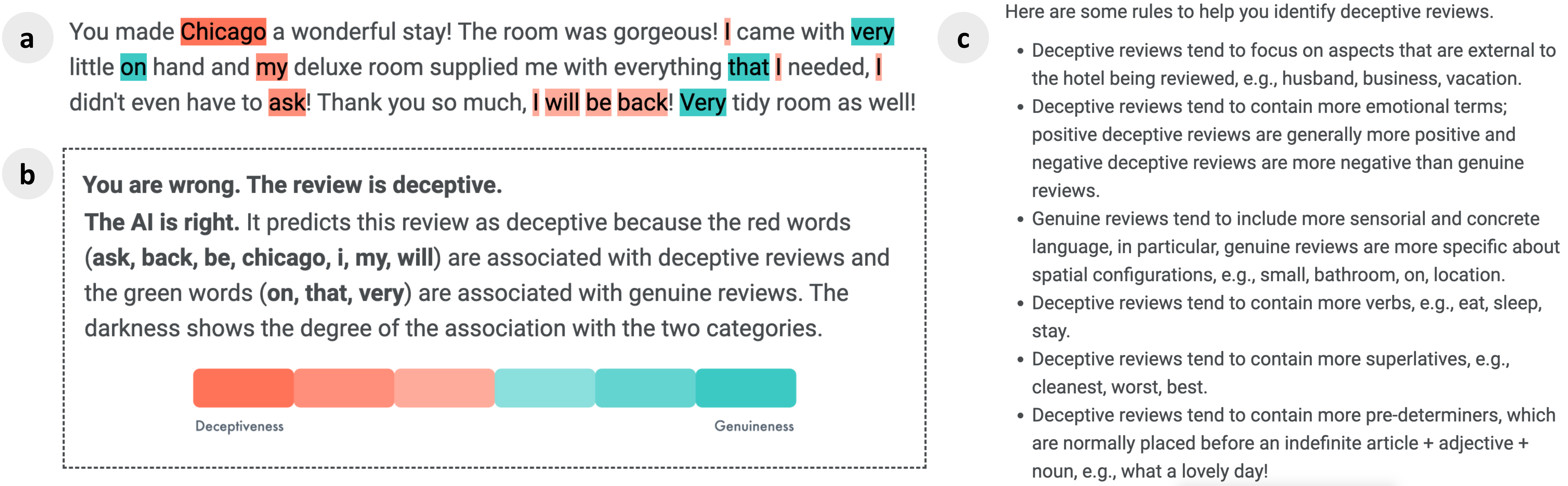}
  \vspace{-0.1in}
  \caption{Illustration of example-driven tutorials and guidelines shown to participants during the training phase: a) top 10 features of the 
  review text are highlighted in green and red ({\em signed highlights}), where green words are associated with genuine reviews
   and red words are associated with deceptive reviews;
b) participants are presented the actual label, the predicted label, and textual explanations for a review after choosing the label of the review in example-driven tutorials;
c) a list of guidelines for identifying deceptive reviews extracted from scientific papers. \label{fig:tutorial}}
\end{figure*}

To further illustrate the difficulty of interpreting explanations in the discovering mode,
\figref{fig:tutorial}(a) shows an example from a deceptive review detection task, where the goal is to distinguish deceptive reviews written by people who %
did not stay at
the hotel from genuine ones.
``Chicago'' is highly associated with deceptive reviews because people are 
more likely to mention the city name instead of specific places when they imagine their experience.
Such a pattern can be hard to comprehend for humans, especially when 
the highlights are shown as real-time assistance without any other information.

Instead of throwing people in at the deep end directly with real-time assistance,
we propose a novel training phase that can help humans understand the nature of a task and the patterns embedded in a model.
This training step is analogous to offline coaching and can be complementary to real-time assistance in explaining machine predictions.
We consider two types of model-driven tutorials:
1) guidelines extracted from scientific papers \citep{li2014towards,ott2013negative,ott2011finding} (\figref{fig:tutorial}(c)), which reflects the current practices of science communication;
2) example-driven tutorials where we 
select examples from the training data and present them along with explanations in the form of highlights (\figref{fig:tutorial}(a)\&(b)).
We also develop a novel algorithm that incorporates spaced repetition to help humans understand the patterns in a 
machine learning 
model,
and conduct an 
in-person user study to refine the design of our tutorials.

Our main contribution in this work is to design large-scale, randomized, pre-registered human-subject experiments to investigate whether tutorials provide useful training to humans, using the aforementioned deceptive review detection task as a testbed.
We choose this 
task because 1) deceptive information including fake news is prevalent on the Internet \citep{allcott2017social,grinberg2019fake,lazer2018science,ott2012estimating} and mechanical turkers can provide a reasonable proxy for humans facing this challenge compared to other tasks such as bail decisions and medical diagnosis that require domain expertise;
2) while humans struggle with detecting deception~\citep{bond2006accuracy}, 
machine learning 
models are able to learn useful patterns in constrained settings (in particular, 
ML
models 
achieve an accuracy of above 85\%
in our deceptive review detection task);
3) full automation might not be desired in this case because the government should not have the authority to automatically block information from individuals, and it is important to {\em enhance} human ability with a machine in the loop.
Specifically, we focus on the following three research questions:
\begin{itemize}[itemsep=-2pt,topsep=-4pt] 
    \item {\bf RQ1:} Do model-driven tutorials improve human performance without any real-time assistance?
    \item {\bf RQ2:} How do varying levels of real-time assistance affect human performance {\em after} training?
    \item {\bf RQ3:} How do model complexity and 
    explanation methods affect human performance with/without training?
\end{itemize}
In all experiments, if training is provided, human subjects first go through a training phase with model-driven tutorials, and then enter the prediction phase to determine whether a review is deceptive or genuine.
The prediction phase allows us to evaluate human performance after training.

Our first experiment aims to compare the effectiveness of different model-driven tutorials.
Ideally, we would hope that these tutorials can help humans understand the patterns embedded in the 
ML models well enough that they can perform decently in the prediction phase without any real-time assistance.
Our results 
show that human performance after tutorials are always better than without training, and the differences are statistically significant for two types of tutorials.
However, the improvement is relatively limited: human performance reaches $\sim$60\%, while the ML models are above 85\%.
Meanwhile, there is no statistically significant difference between human performance after any type of tutorial, which suggests that all model-driven tutorials are similarly effective.

One possible reason for the limited improvement of human performance in Experiment 1 is that the patterns might be too complicated for humans to apply in the prediction phase without any real-time assistance.
Therefore, our second experiment is designed to understand the effect of tutorials with real-time assistance.
Inspired by Lai and Tan \cite{lai+tan:19}, we develop a spectrum 
with varying levels of real-time assistance between full human agency and full automation (\figref{fig:spectrum}).
Our results demonstrate that real-time assistance can indeed significantly improve human performance to above 70\%.
However, compared to Lai and Tan \cite{lai+tan:19}, the best human performance is not significantly improved.\footnote{We only discuss qualitative differences from 
\cite{lai+tan:19}, 
as these are 
separate experiments subject to different randomization processes.}
It suggests that given real-time assistance, tutorials are mainly useful in %
that humans can perform similarly well in the prediction phase with only signed highlights, 
thus retaining a 
higher level of human agency.

Finally, in order to understand how our results generalize to different kinds of 
models, we would like to examine the effect of model complexity and methods of deriving explanations.
Our first two experiments use a linear SVM classifier because linear models are typically deemed interpretable, but deep learning models are 
increasingly prevalent because of their superior predictive power.
While it is well recognized that deep learning models are more complex,
it remains an open question how human performance changes with assistance from deep learning models (e.g., BERT) vs. simple models (e.g., linear SVM).
Our results show that tutorials and explanations of simple models lead to better human performance than deep learning models, which 
highlights the tradeoff between {\em model complexity} and 
{\em interpretability}.
We also show that for BERT, post-hoc signed explanations from LIME are more effective than built-in explanations derived from attention mechanisms.
Moreover, tutorials are effective in improving human performance for both kinds of models compared to without training.

Overall, our results show that model-driven tutorials can somewhat improve human performance with and without real-time assistance, and humans also find these tutorials useful.
However, the limited improvement also points to future directions of human-centered interpretable machine learning.
We highlight two implications here and present further discussions in \thediscussionsection.
First, it is important to explain beyond the surface patterns and facilitate humans in reasoning about why a feature is important.
A strategy is to develop interactive explanations that allow humans to explore the patterns 
in both the training and the prediction phase.
Second, it is useful to bridge the gap between training and generalization in developing tutorials because the model behavior and performance in training data might differ from that on unseen data.
The ability to understand this difference is crucial for humans to calibrate trust and generalize beyond the constrained dataset.

\section{Related Work}

We 
start by introducing recent methods for interpretable ML, and then discuss experimental studies on human interaction with explanations and predictions derived from ML models.
We end by summarizing related work on deception detection.

\subsection{Methods for interpretable machine learning}
A battery of studies propose various algorithms to explain a machine prediction by uncovering model internals (also known as local explanations) \citep{guidotti2019survey}.
Most relevant to our work is feature attribution that assigns an importance weight to each feature \citep{lei2016rationalizing,lundberg2017unified,ribeiro2016should,anchors:aaai18}.
For instance, Ribeiro et al. \citep{ribeiro2016should} propose LIME that fits a sparse linear model to approximate local machine predictions, and coefficients in this linear model are used as explanations.
Lai et al. \citep{lai+cai+tan:19} compare the built-in and post-hoc explanations methods in text classification and show that different methods lead to very different explanations, in particular, deep learning models lead to explanations with less consistency than simple models such as linear SVM.
Other popular approaches include 1) example-based \citep{kim2016examples,kim2014bayesian,mothilal2019explaining,russell2019efficient,wachter2017counterfactual}, e.g., counterfactual explanations find alternative examples that would have obtained a different prediction, and 
2) rule-based \citep{andrews1995survey,guidotti2018lore} that summarizes local rules (e.g., via decision trees).
Notably, 
SP-LIME is an algorithm that selects examples to provide a global understanding of the model \citep{ribeiro2016should}, which aligns with our goal of generating tutorials.
However,
to the best of our knowledge, there have not been any human-subject experiments with such example-driven tutorials.

\subsection{Human interaction with explanations and models}

The importance of human-subject experiments is increasingly recognized in understanding the effectiveness of explanations because they are ultimately used by humans.
In addition to 
studies mentioned in the introduction, researchers have investigated other desiderata of explanations 
\citep{binns2018s,cai2019human,green2019disparate,green2019principles,kunkel2019let,poursabzi2018manipulating,yin2019understanding}.
For instance, Binns et al. \citep{binns2018s} examine 
perception of justice given multiple styles of explanations and conclude that there is no best approach to explaining algorithmic decisions.
Cai et al. \citep{cai2019human} show that a user-centered design improves human perception of an image-search tool's usefulness, but does not improve human performance.
Green and Chen \citep{green2019disparate} find that humans underperformed a risk assessment tool even when presented with its predictions, and exhibited behaviors that could exacerbate biases against minority groups.
Yin et al. \citep{yin2019understanding} examine the effect of stated accuracy and observed accuracy on humans' trust in models,
while Kunkel et al. \cite{kunkel2019let} study the effect of explanations on trust in recommender systems.
This line of work on trust also relates to the 
literature on appropriate reliance 
with general automation \citep{lee2004trust,lewandowsky2000dynamics}.
Retaining human agency is particularly important in societally critical domains where consequences can be dire.
Finally, Bansal et al. \cite{bansal2019beyond} provide feedback during decision making, which can be seen as a form of continuous learning.
Our focus 
is to understand the effect of offline tutorials, which can be potentially combined with real-time assistance/feedback in practice.\footnote{Although feedback (e.g., true labels) on real decisions such as bail decisions can take a long time to observe.}

\subsection{Deception detection}
Deception is a ubiquitous phenomenon and has been studied in  many disciplines \citep{vrij2000detecting}.
In psychology, deception is defined as an act that is intended to foster in another person a belief or understanding which the deceiver considers false \citep{krauss1976modalities}.
Computer scientists have been developing machine learning models to identify deception in texts, images, and videos \citep{abouelenien2014deception,feng2012syntactic,feng2013detecting,jindal2008opinion,ott2011finding,perez2015verbal,wu2010distortion,yoo2009comparison}.
An important challenge in studying deception is to obtain groundtruth labels because it is well recognized that humans struggle at detecting deception \citep{bond2006accuracy}.
Ott et al. \citep{ott2011finding} created the first sizable dataset in deception detection by employing workers on Amazon Mechanical Turk to write imagined experiences in hotels.

As people increasingly rely on information on the Internet (e.g., online reviews for making purchase decisions \citep{chevalier2006effect,trusov2009effects,ye2011influence,zhang2010impact}),
deceptive information also becomes prevalent \citep{caspi2006online,ott2012estimating,shin2011prevalence}.
The issue of misinformation and fake news has also attracted significant attention from both the public and the research community \citep{farsetta2006fake,grinberg2019fake,lazer2018science,vosoughi2018spread,zhang2018structured}.
Our work employs the deceptive review detection task in Ott et al. \citep{ott2013negative,ott2011finding} to investigate the effectiveness of model-driven tutorials.
While this task is a constrained case of deception and may differ from intentionally malicious deception, it represents an important issue that people face on a daily basis and can potentially benefit from assistance from ML models.

\section{Methods}

In this section, we introduce the preliminaries for our prediction task, machine learning models, and explanation methods.
We then develop tutorials to help humans understand the embedded patterns in the 
models in the training phase.
Finally, we present types of real-time assistance in the prediction phase.
A demo is available at \url{https://deception.machineintheloop.com}.

\subsection{Dataset, models, and explanations}

\para{Dataset and prediction task.} 
We employ the deceptive review detection task developed by Ott et al. \citep{ott2013negative,ott2011finding}, consisting of
800 genuine and 800 deceptive hotel reviews for 20 hotels in Chicago.
The genuine reviews were extracted from TripAdvisor and the deceptive ones were written by turkers who were asked to imagine their experience. 
We use 80\% of the reviews as the training set and the remaining 20\% as 
the 
test set.
We evaluate human performance based on their accuracy on sampled reviews from the test set.
The task 
for both humans and ML models 
is to determine whether a review is deceptive or genuine based on the text.

\para{Models.} 
We consider 
a linear SVM classifier with unigram bag-of-words as features, which represents a simple model, and BERT \citep{devlin2018bert}, which represents a deep learning model with state-of-the-art performance in many NLP tasks.
The hyperparameter for linear SVM was selected via 5-fold cross validation with the training set; BERT was fine-tuned on 70\% of the reviews 
and 
the other 
10\% of the reviews 
in the training set 
were used as the development set for selecting hyperparameters.
Table~\ref{tb:accuracy} shows their accuracy 
on the test set.

\begin{table}[h]
\vspace{-0.1in}
\small
\centering
\begin{tabular}{p{0.2\textwidth}r}
\toprule
Model &  Accuracy (\%) \\
\midrule
SVM & 86.3 \\
BERT & 90.9 \\
\bottomrule
\end{tabular}
\caption{Accuracy of machine learning models on the test set.}
\label{tb:accuracy}
\vspace{-0.1in}
\end{table}

\para{Methods of deriving explanations.}
We explain a machine prediction by highlighting the most important 10 words.
For linear SVM, we use the absolute value of coefficients to determine feature importance, and the highlights are signed because coefficients are either positive or negative.
For BERT, we consider two methods following Lai et al. \citep{lai+cai+tan:19}:
1) BERT attention based on the built-in mechanism of Transformer~\citep{vaswani2017attention} (specifically, feature importance is calculated using the average attention values of 12 heads used by the first token at the final layer; these highlights are unsigned because attention values 
range between 0 and 1);
2) BERT LIME, where feature importance comes from LIME by fitting a sparse
linear model to approximate local model predictions (these highlights are signed as they come from coefficients in a linear model).

\subsection{Tutorial generation}

Our main innovation in this work is to introduce a training phase with {\em model-driven} tutorials before humans interact with ML models.
We consider the following two types of tutorials.

\noindent {\em Guidelines.}
We follow the current practice of science communication and summarize findings in scientific papers \cite{ott2013negative,ott2011finding,li2014towards} as a list of guidelines.
These guidelines are observations derived from the 
ML model
(see ``\figref{fig:tutorial}(c)'') and paraphrased by us.
A ``Next'' button is enabled after a 30-second timer.

\noindent {\em Example-driven tutorials.}
Inspired by Ribeiro et al. \citep{ribeiro2016should}, another way to give humans a global sense of a model is to present a sequence of examples along with predicted labels and explanations of predictions.
For each example in our tutorial, informed by our in-person user study, we 
first ask participants to determine the label of the example, and then reveal the actual label and the predicted label along with explanations in the form of highlights.
The algorithm selects 
10 examples that are representative of the patterns that the ML model identifies from the training set.\footnote{We chose 10 so that an experiment session finishes within a reasonable amount of time (30 minutes), and all examples happened to be classified correctly by the model (since machine performance is even better on the training set).}
There could be genuine insights as well as spurious patterns.
Ideally, these examples allow participants to understand the problem at hand and then apply the patterns, including correcting spurious ones, in the prediction phase.
\figref{fig:tutorial}(a)\&(b) presents an example review after 
the label is chosen and the predicted label and its explanations are shown.
A ``Continue'' button is enabled after a 10-second timer.
See the supplementary material for screenshots.

\begin{figure*}[t]
  \centering
  \includegraphics[width=0.8\textwidth]{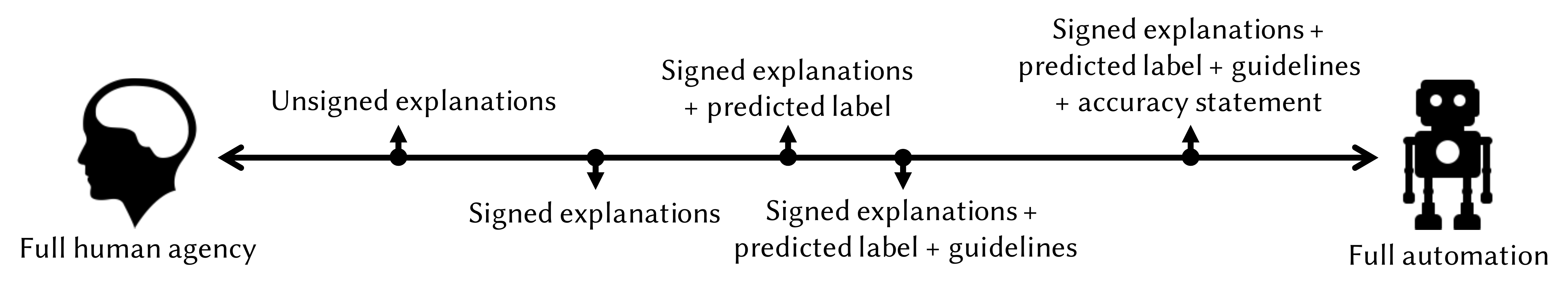}
  \vspace{-0.1in}
  \caption{An adapted spectrum between full human agency and full automation from Lai and Tan [32]. 
  The order approximates our intuition, but the distance does not reflect linear changes in machine influence.
  In particular, guidelines do not necessarily increase the influence of predicted labels.
  }
  \label{fig:spectrum}
\end{figure*}

We consider the following algorithms for example selection:

\begin{itemize}[itemsep=-2pt,topsep=-2pt,leftmargin=*]
  \item {\bf Random.} 10 random examples are chosen.
  \item {\bf SP-LIME.} 
  Ribeiro et al. \citep{ribeiro2016should} propose SP-LIME to select examples with features that provide great coverage in the training set.
  To do that, the global importance of each feature is defined as $I_j = \sqrt{\sum_{i=1}^n W_{ij}}$, where $W_{ij}$ is the importance of feature $j$ in the $i$-th instance.
  Since we only highlight the top 10 features, $W_{ij}=0$ for any other features.
  Then, 10 examples are selected to maximize the following objective function:
$\argmax_{S, |S| \leq B}  \sum_{j=1}^{d} \mathbb{1}(\exists i \in S: W_{ij} > 0) I_{j},$
where $B=10$ and $d$ represents the dimension of features.
This objective function presents a weighted coverage problem over all features, and is thus submodular.
A greedy algorithm provides a solution with a constant-factor approximation guarantee of $1 - 1/e$ to the optimum \citep{krause2014submodular}.
  \item {\bf Spaced repetition (SR).} 
  We propose this algorithm to leverage insights from the education literature regarding the effectiveness of spaced repetition (e.g., on long-term retention) \citep{kang2016spaced,tabibian2019enhancing}.
  Specifically, 
  we develop the following novel objective function so that users can be exposed to important features repeatedly:
  $\argmax_{S, |S| \leq B} \sum_{j=1}^d U(\{W_{kj}\}_{1 \leq k \leq |S|})I_{j},$
  where $U(\{w_{kj}\}_{1 \leq k \leq |S|}) = \mathbb{1}(\max(\{k, W_{kj} > 0\}) - \min(\{k, W_{kj} > 0\}) \geq 3)$.
The key difference from SP-LIME is that the weight of a feature is included only if it is repeated in two examples with a gap of at least three.
\end{itemize}

Finally, we consider the combination of guidelines and examples selected with spaced repetition by first showing the guidelines for 15 seconds, 10 examples selected with spaced repetition, and the guidelines again for 15 seconds.

\subsection{Real-time assistance}

\begin{figure}[t]
  \centering
  \includegraphics[width=0.45\textwidth]{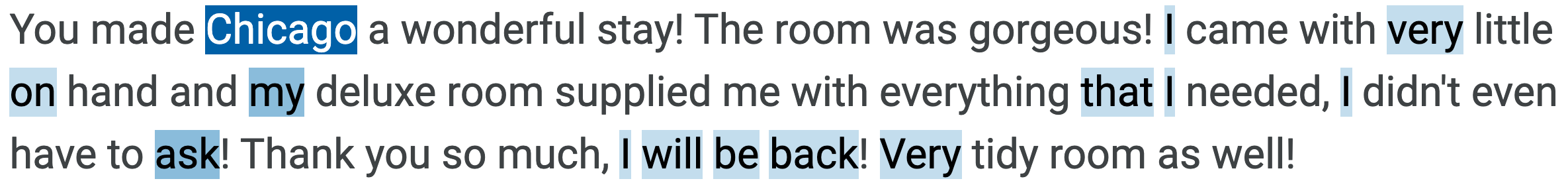}
  \vspace{-0.05in}
  \caption{Unsigned highlights for the example review in \figref{fig:tutorial}(a).}
  \vspace{-0.2in}
  \label{fig:unsigned-highlights}
\end{figure}

In addition to tutorials in the training phase, we introduce varying levels of real-time assistance in the prediction phase.
Inspired by Lai and Tan \citep{lai+tan:19}, we design six levels of real-time assistance, as illustrated in \figref{fig:spectrum}.

\begin{itemize}[itemsep=-2pt,topsep=-2pt]
  \item {\bf No machine assistance.} Participants are not exposed to any real-time machine assistance.
  \item {\bf Unsigned highlights.} Top 10 features are highlighted in shades of blue. The darker the color, the more important the feature.
  See \figref{fig:unsigned-highlights} for an example.
  \item {\bf Signed highlights.} Top 10 features are highlighted in shades of green and red: green words are associated with genuine reviews, while red words are associated with deceptive reviews. 
  The darker the color, the more important the feature. See \figref{fig:tutorial}(a) for an example.\footnote{We use an attention check question to make sure that participants can distinguish red from green.}
  \item {\bf Signed highlights + predicted label.} 
  In addition to signed highlights, we display the predicted label.
  \item {\bf Signed highlights + predicted label + guidelines.} 
  We additionally provide the option of revealing guidelines.
  \item {\bf Signed highlights + predicted label + guidelines + accuracy statement}. 
  We further add an accuracy statement, ``It has an accuracy of approximately 86\%'', emphasizing the strong performance of the ML model.
\end{itemize}

These six levels gradually increase the amount of information and prime users towards machine predictions.
Ideally, we 
hope to retain human agency 
as much as possible 
while achieving strong human performance.

\section{In-person User Study}

To obtain a qualitative understanding of human interaction with model-driven tutorials,
we conduct an in-person semi-structured user study.
This user study allows us to gather in-depth insights on how humans learn and apply our tutorials through interviews, 
as well as feedback on the 
interface before conducting large-scale, randomized experiments.

\subsection{Experimental design}
We employ a concurrent think-aloud process with participants~%
\citep{nielsen2002getting}. 
Each participant went through a tutorial and determined the label of 20 reviews from the test set.
They were told to verbalize the reason before deciding on the label 
both in the training and the prediction phase with the following syntax: I think the review is {\em predicted label} because {\em reason}.
After the prediction phase, we conducted an interview to gather general feedback on tutorials.
We manually transcribed the audio recordings after an initial pass with the Google Cloud API.

A total of 16 participants were recruited from mailing lists in our department:
3 were female and 13 were male, ranging between age 20 and 35.
All participants were engineering graduate students and most of them studied computer science. 
Participants were invited to the lab where the study occurred. 
Either a personal or a provided laptop was used.
Participants were compensated between \$15 and \$20 for \$10 every 30 minutes.
Four types of tutorials (guidelines, examples selected with \limeglobal, examples selected with SR, guidelines + examples selected with SR) were randomly assigned to participants and each tutorial type had a sample size of 4.
Thematic analysis was undertaken to identify common themes in participants' think-aloud processes.
Thematic codes were collectively coded by the first two authors.

\subsection{Results}

We summarize the key themes into the following three parts.

\para{Tutorial training and application.}
8 out of 8 participants with access to guidelines remembered a couple of ``rules'' and applied them in the prediction phase.
P13 said (the number is randomly assigned), ``I believe it is deceptive based on rule No. one and No. three, if I remembered them correctly, it just describes its experience, and does not have a lot of details''.

7 out of 12 participants exposed to selected examples 
adopted pure memorization or pattern-matching during the prediction phase.
Participants remembered key deceptive words such as ``chicago'' 
to help them 
decide the review label:
P2 said, ``My husband is deceptive, I is deceptive, Chicago is deceptive''.
Some participants were even able to generate similar theories to our guidelines without 
exposure to it.
P14 commented, ``The review didn't have anything specific to offer'' before deciding that the respective review was deceptive.
However, reasoning about the patterns is generally challenging.
Quoting from P2, this is mainly because they ``can't seem to find a rhyme or reason for those words being genuine or deceptive''.

Participants 
also created
theories such as length of review
when predicting.
P8 remarked, ``no one would take that much time to write a review so it won't cross more than 5 lines''.

\para{Improvements on tutorials.}
Participants 
thought that the guidelines should be available during the prediction phase 
to better assist them.
4 out of 4 participants felt that they were unable to remember 
as
there were too many guidelines to be memorized.
P11 felt that ``the tutorial is helpful but it's just hard not being able to reference it''
and P9 said that he could ``keep checking if it is on the top right corner''.

12 out of 12 participants exposed to selected examples 
expressed confusion about why the features were highlighted as deceptive or genuine but made up their own reasonings for ease of memory.
They felt that they would have learned better if some form of explanations were given to justify each feature's indication.
P16 remarked that ``it would be nice if it can let me know why exactly it thinks the word is deceptive''
and P10 commented that on top of the current explanations in selected examples, ``more detailed explanation would be helpful'' to help understand.

\para{Improvements on the interface.}
We found that
some participants thought that deceptive reviews are written by an AI without reading the instructions, which is false.
We thus introduced three additional questions for our large-scale experiments:
1) how are deceptive reviews defined in this study?; 2) identify the color that highlights a word; 3) reiterate the training process and ask user to answer true or false to ensure that the participants know which treatment they are exposed to.
We also changed the flow of showing explanations in the training phase: users need to first determine the label for a review before the explanations, the actual label, and the predicted label are shown for at least 10 seconds.
Refer to 
the video 
and detailed feedback 
in the supplementary material.

\section{Experiment 1: Do Tutorials Improve Human Performance without any Real-time Assistance?}

As introduced in \themethodssection, we hope to build tutorials that can help humans understand the embedded patterns in 
ML models, which can sometimes be unsalient, unknown, or even counterintuitive to humans.
Ideally, humans reflect on these patterns from our tutorials and can apply them in 
their decision making
without any further real-time assistance from 
ML models.
Therefore, we start with 
RQ1: do tutorials improve human performance without any real-time assistance?

\subsection{Experimental treatments \& hypotheses}

We consider the following treatments to examine the effectiveness of various tutorials proposed in \themethodssection:
1) guidelines;
2) random examples;
3) examples selected with \limeglobal;
4) examples selected with SR;
5) guidelines + examples selected with SR.
All the tutorials and explanations in the tutorials are based on the linear SVM classifier in \themethodssection.
After a training phase, participants will then decide whether a review is deceptive or genuine based on the text.
Note that ML models also rely exclusively on textual information.
In addition to these tutorials, we include a control setup where no training was provided to humans.

We hypothesize that 1) training is important for humans to understand this task, since it has been shown that humans struggle with deception detection \citep{bond2006accuracy};
2) it would be easier for participants to understand the patterns embedded in the ML model situated with examples;
3) carefully chosen examples provide more comprehensive coverage and can better familiarize participants with the patterns \citep{kang2016spaced,tabibian2019enhancing};
4) guidelines and examples have complementary effects in the training phase.
To summarize, our hypotheses in Experiment 1 are as follows:

\begin{itemize}[itemsep=-2pt,topsep=-2pt]
    \item ({\bf H1a}) Any tutorial treatment leads to better human performance than the control setup.
    \item ({\bf H1b}) Examples (including {\em random examples, examples selected with \limeglobal and SR}) lead to better human performance than {\em guidelines}.
    \item ({\bf H1c}) {\em Selected examples (with SP-LIME or SR)} lead to better human performance than {\em random examples}.
    \item ({\bf H1d}) {\em Examples selected with spaced repetition} lead to better human performance those selected with {\em \limeglobal}.
    \item ({\bf H1e}) {\em Guidelines + examples selected with SR} lead to the best performance.
\end{itemize}
These five hypotheses
were pre-registered on AsPredicted.\footnote{The anonymized pre-registration document is available at \url{https://aspredicted.org/blind.php?x=v8f7zh}.
A minor inconsistency 
is that we did not experiment with ``guidelines + examples selected from SP-LIME'' as we hypothesized that SR is better.
}

\subsection{Experimental design}

To evaluate human performance under different experimental setups, participants were recruited via Amazon Mechanical Turk and filtered to include only individuals residing in the United States, with at least 50 Human Intelligence Tasks (HITs)  completed and 99\% of HITs approved.
Each participant is randomly assigned to one of the six conditions (five types of tutorials + control).
We did not allow any repeated participation.
We adopted this between-subject design
because exposure to any type of tutorial cannot be undone.

In our experiment, each participant finishes the following steps sequentially:
1) reading an explanation of the task and a consent form;
2) answering a few attention-check questions depending on the experimental condition assigned;
3) undergoing a set of tutorials if applicable (training phase);
4) predicting the labels of 20 randomly selected reviews in the test set (prediction phase);
5) completing an exit survey.
Participants who failed the attention-check questions are automatically disqualified from the study.
Based on feedback from our in-person user study, for each example in the tutorials, a participant 
first chooses genuine or deceptive without any assistance, and then the answer is revealed and the predicted label and explanations are shown (\figref{fig:tutorial}(a)\&(b)).
In the exit survey, participants were asked to report basic demographic information, if the tutorial was helpful (yes or no), and feedback in free responses.\footnote{Feedback from Turkers generally confirmed findings in the in-person user study. See the supplementary material for an analysis.}

Each participant was compensated \$2.50 and an additional \$0.05 bonus for each correctly labeled test review.
80 subjects were recruited for each condition so that each review in the test set was labeled five times.
In total 480 subjects completed Experiment 1.
They were balanced on gender (224 females, 251 males, and 5 preferred not to answer).
Refer to the supplementary material for additional information about experiments (e.g., education background, time taken).

To quantify human performance, we measure it by the percentage of correctly labeled instances by humans.
In other words, the prediction phase provides an estimate of human accuracy through 20 samples.
In addition to this objective metric, we also report subject perception of tutorial usefulness reported in the exit surveys.

\begin{figure}[t]
  \centering
  \includegraphics[width=0.35\textwidth]{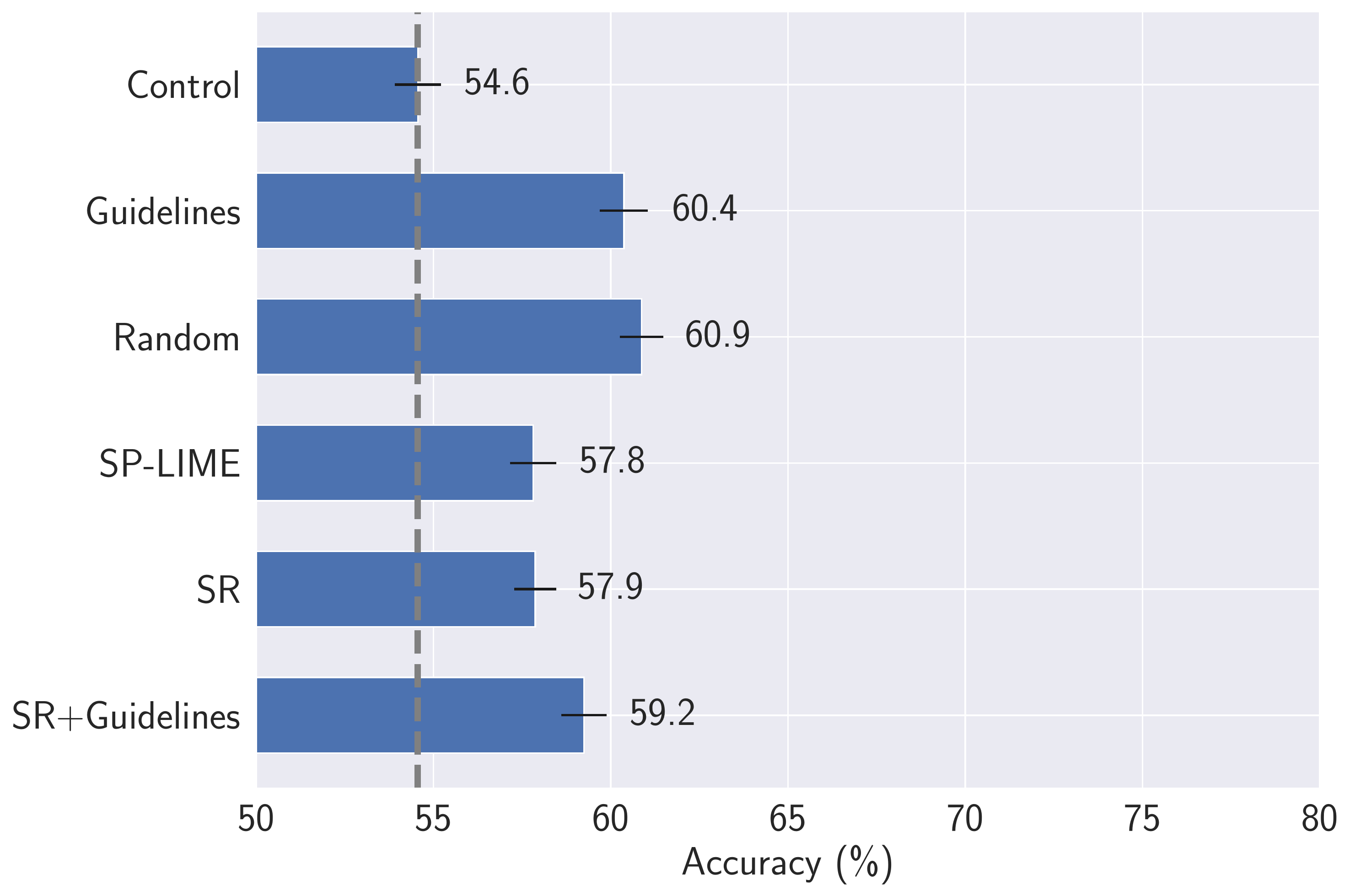}
  \caption{Human accuracy without any real-time assistance after different types of tutorials.
    Error bars represent standard errors.
  Human accuracy after tutorials is always better than that without any training.
  Differences are statistically significant between random and control, and guidelines and control based on post-hoc Tukey's HSD test.
  }
  \label{fig:exp1-acc}
\end{figure}

\begin{figure}[t]
  \centering
  \includegraphics[width=0.35\textwidth]{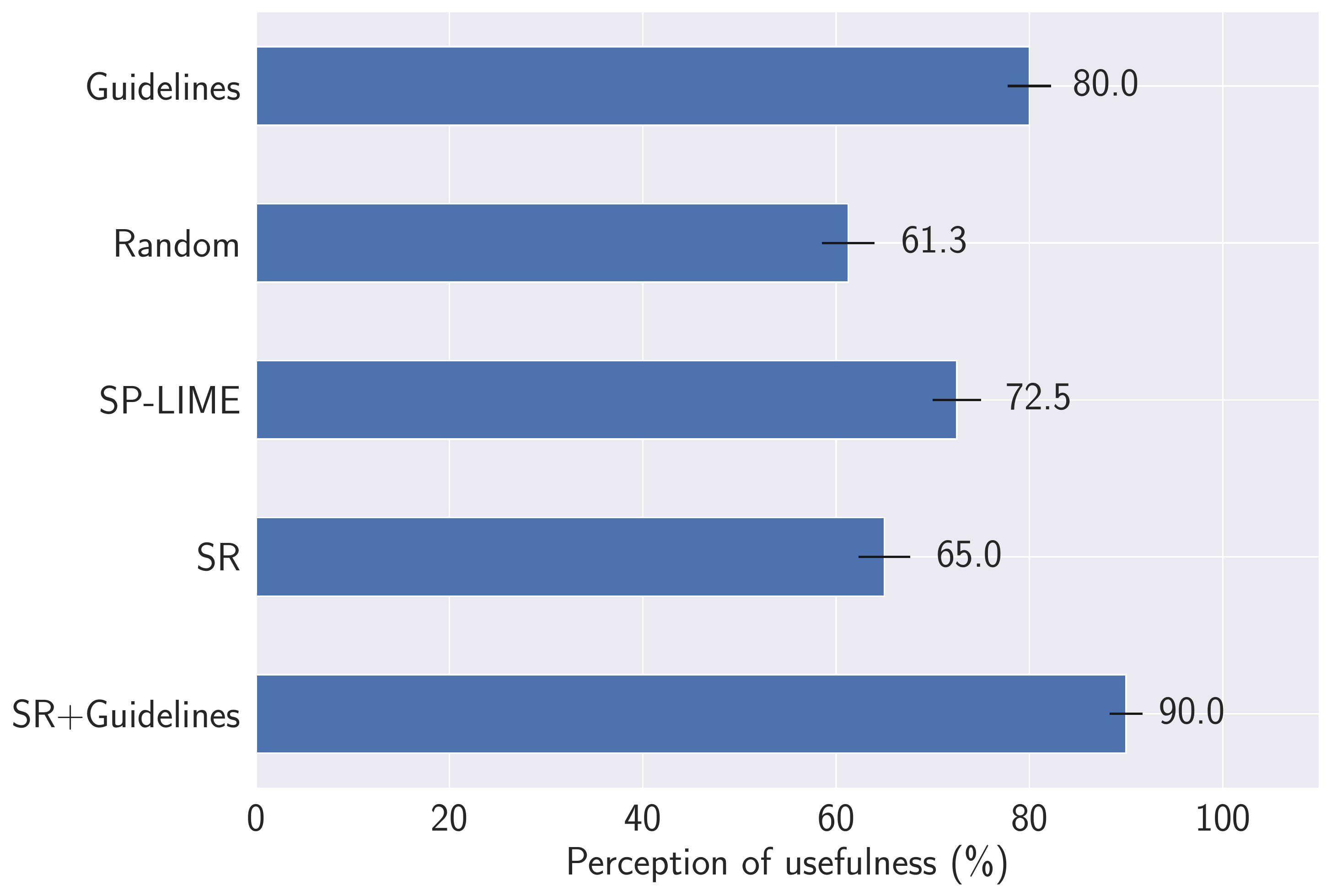}
  \caption{Subjective perception of tutorial usefulness.
    Error bars represent standard errors.
    Differences are statistically different in the following pairs based on post-hoc Tukey's HSD test: guidelines vs. random, random vs. SR+guidelines, and SR vs. SR+guidelines.}
  \label{fig:exp1-use}
  \vspace{-0.1in}
\end{figure}

\subsection{Results}
We first
present human accuracy in the prediction phase, an objective measurement of tutorial effectiveness.
Our results suggest that tutorials are useful to some extent:
all tutorials lead to better human performance ($\sim$60\%) than the control setup without any training (\figref{fig:exp1-acc}).
To formally compare the treatments, we conduct an one-way ANOVA 
 and find
a statistically significant 
effect 
($\eta^2 = 0.033$; $p = \num{7.70e-3}$).
We further use post-hoc Tukey's HSD test to identify pairs of experimental conditions in which human performance exhibits significant differences.
The only statistically significant differences are {\em guidelines} vs. {\em control} ($p = \num{1.75e-2}$) and {\em random} vs. {\em control} ($p = \num{7.0e-3}$) (the difference between {\em guidelines+SR} and {\em control} is borderline significant with $p = 0.10$).

In other words, our experiment results provide partial support to {\bf H1a}, and reject all other hypotheses in Experiment 1.
These results suggest that although tutorials provide somewhat useful training, different tutorials are similarly effective.
The limited improvement in human performance across all tutorials indicates that the utility of tutorials is small.
We hypothesized that it is too challenging for humans to remember all the patterns after a short tutorial (supported by feedback from in-person user study), which motivated Experiment 2 to understand the effect of real-time assistance in conjunction with tutorials.
Another contributing factor certainly lies in the design of tutorials, which we will further discuss in \thediscussionsection.

As for subjective perception of tutorial usefulness, 
we find that participants generally find our tutorials useful:
73.8\% of 400 participants reported that the tutorial was useful (excluding 80 participants in the control setup).
\figref{fig:exp1-use} shows the results by types of tutorials.
Among different treatments, participants in {\em guidelines} and {\em guidelines + examples selected with SR} find the tutorials most useful, as high as 90\% in {\em guidelines + examples selected with SR}.
Formally, post-hoc Tukey's HSD test shows that the differences between the following pairs are statistically different: {\em guidelines} vs. {\em random} ($p = 0.048$), {\em random} vs. {\em SR+guidelines} ($p < 0.001$), and {\em SR} vs. {\em SR+guidelines} ($p = 0.003$).
The difference between {\em SP-LIME} and {\em SR+guidelines} is borderline significant with $p = 0.078$.
These results suggest that tutorials provide strong positive effects in humans' subjective perception.
\section{Experiment 2: Human Performance with Varying Real-time Assistance after Tutorials}

Our second experiment is concerned with human performance with varying levels of real-time assistance after going through the training phase.
While Experiment 1 suggests that tutorials provide somewhat useful training, the improvement is limited without any real-time assistance.
We hypothesize that human performance could be further improved by introducing real-time assistance.
We adapt a spectrum with varying levels of real-time assistance from Lai and Tan \citep{lai+tan:19} (\figref{fig:spectrum}).
Moving along the spectrum, the influence of the machine generally becomes greater on the human
as more information from the model is presented.
For instance, a statement of strong machine performance is likely to bias humans towards machine predictions.
Lai and Tan \citep{lai+tan:19} find that there exists a tradeoff between human performance and human agency, i.e., as the real-time assistance gives stronger priming along the spectrum, human performance improves and human agency decreases.
Explanations such as highlighting important words can moderate this tradeoff {\em when predicated labels are given}.
It remains an open question how this tradeoff unfolds after training.

\subsection{Experimental treatments \& hypotheses}

All conditions in Experiment 2 used the {\em guidelines + selected examples with spaced repetition} tutorial in the training phase
because all tutorials are similarly effective and our participants find this one most useful in subjective perception.
To examine how humans perform under different levels of real-time assistance from machine learning models, we consider
the spectrum in \figref{fig:spectrum},
inspired by Lai and Tan \citep{lai+tan:19}.

We hypothesize that 1) real-time assistance results in 
improved human performance, since it has been shown that highlights and predicted labels improve human performance \citep{lai+tan:19}; 2) signed highlights result in better human performance compared to unsigned highlights because signed highlights reveal information about directionality; 3) predicted labels result in better human performance compared to highlights alone; 4) guidelines and signed highlights might moderate the tradeoff between human performance and human agency while achieving the same effect as when 
an accuracy statement is shown.
To summarize, our hypotheses 
are as follows:
\begin{itemize}[itemsep=-2pt,topsep=-2pt]
    \item ({\bf H2a}) Real-time assistance leads to better human performance than no assistance.
    \item ({\bf H2b}) {\em Signed highlights} lead to better human performance than {\em unsigned highlights}.
    \item ({\bf H2c}) {\em Predicted label} leads to better human performance than highlights alone.
    \item ({\bf H2d}) {\em Signed highlights + predicted label + guidelines + accuracy statement} leads to the best performance.
    \item ({\bf H2e}) {\em Signed highlights + predicted label + guidelines}
    and {\em Signed highlights + predicted label} perform as well as {\em Signed highlights + predicted label + guidelines + accuracy statement}.
\end{itemize}
These five hypotheses were pre-registered on AsPredicted.\footnote{The anonymized pre-registration document is available at \url{http://aspredicted.org/blind.php?x=fi8kz8}.}

\subsection{Experimental design}
We adopted the same experimental design as stated in Experiment 1 except that real-assistance is provided in the prediction phase when applicable.
In total 480 subjects completed the experiment (80 participants in each type of real-time assistance).
They were balanced on gender (238 females, 237 males, and 5 preferred not to answer).
Refer to the supplementary material for additional information about experiments (e.g., education background, time taken).

Human performance is measured by the percentage of correctly predicted instances by humans, which provides an objective measure of human performance with real-time assistance.
We also consider the percentage of humans whose performance exceeds machine performance for the corresponding 20 reviews in the prediction phase.\footnote{We also pre-registered trust as a measure and present the results in the supplementary material for space reasons.}

\begin{figure}[t]
  \centering
  \includegraphics[width=0.4\textwidth]{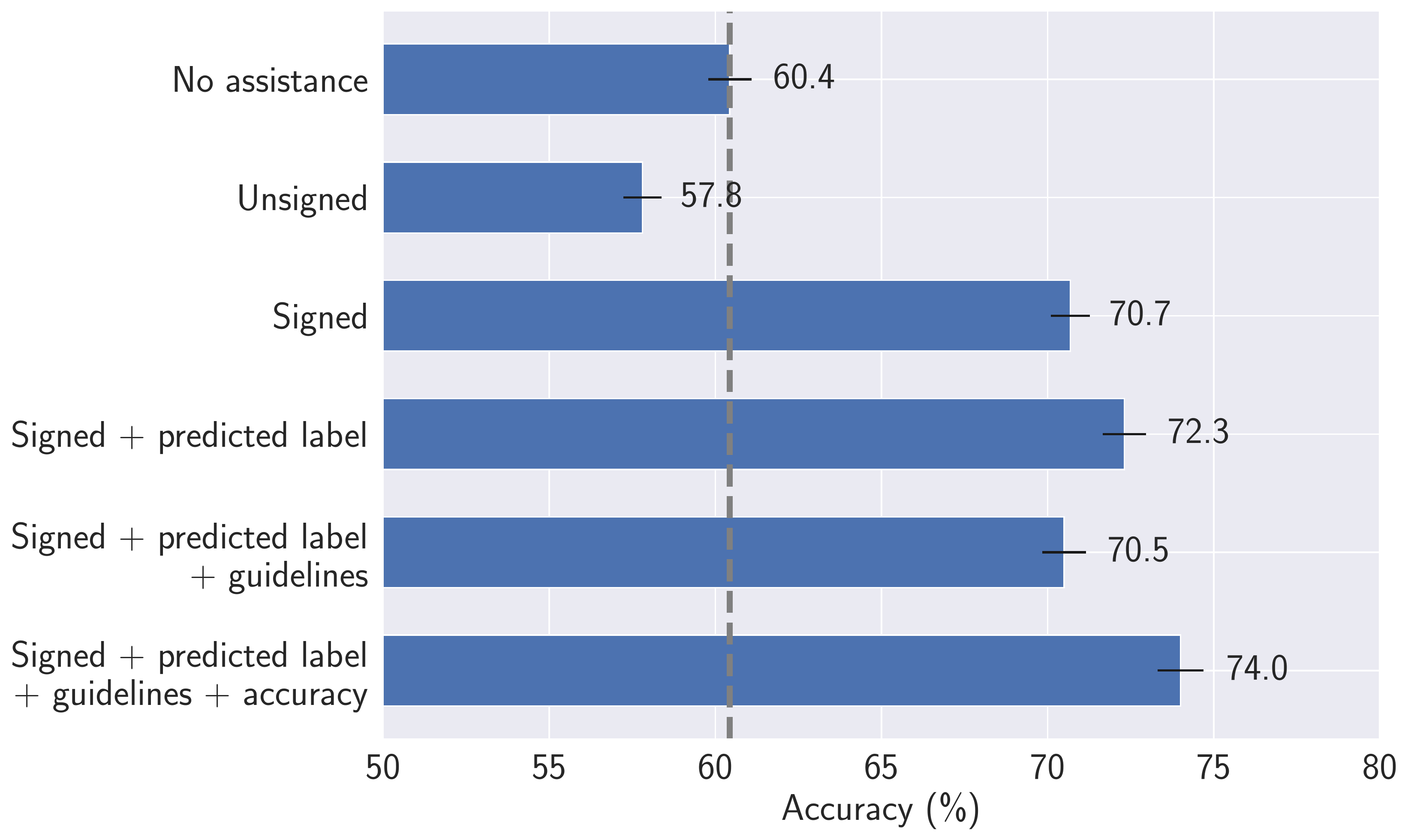}
  \caption{Human accuracy with varying levels of real-time assistance after training. 
  Error bars represent standard errors. 
  With the exception of {\em unsigned highlights}, human accuracy with real-time assistance is better than without real-time assistance.
  Differences between {\em no assistance} and any assistance with signed highlights are statistically significant based on post-hoc Tukey's HSD test.
  }
  \label{fig:exp2-acc}
  \vspace{-0.1in}
\end{figure}

\subsection{Results}
We first present human accuracy in the prediction phase.
Our results suggest that real-time assistance is indeed effective: all the levels of real-time assistance except unsigned highlights lead to better human performance than the setup without machine assistance in \figref{fig:exp2-acc}.
To formally compare the treatments, we conduct an one-way ANOVA
and find a statistically significant 
effect 
($\eta^2 = 0.23$; $p = \num{5.15e-25}$).
We further use post-hoc Tukey's HSD test to identify pairs of experimental conditions in which human performance exhibits significant differences.
With the exception of {\em no assistance} vs. {\em unsigned highlights} ($p = 0.67$), differences in remaining setups compared to {\em no assistance} are all statistically significant ($p < 0.001$).
Moreover, the difference between {\em unsigned highlights} and {\em signed highlights} is significant ($p < 0.001$),
demonstrating the effectiveness of signed highlights.
Finally, the difference between {\em signed highlights} and any other real-time assistance with stronger priming ({\em signed highlights + predicted labels, signed highlights + predicted labels + guidelines, signed highlights + predicted labels + guidelines + accuracy statement}) is not significant.

In summary, our experimental results support {\bf H2a} with the exception of {\em unsigned highlights}, {\bf H2b}, {\bf H2e}, and reject {\bf H2c} and {\bf H2d} in Experiment 2 (note that {\em signed highlights + predicted label + guidelines + accuracy statement} indeed leads to the best performance but the difference with other methods is not always statistically significant).
These results suggest that {\em signed highlights}
provide sufficient information for improving human performance, and we do not gain much from presenting additional information with stronger priming.
While there is significant improvement in human performance with real-time assistance (from $\sim$60\% to $\sim$70\%),
the improvement is still limited compared to the machine performance, which is above 85\%.
This improvement is similar to results reported in Lai and Tan \citep{lai+tan:19}, which did not use any tutorials other than minimal examples to introduce the task.
These observations taken together suggest that the utility of our tutorials mainly lies in that humans can perform well with only signed highlights, a type of real-time assistance with relatively weak priming.

Another ambitious measurement 
is 
how frequent humans outperform the ML model.
It was rare in Experiment 1 (2 of 480, 0.4\%).
With effective real-time assistance (i.e., signed highlights included),
we find that 26 of 320 (8.1\%, 20 times the percentage in Experiment 1) of our participants are able to outperform the ML model.
The difference between 8.1\% and 0.4\% is statistically significant using chi-squared tests ($p < 0.001$).
This observation suggests that with the help of tutorial and real-time assistance, there exists hope for a synergy of {\em humans and AI} outperforming AI alone.
We hypothesize that facilitating hypothesis generation is important 
and present detailed discussions in \thediscussionsection.

\section{Experiment 3: The Effect of Model Complexity and Methods of Deriving Explanations}

Our experiments so far are based on explanations (coefficients) from a linear SVM classifier.
Meanwhile, deep learning models are being widely adopted because of their superior predictive power.
However, it is also increasingly recognized that they might be more complex and harder to interpret for humans.
Our final experiment investigates how model complexity and methods of deriving explanations relate to human performance and effect of training.

\subsection{Experimental treatments \& hypotheses}
Participants are exposed to two different treatments: presence of training and methods of deriving highlights.
Where training is present,
we use the {\em selected examples with spaced repetition} tutorial in this experiment.
Note that example selection depends on the model and the explanation method (i.e., which features are considered important).
In comparison, guidelines are static and are extracted from papers based on linear SVM, so they are not appropriate here.
Based on results from Experiment 2, we adopted {\em signed highlights} as our real-time assistance in the prediction phase when applicable.\footnote{Since BERT performs better than linear SVM, only showing signed highlights also avoids the potential effect of predicted labels.}
To summarize, we consider the following setups to examine how humans perform when exposed to training and different methods of deriving explanations:
1) no training + SVM coefficients;
2) no training + BERT attention;
3) no training + BERT LIME;
4) training + SVM coefficients;
5) training + BERT attention;
6) training + BERT LIME.

Note that the deep learning model (BERT) leads to both different real-time assistance and examples selected for tutorials
because they consider different words important.
We can only use unsigned highlights for BERT attention because attention values range between 0 and 1.
Refer to \themethodssection for details of BERT attention and BERT LIME.

We hypothesize that 1) SVM results in better performance compared to BERT, since it is a common assumption that linear models are more interpretable and it has been shown that SVM results in important features with lower entropy~\citep{lai+cai+tan:19}; 2) BERT LIME results in better performance compared to BERT attention because signed highlights can reveal more information about the underlying decision; 3) participants would perform better with training than without training.
To summarize, our hypotheses in Experiment 3 are as follows:
\begin{itemize}[itemsep=-4pt,topsep=-5pt]
    \item ({\bf H3a}) The simple model ({\em SVM}) leads to better human performance than the deep learning model ({\em BERT}).
    \item ({\bf H3b}) {\em BERT LIME} leads to better human performance than {\em BERT attention}.
    \item ({\bf H3c}) {\em Training} leads to better human performance than {\em without training}.
\end{itemize}
These three hypotheses were pre-registered on AsPredicted.\footnote{The anonymized pre-registration document is available at \url{http://aspredicted.org/blind.php?x=vy794a}.}

\subsection{Experimental design}
We adopted the same experimental design as
in Experiment 1.
In total 480 subjects completed the experiment (80 participants in each experimental setup).
They were balanced on gender (239 females, 240 males, and 1 preferred not to answer).
Refer to the supplementary material for additional information about experiments (e.g., education background, time taken).

To quantify human performance, we measure it by the percentage of correctly predicted instances by humans.
In addition to this objective metric, we also report subject perception of tutorial usefulness reported in the exit surveys (note that this is only applicable for the experimental setups with training).

\begin{figure}[t]
  \centering
  \includegraphics[width=0.4\textwidth]{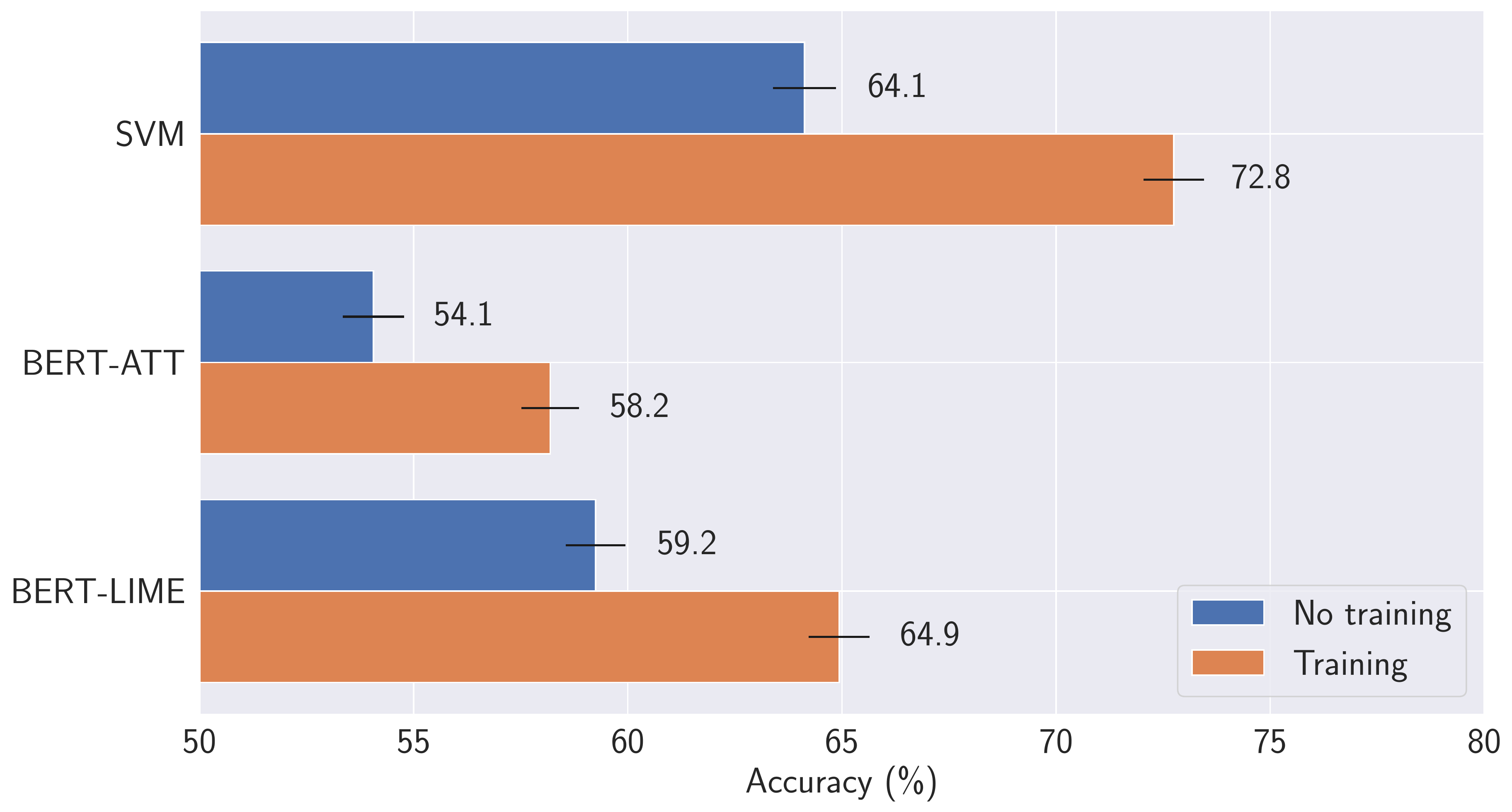}
  \caption{Human accuracy grouped by methods of deriving explanations.
  Error bars represent standard errors.
  SVM explanations lead to
  better human performance than explanations based on BERT.
  Training (second bar from the top in each method) also consistently improves human performance for all explanation methods.
  }
  \label{fig:exp3-acc}
\end{figure}

\begin{figure}[t]
  \centering
  \includegraphics[width=0.4\textwidth]{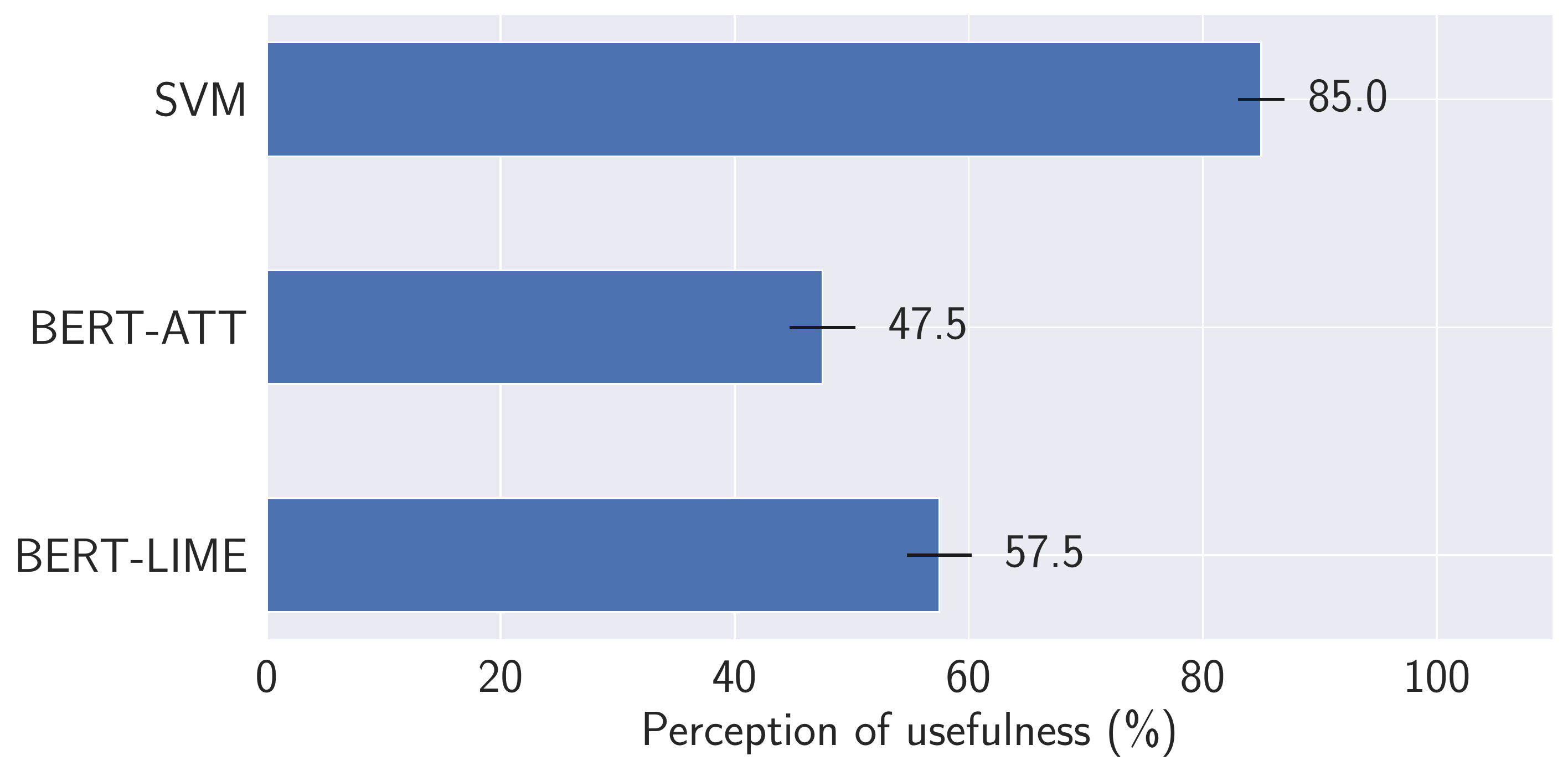}
  \caption{Human perception of tutorial usefulness.
    Error bars represent standard errors.
Participants are more likely to find SVM tutorials useful
 (differences between (SVM, BERT attention) and (SVM, BERT LIME) are statistically significant using post-hoc Tukey's HSD test).
}
  \label{fig:exp3-use}
  \vspace{-0.05in}
\end{figure}

\subsection{Results}
We first present human accuracy in the prediction phase.
Our results suggest that
methods of deriving explanations make a significant difference (\figref{fig:exp3-acc}): 1) human performance is consistently better when important words derived from the linear SVM
are highlighted as compared to deep models; 2)
BERT LIME leads to better human performance than BERT attention.
It also reinforces the point that training leads to better human performance as compared to no training: humans achieve better performance with training with any kind of explanation methods.
To formally compare the treatments, we conduct a two-way ANOVA
and find a statistically significant
effect of tutorials ($\eta^2 = 0.049$; $p = \num{1.50e-7}$) and methods of deriving explanations ($\eta^2 = 0.13$; $p = \num{4.66e-16}$).
Differences among all pairs of treatments are also statistically significant using post-hoc Tukey's HSD test ($p < 0.001$).\footnote{It is reduced to $t$-test for the training/no training treatment since the degree of freedom is 1.}

In other words, our experiment results provide support to all hypotheses in Experiment 3.
These results suggest that tutorials  are  indeed useful in improving human performance, albeit improvement is still limited in the sense that human performance is $\sim$70\% after training with real-time assistance, echoing results in Experiment 2.
It also suggests that simple models are preferred to deep learning models when serving as explanations to support human decision making.
Between explanations derived from post-hoc and built-in methods from BERT, attention
provides the least
value for humans,
again demonstrating the importance of signed highlights.
The effectiveness of training for simple models is further validated by subjective perception of tutorial usefulness.
\figref{fig:exp3-use} shows that participants are much more likely to find the tutorials derived from SVM explanations useful: 85\% of our participants find it useful.
The differences between the following pairs are statistically different using post-hoc Tukey's HSD test: {\em SVM} vs. {\em BERT attention} ($p < 0.001$) and {\em SVM} vs. {\em BERT LIME} ($p < 0.001$).
Interestingly, with real-time assistance, humans also find the tutorials more useful compared to the same tutorial in \figref{fig:exp1-use}.
These results underscore our findings in Experiment 3 that simple models provide more interpretable tutorials and explanations than deep models.

\section{Discussion}
In this paper, 
we conduct the first large-scale, randomized, pre-registered human-subject experiments to investigate whether model-driven tutorials can help humans understand the patterns embedded in ML models and improve human performance.
We find that tutorials can indeed improve human performance to some extent,
with and without real-time assistance, and humans also find them useful.
Moreover, real-time assistance is crucial for further improving human performance in such challenging tasks.
Finally, we show that simple models like linear SVM generate more useful tutorials and explanations for humans than complex deep learning models.

\para{Towards human-centered tutorials.}
Both quantitative results from our randomized experiments and qualitative feedback from in-person user study demonstrate that humans can benefit from model-driven tutorials, which suggests that developing model-driven tutorials is a promising direction for future work in human-centered interpretable machine learning.

However, the improvement in human performance remains limited compared to machine performance in the deceptive review detection task.
In order to further advance the synergy between humans and AI,
we need to develop human-centered tutorials.
Many participants 
commented that they could not understand why certain words were deceptive or genuine (an example reason could be that imaginative writing does not cover specific details).
These results highlight the importance of {\em facilitating hypothesis generation} in the tutorials.
It is insufficient to highlight important features via feature attribution methods,
and these tutorials need to also explain why some features are useful.
While it is challenging to develop automatic methods that can propose theories about particular features, 
we might prompt humans to propose theories and evaluate them through the ML model.

Another reason that tutorials had limited improvement in human performance is that the tutorials failed to establish proper trust in machine predictions. %
It is important to highlight both strengths and caveats of ML models in the tutorials, echoing 
recent work on 
understanding trust 
 \citep{kunkel2019let,yin2019understanding}.
A challenge lies in how to bridge the gap between training and generalization in tutorials, i.e., model behavior and performance in the tutorials might 
differ from that in unseen data.

\para{Beyond static explanations.}
Another important direction is to design interactive explanations beyond static explanations such as simply highlighting important words.
Interactive explanations allow humans to experiment with their hypothesis about feature importance.
One strategy is to enable humans to inquire about the importance of any word in a review.
An alternative strategy is to assess model predictions of counterfactual examples.
For instance, humans can remove or add words/sentences in a review, which can help humans understand model behavior in new scenarios.

\para{Choice of tasks.} 
We would like to highlight the importance of task choice in understanding human-AI interaction.
Deception detection might simply be too challenging a task for humans, and a short tutorial is insufficient to help humans understand the patterns embedded in ML models.
There may also exist significant variation between understanding text and interpreting images, because the former depends on culture
and life experience, while the latter relies on basic visual cognition.

We believe that it is important to study human-AI interaction in challenging tasks where human agency is important because the nature of explanations in decision making is distinct from that in debugging.
While machines excel at identifying patterns from existing datasets, humans might be able to complement ML models by deriving theories and appropriately correcting machine predictions in unseen data, e.g., spotting mistakes when machines apply patterns (``chicago'' becomes a specific comparison point for reviews about a hotel in New York City).
So there exists hope for further advancing human performance in these challenging tasks.

\para{Limitation of our samples.}
Our study is limited by our samples of human subjects.
The in-person user study was conducted with university students who tend to have a computer science education, and large-scale, randomized, pre-registered experiments were conducted with 
Mechanical Turkers from the United States.
While our samples are likely to face the challenges of deception on the Internet and would benefit from enhancements in deception detection, they may not be representative of the general population.
The effectiveness of model-driven tutorials can also potentially depend on properties of the sample population.
In general, we did not find any consistent differences between demographic groups 
based on age, gender, education background, and review experience 
(see the supplementary material).
It is certainly possible that other demographic information could affect the effectiveness of tutorials.
We leave that 
for future studies.

It is important to point out that our setup employs a random split to obtain training and testing data, which is a standard assumption in supervised machine learning.
While humans can ideally improve generalization in this case, 
humans might be more likely to correct generalization errors in machine learning models when the testing distribution differs from training.
In that case, understanding the embedded patterns, especially spotting spurious ones, can help humans generalize these data-driven insights.

In summary, our work highlights the promise of (automatically) building model-driven tutorials to help humans understand the patterns embedded in ML models, especially in challenging tasks.
We hope to encourage future work on human-centered tutorials and explanations beyond static real-time assistance towards a synergy between humans and AI.

\para{Acknowledgments.}
We thank helpful comments from anonymous reviewers.
All experiments were approved by the University of Colorado IRB (18-0558).
This work was supported in part by NSF grants IIS-1837986, 1849931, and 1927322.

\balance{}

\bibliographystyle{SIGCHI-Reference-Format}
\bibliography{refs}


\begin{thebibliography}{00}


\ifx \showCODEN    \undefined \def \showCODEN     #1{\unskip}     \fi
\ifx \showDOI      \undefined \def \showDOI       #1{{\tt DOI:}\penalty0{#1}\ }
  \fi
\ifx \showISBNx    \undefined \def \showISBNx     #1{\unskip}     \fi
\ifx \showISBNxiii \undefined \def \showISBNxiii  #1{\unskip}     \fi
\ifx \showISSN     \undefined \def \showISSN      #1{\unskip}     \fi
\ifx \showLCCN     \undefined \def \showLCCN      #1{\unskip}     \fi
\ifx \shownote     \undefined \def \shownote      #1{#1}          \fi
\ifx \showarticletitle \undefined \def \showarticletitle #1{#1}   \fi
\ifx \showURL      \undefined \def \showURL       #1{#1}          \fi

\bibitem{abouelenien2014deception}
{Mohamed Abouelenien}, {Veronica P{\'e}rez-Rosas}, {Rada Mihalcea}, {and}
  {Mihai Burzo}. 2014.
\newblock \showarticletitle{Deception detection using a multimodal approach}.
  In {\em Proceedings of ICMI}.
\newblock


\bibitem{allcott2017social}
{Hunt Allcott} {and} {Matthew Gentzkow}. 2017.
\newblock \showarticletitle{{Social media and fake news in the 2016 election}}.
\newblock {\em Journal of Economic Perspectives\/} {31}, 2 (2017), 211--236.
\newblock


\bibitem{andrews1995survey}
{Robert Andrews}, {Joachim Diederich}, {and} {Alan~B Tickle}. 1995.
\newblock \showarticletitle{Survey and critique of techniques for extracting
  rules from trained artificial neural networks}.
\newblock {\em Knowledge-based systems\/} {8}, 6 (1995), 373--389.
\newblock


\bibitem{bansal2019beyond}
{Gagan Bansal}, {Besmira Nushi}, {Ece Kamar}, {Walter~S Lasecki}, {Daniel~S
  Weld}, {and} {Eric Horvitz}. 2019.
\newblock \showarticletitle{Beyond Accuracy: The Role of Mental Models in
  Human-AI Team Performance}. In {\em Proceedings of the AAAI Conference on
  Human Computation and Crowdsourcing}, Vol.~7. 2--11.
\newblock


\bibitem{binns2018s}
{Reuben Binns}, {Max Van~Kleek}, {Michael Veale}, {Ulrik Lyngs}, {Jun Zhao},
  {and} {Nigel Shadbolt}. 2018.
\newblock \showarticletitle{'It's Reducing a Human Being to a Percentage':
  Perceptions of Justice in Algorithmic Decisions}. In {\em Proceedings of the
  2018 CHI Conference on Human Factors in Computing Systems}. ACM, 377.
\newblock


\bibitem{bond2006accuracy}
{Charles~F Bond~Jr} {and} {Bella~M DePaulo}. 2006.
\newblock \showarticletitle{Accuracy of deception judgments}.
\newblock {\em Personality and social psychology Review\/} {10}, 3 (2006),
  214--234.
\newblock


\bibitem{bussone2015role}
{Adrian Bussone}, {Simone Stumpf}, {and} {Dympna O'Sullivan}. 2015.
\newblock \showarticletitle{The role of explanations on trust and reliance in
  clinical decision support systems}. In {\em Healthcare Informatics (ICHI),
  2015 International Conference on}. IEEE, 160--169.
\newblock


\bibitem{cai2019human}
{Carrie~J Cai}, {Emily Reif}, {Narayan Hegde}, {Jason Hipp}, {Been Kim},
  {Daniel Smilkov}, {Martin Wattenberg}, {Fernanda Viegas}, {Greg~S Corrado},
  {Martin~C Stumpe}, {and} {others}. 2019.
\newblock \showarticletitle{Human-centered tools for coping with imperfect
  algorithms during medical decision-making}. In {\em Proceedings of the 2019
  CHI Conference on Human Factors in Computing Systems}. ACM, 4.
\newblock


\bibitem{caspi2006online}
{Avner Caspi} {and} {Paul Gorsky}. 2006.
\newblock \showarticletitle{Online deception: Prevalence, motivation, and
  emotion}.
\newblock {\em CyberPsychology \& Behavior\/} {9}, 1 (2006), 54--59.
\newblock


\bibitem{chevalier2006effect}
{Judith~A Chevalier} {and} {Dina Mayzlin}. 2006.
\newblock \showarticletitle{{The effect of word of mouth on sales: Online book
  reviews}}.
\newblock {\em Journal of marketing research\/} {43}, 3 (2006), 345--354.
\newblock


\bibitem{deng2009imagenet}
{Jia Deng}, {Wei Dong}, {Richard Socher}, {Li-Jia Li}, {Kai Li}, {and} {Li
  Fei-Fei}. 2009.
\newblock \showarticletitle{Imagenet: A large-scale hierarchical image
  database}. In {\em 2009 IEEE conference on computer vision and pattern
  recognition}. 248--255.
\newblock


\bibitem{devlin2018bert}
{Jacob Devlin}, {Ming-Wei Chang}, {Kenton Lee}, {and} {Kristina Toutanova}.
  2019.
\newblock \showarticletitle{BERT: Pre-training of Deep Bidirectional
  Transformers for Language Understanding}. In {\em Proceedings of NAACL}.
\newblock


\bibitem{doshi2017towards}
{Finale Doshi-Velez} {and} {Been Kim}. 2017.
\newblock \showarticletitle{Towards a rigorous science of interpretable machine
  learning}.
\newblock {\em arXiv preprint arXiv:1702.08608\/} (2017).
\newblock


\bibitem{farsetta2006fake}
{Diane Farsetta} {and} {Daniel Price}. 2006.
\newblock \showarticletitle{Fake TV news: Widespread and undisclosed}.
\newblock {\em Center for Media and Democracy\/}  {6} (2006).
\newblock


\bibitem{feng2012syntactic}
{Song Feng}, {Ritwik Banerjee}, {and} {Yejin Choi}. 2012.
\newblock \showarticletitle{Syntactic stylometry for deception detection}. In
  {\em Proceedings of ACL (short papers)}.
\newblock


\bibitem{feng2013detecting}
{Vanessa~Wei Feng} {and} {Graeme Hirst}. 2013.
\newblock \showarticletitle{Detecting deceptive opinions with profile
  compatibility}. In {\em Proceedings of IJCNLP}.
\newblock


\bibitem{green2019disparate}
{Ben Green} {and} {Yiling Chen}. 2019a.
\newblock \showarticletitle{Disparate interactions: An algorithm-in-the-loop
  analysis of fairness in risk assessments}. In {\em Proceedings of the
  Conference on Fairness, Accountability, and Transparency}. ACM, 90--99.
\newblock


\bibitem{green2019principles}
{Ben Green} {and} {Yiling Chen}. 2019b.
\newblock \showarticletitle{The principles and limits of algorithm-in-the-loop
  decision making}.
\newblock {\em Proceedings of the ACM on Human-Computer Interaction\/} {3},
  CSCW (2019), 50.
\newblock


\bibitem{grinberg2019fake}
{Nir Grinberg}, {Kenneth Joseph}, {Lisa Friedland}, {Briony Swire-Thompson},
  {and} {David Lazer}. 2019.
\newblock \showarticletitle{Fake news on Twitter during the 2016 US
  presidential election}.
\newblock {\em Science\/} {363}, 6425 (2019), 374--378.
\newblock


\bibitem{guidotti2018lore}
{Riccardo Guidotti}, {Anna Monreale}, {Salvatore Ruggieri}, {Dino Pedreschi},
  {Franco Turini}, {and} {Fosca Giannotti}. 2018.
\newblock \showarticletitle{Local rule-based explanations of black box decision
  systems}.
\newblock {\em arXiv preprint arXiv:1805.10820\/} (2018).
\newblock


\bibitem{guidotti2019survey}
{Riccardo Guidotti}, {Anna Monreale}, {Salvatore Ruggieri}, {Franco Turini},
  {Fosca Giannotti}, {and} {Dino Pedreschi}. 2019.
\newblock \showarticletitle{A survey of methods for explaining black box
  models}.
\newblock {\em ACM computing surveys (CSUR)\/} {51}, 5 (2019), 93.
\newblock


\bibitem{he2015delving}
{Kaiming He}, {Xiangyu Zhang}, {Shaoqing Ren}, {and} {Jian Sun}. 2015.
\newblock \showarticletitle{Delving deep into rectifiers: Surpassing
  human-level performance on imagenet classification}. In {\em Proceedings of
  ICCV}.
\newblock


\bibitem{horne2019rating}
{Benjamin~D Horne}, {Dorit Nevo}, {John O'Donovan}, {Jin-Hee Cho}, {and} {Sibel
  Adali}. 2019.
\newblock \showarticletitle{Rating Reliability and Bias in News Articles: Does
  AI Assistance Help Everyone?}. In {\em Proceedings of ICWSM}.
\newblock


\bibitem{jindal2008opinion}
{Nitin Jindal} {and} {Bing Liu}. 2008.
\newblock \showarticletitle{Opinion spam and analysis}. In {\em Proceedings of
  WSDM}.
\newblock


\bibitem{kang2016spaced}
{Sean~HK Kang}. 2016.
\newblock \showarticletitle{Spaced repetition promotes efficient and effective
  learning: Policy implications for instruction}.
\newblock {\em Policy Insights from the Behavioral and Brain Sciences\/} {3}, 1
  (2016), 12--19.
\newblock


\bibitem{kim2016examples}
{Been Kim}, {Rajiv Khanna}, {and} {Oluwasanmi~O Koyejo}. 2016.
\newblock \showarticletitle{Examples are not enough, learn to criticize!
  criticism for interpretability}. In {\em Proceedings of NIPS}.
\newblock


\bibitem{kim2014bayesian}
{Been Kim}, {Cynthia Rudin}, {and} {Julie~A Shah}. 2014.
\newblock \showarticletitle{The bayesian case model: A generative approach for
  case-based reasoning and prototype classification}. In {\em Proceedings of
  NIPS}.
\newblock


\bibitem{kleinberg2017human}
{Jon Kleinberg}, {Himabindu Lakkaraju}, {Jure Leskovec}, {Jens Ludwig}, {and}
  {Sendhil Mullainathan}. 2017.
\newblock \showarticletitle{Human decisions and machine predictions}.
\newblock {\em The Quarterly Journal of Economics\/} {133}, 1 (2017), 237--293.
\newblock


\bibitem{kleinberg2015prediction}
{Jon Kleinberg}, {Jens Ludwig}, {Sendhil Mullainathan}, {and} {Ziad Obermeyer}.
  2015.
\newblock \showarticletitle{Prediction policy problems}.
\newblock {\em American Economic Review\/} {105}, 5 (2015), 491--95.
\newblock


\bibitem{krause2014submodular}
{Andreas Krause} {and} {Daniel Golovin}. 2014.
\newblock Submodular function maximization.
\newblock   (2014).
\newblock


\bibitem{krauss1976modalities}
{Robert~M Krauss}, {Valerie Geller}, {and} {Christopher Olson}. 1976.
\newblock \showarticletitle{Modalities and cues in the detection of deception}.
  In {\em Meeting of the American Psychological Association, Washington, DC}.
\newblock


\bibitem{kunkel2019let}
{Johannes Kunkel}, {Tim Donkers}, {Lisa Michael}, {Catalin-Mihai Barbu}, {and}
  {J{\"u}rgen Ziegler}. 2019.
\newblock \showarticletitle{Let Me Explain: Impact of Personal and Impersonal
  Explanations on Trust in Recommender Systems}. In {\em Proceedings of the
  2019 CHI Conference on Human Factors in Computing Systems}. ACM, 487.
\newblock


\bibitem{lai+cai+tan:19}
{Vivian Lai}, {Jon~Z. Cai}, {and} {Chenhao Tan}. 2019.
\newblock \showarticletitle{Many Faces of Feature Importance: Comparing
  Built-in and Post-hoc Feature Importance in Text Classification}. In {\em
  Proceedings of EMNLP}.
\newblock


\bibitem{lai+tan:19}
{Vivian Lai} {and} {Chenhao Tan}. 2019.
\newblock \showarticletitle{On Human Predictions with Explanations and
  Predictions of Machine Learning Models: A Case Study on Deception Detection}.
  In {\em Proceedings of FAT*}.
\newblock


\bibitem{lazer2018science}
{David~MJ Lazer}, {Matthew~A Baum}, {Yochai Benkler}, {Adam~J Berinsky},
  {Kelly~M Greenhill}, {Filippo Menczer}, {Miriam~J Metzger}, {Brendan Nyhan},
  {Gordon Pennycook}, {David Rothschild}, {Michael Schudson}, {Steven~A.
  Sloman}, {Cass~R. Sunstein}, {Emily~A. Thorson}, {Duncan~J. Watts}, {and}
  {Jonathan~L. Zittrain}. 2018.
\newblock \showarticletitle{The science of fake news}.
\newblock {\em Science\/} {359}, 6380 (2018), 1094--1096.
\newblock


\bibitem{lee2004trust}
{John~D Lee} {and} {Katrina~A See}. 2004.
\newblock \showarticletitle{Trust in automation: Designing for appropriate
  reliance}.
\newblock {\em Human factors\/} {46}, 1 (2004), 50--80.
\newblock


\bibitem{lei2016rationalizing}
{Tao Lei}, {Regina Barzilay}, {and} {Tommi Jaakkola}. 2016.
\newblock \showarticletitle{Rationalizing neural predictions}. {\em Proceedings
  of EMNLP\/} (2016).
\newblock


\bibitem{lewandowsky2000dynamics}
{Stephan Lewandowsky}, {Michael Mundy}, {and} {Gerard Tan}. 2000.
\newblock \showarticletitle{The dynamics of trust: Comparing humans to
  automation.}
\newblock {\em Journal of Experimental Psychology: Applied\/} {6}, 2 (2000),
  104.
\newblock


\bibitem{li2014towards}
{Jiwei Li}, {Myle Ott}, {Claire Cardie}, {and} {Eduard Hovy}. 2014.
\newblock \showarticletitle{Towards a general rule for identifying deceptive
  opinion spam}. In {\em Proceedings of the 52nd Annual Meeting of the
  Association for Computational Linguistics (Volume 1: Long Papers)}.
  1566--1576.
\newblock


\bibitem{nytimes}
{Adam Liptak}. 2017.
\newblock Sent to Prison by a Software Program's Secret Algorithms.
\newblock   (2017).
\newblock


\bibitem{lipton2016mythos}
{Zachary~C Lipton}. 2016.
\newblock \showarticletitle{The mythos of model interpretability}.
\newblock {\em arXiv preprint arXiv:1606.03490\/} (2016).
\newblock


\bibitem{lundberg2017unified}
{Scott~M Lundberg} {and} {Su-In Lee}. 2017.
\newblock \showarticletitle{A unified approach to interpreting model
  predictions}. In {\em Proceedings of NIPS}.
\newblock


\bibitem{mothilal2019explaining}
{Ramaravind~Kommiya Mothilal}, {Amit Sharma}, {and} {Chenhao Tan}. 2020.
\newblock \showarticletitle{Explaining Machine Learning Classifiers through
  Diverse Counterfactual Explanations}. In {\em Proceedings of FAT*}.
\newblock


\bibitem{nielsen2002getting}
{Janni Nielsen}, {Torkil Clemmensen}, {and} {Carsten Yssing}. 2002.
\newblock \showarticletitle{Getting access to what goes on in people's heads?:
  reflections on the think-aloud technique}. In {\em Proceedings of the second
  Nordic conference on Human-computer interaction}. ACM, 101--110.
\newblock


\bibitem{ott2012estimating}
{Myle Ott}, {Claire Cardie}, {and} {Jeff Hancock}. 2012.
\newblock \showarticletitle{Estimating the prevalence of deception in online
  review communities}. In {\em Proceedings of WWW}.
\newblock


\bibitem{ott2013negative}
{Myle Ott}, {Claire Cardie}, {and} {Jeffrey~T Hancock}. 2013.
\newblock \showarticletitle{Negative deceptive opinion spam}. In {\em
  Proceedings of NAACL}.
\newblock


\bibitem{ott2011finding}
{Myle Ott}, {Yejin Choi}, {Claire Cardie}, {and} {Jeffrey~T Hancock}. 2011.
\newblock \showarticletitle{Finding deceptive opinion spam by any stretch of
  the imagination}. In {\em Proceedings of ACL}.
\newblock


\bibitem{perez2015verbal}
{Ver{\'o}nica P{\'e}rez-Rosas}, {Mohamed Abouelenien}, {Rada Mihalcea}, {Yao
  Xiao}, {CJ Linton}, {and} {Mihai Burzo}. 2015.
\newblock \showarticletitle{Verbal and nonverbal clues for real-life deception
  detection}. In {\em Proceedings of the 2015 Conference on Empirical Methods
  in Natural Language Processing}. 2336--2346.
\newblock


\bibitem{poursabzi2018manipulating}
{Forough Poursabzi-Sangdeh}, {Daniel~G Goldstein}, {Jake~M Hofman},
  {Jennifer~Wortman Vaughan}, {and} {Hanna Wallach}. 2018.
\newblock \showarticletitle{Manipulating and measuring model interpretability}.
\newblock {\em arXiv preprint arXiv:1802.07810\/} (2018).
\newblock


\bibitem{ribeiro2016should}
{Marco~Tulio Ribeiro}, {Sameer Singh}, {and} {Carlos Guestrin}. 2016.
\newblock \showarticletitle{Why should i trust you?: Explaining the predictions
  of any classifier}. In {\em Proceedings of KDD}.
\newblock


\bibitem{anchors:aaai18}
{Marco~Tulio Ribeiro}, {Sameer Singh}, {and} {Carlos Guestrin}. 2018.
\newblock \showarticletitle{Anchors: High-Precision Model-Agnostic
  Explanations}. In {\em Proceedings of AAAI}.
\newblock


\bibitem{russell2019efficient}
{Chris Russell}. 2019.
\newblock \showarticletitle{Efficient Search for Diverse Coherent
  Explanations}. In {\em Proceedings of FAT*}.
\newblock


\bibitem{shin2011prevalence}
{Youngsang Shin}, {Minaxi Gupta}, {and} {Steven Myers}. 2011.
\newblock \showarticletitle{Prevalence and mitigation of forum spamming}. In
  {\em 2011 Proceedings IEEE INFOCOM}. IEEE, 2309--2317.
\newblock


\bibitem{wicourts}
{{Supreme Court of Wisconsin}}. 2016.
\newblock {State of Wisconsin, Plaintiff-Respondent, v. Eric L. Loomis,
  Defendant-Appellant}.
\newblock   (2016).
\newblock


\bibitem{tabibian2019enhancing}
{Behzad Tabibian}, {Utkarsh Upadhyay}, {Abir De}, {Ali Zarezade}, {Bernhard
  Sch{\"o}lkopf}, {and} {Manuel Gomez-Rodriguez}. 2019.
\newblock \showarticletitle{Enhancing human learning via spaced repetition
  optimization}.
\newblock {\em Proceedings of the National Academy of Sciences\/} {116}, 10
  (2019), 3988--3993.
\newblock


\bibitem{tan2014effect}
{Chenhao Tan}, {Lillian Lee}, {and} {Bo Pang}. 2014.
\newblock \showarticletitle{The effect of wording on message propagation:
  Topic-and author-controlled natural experiments on Twitter}. In {\em
  Proceedings of ACL}.
\newblock


\bibitem{trusov2009effects}
{Michael Trusov}, {Randolph~E Bucklin}, {and} {Koen Pauwels}. 2009.
\newblock \showarticletitle{Effects of word-of-mouth versus traditional
  marketing: findings from an internet social networking site}.
\newblock {\em Journal of marketing\/} {73}, 5 (2009), 90--102.
\newblock


\bibitem{vaswani2017attention}
{Ashish Vaswani}, {Noam Shazeer}, {Niki Parmar}, {Jakob Uszkoreit}, {Llion
  Jones}, {Aidan~N Gomez}, {{\L}ukasz Kaiser}, {and} {Illia Polosukhin}. 2017.
\newblock \showarticletitle{Attention is all you need}. In {\em Proceedings of
  NeurIPS}.
\newblock


\bibitem{vosoughi2018spread}
{Soroush Vosoughi}, {Deb Roy}, {and} {Sinan Aral}. 2018.
\newblock \showarticletitle{The spread of true and false news online}.
\newblock {\em Science\/} {359}, 6380 (2018), 1146--1151.
\newblock


\bibitem{vrij2000detecting}
{Aldert Vrij}. 2000.
\newblock {\em Detecting lies and deceit: The psychology of lying and
  implications for professional practice}.
\newblock Wiley.
\newblock


\bibitem{wachter2017counterfactual}
{Sandra Wachter}, {Brent Mittelstadt}, {and} {Chris Russell}. 2017.
\newblock \showarticletitle{Counterfactual explanations without opening the
  black box: Automated decisions and the GDPR}.
\newblock


\bibitem{weerts2019human}
{Hilde~JP Weerts}, {Werner van Ipenburg}, {and} {Mykola Pechenizkiy}. 2019.
\newblock \showarticletitle{A Human-Grounded Evaluation of SHAP for Alert
  Processing}.
\newblock {\em arXiv preprint arXiv:1907.03324\/} (2019).
\newblock


\bibitem{wu2010distortion}
{Guangyu Wu}, {Derek Greene}, {Barry Smyth}, {and} {P{\'a}draig Cunningham}.
  2010.
\newblock \showarticletitle{Distortion as a validation criterion in the
  identification of suspicious reviews}. In {\em Proceedings of the First
  Workshop on Social Media Analytics}.
\newblock


\bibitem{ye2011influence}
{Qiang Ye}, {Rob Law}, {Bin Gu}, {and} {Wei Chen}. 2011.
\newblock \showarticletitle{The influence of user-generated content on traveler
  behavior: An empirical investigation on the effects of e-word-of-mouth to
  hotel online bookings}.
\newblock {\em Computers in Human behavior\/} {27}, 2 (2011), 634--639.
\newblock


\bibitem{yin2019understanding}
{Ming Yin}, {Jennifer Wortman~Vaughan}, {and} {Hanna Wallach}. 2019.
\newblock \showarticletitle{Understanding the Effect of Accuracy on Trust in
  Machine Learning Models}. In {\em Proceedings of the 2019 CHI Conference on
  Human Factors in Computing Systems}. ACM, 279.
\newblock


\bibitem{yoo2009comparison}
{Kyung-Hyan Yoo} {and} {Ulrike Gretzel}. 2009.
\newblock \showarticletitle{Comparison of deceptive and truthful travel
  reviews}.
\newblock {\em Information and communication technologies in tourism 2009\/}
  (2009), 37--47.
\newblock


\bibitem{zhang2018structured}
{Amy~X Zhang}, {Aditya Ranganathan}, {Sarah~Emlen Metz}, {Scott Appling},
  {Connie~Moon Sehat}, {Norman Gilmore}, {Nick~B Adams}, {Emmanuel Vincent},
  {Martin Robbins}, {Ed Bice}, {Sandro Hawke}, {David Karger}, {and} {An~Xiao
  Mina}. 2018.
\newblock \showarticletitle{A Structured Response to Misinformation: Defining
  and Annotating Credibility Indicators in News Articles}. In {\em Proceedings
  of WWW (Companion)}.
\newblock


\bibitem{zhang2010impact}
{Ziqiong Zhang}, {Qiang Ye}, {Rob Law}, {and} {Yijun Li}. 2010.
\newblock \showarticletitle{The impact of e-word-of-mouth on the online
  popularity of restaurants: A comparison of consumer reviews and editor
  reviews}.
\newblock {\em International Journal of Hospitality Management\/} {29}, 4
  (2010), 694--700.
\newblock


\end{thebibliography}

\newpage
\appendix

\section{Preview of the supplementary video}

You can skip the screen shots of tutorial interfaces if you choose to watch the supplementary video.
To help you skim the video, here are the starting time for each type of tutorials:

\begin{itemize}[itemsep=0pt]
    \item {Guidelines}: 00:08
    \item {Random:} 00:19
    \item {SP-LIME:} 00:53
    \item {Spaced repetition}: 01:27
    \item {SR + guidelines:} 02:01
    \item {BERT + attention:} 02:41
    \item {BERT + LIME:} 03:15
\end{itemize}

\vspace{1in}

\section{Experiment Interfaces}

\begin{figure}[H]
  \centering
  \includegraphics[width=0.47\textwidth]{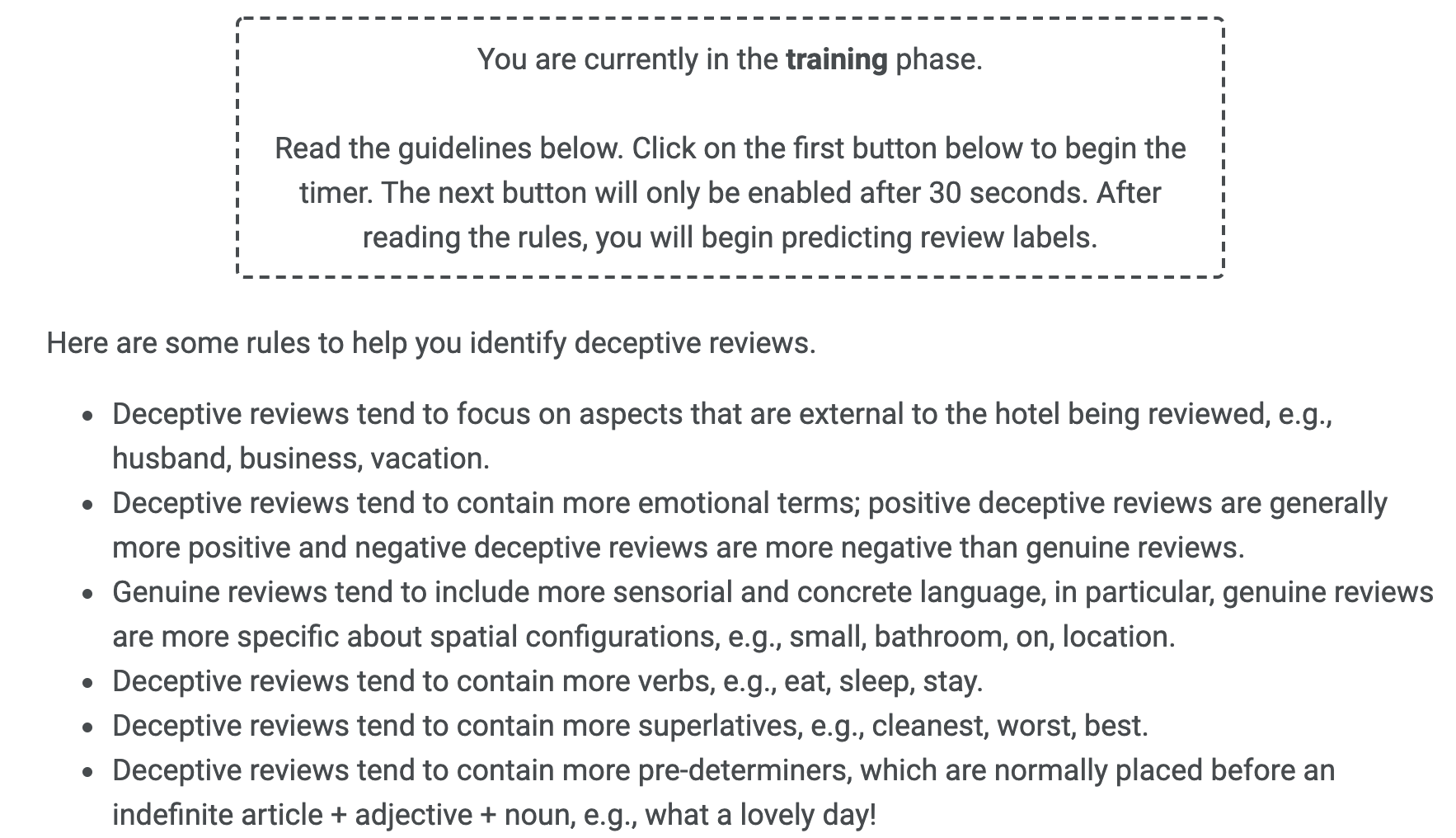}
  \caption{Experiment 1 tutorial: guidelines.}
  \label{fig:exp1-guidelines}
\end{figure}

\begin{figure}[H]
  \centering
  \includegraphics[width=0.47\textwidth]{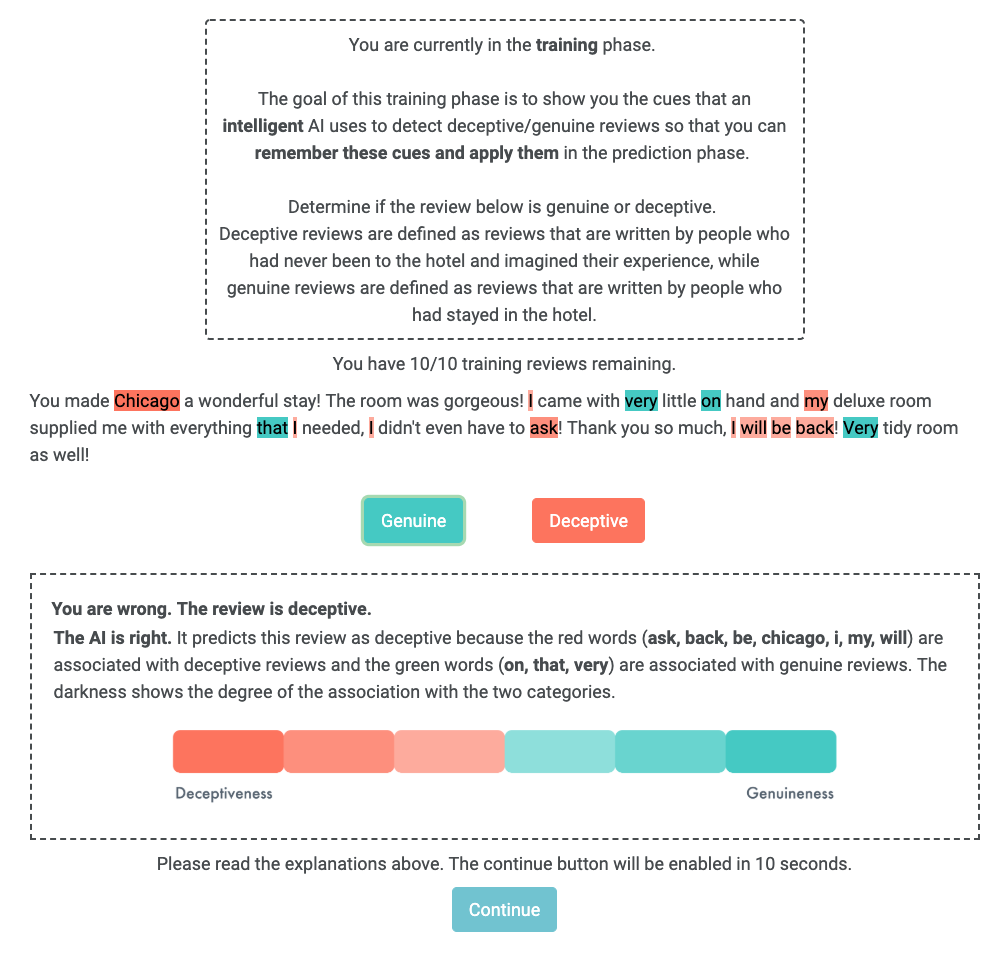}
  \caption{Experiment 1 tutorial: selected examples. Selected examples of {\em random}, {\em SP-LIME}, and {\em SR} are captured in video submission.}
  \label{fig:exp1-selected-examples}
\end{figure}

\begin{figure}[H]
  \centering
  \includegraphics[width=0.47\textwidth]{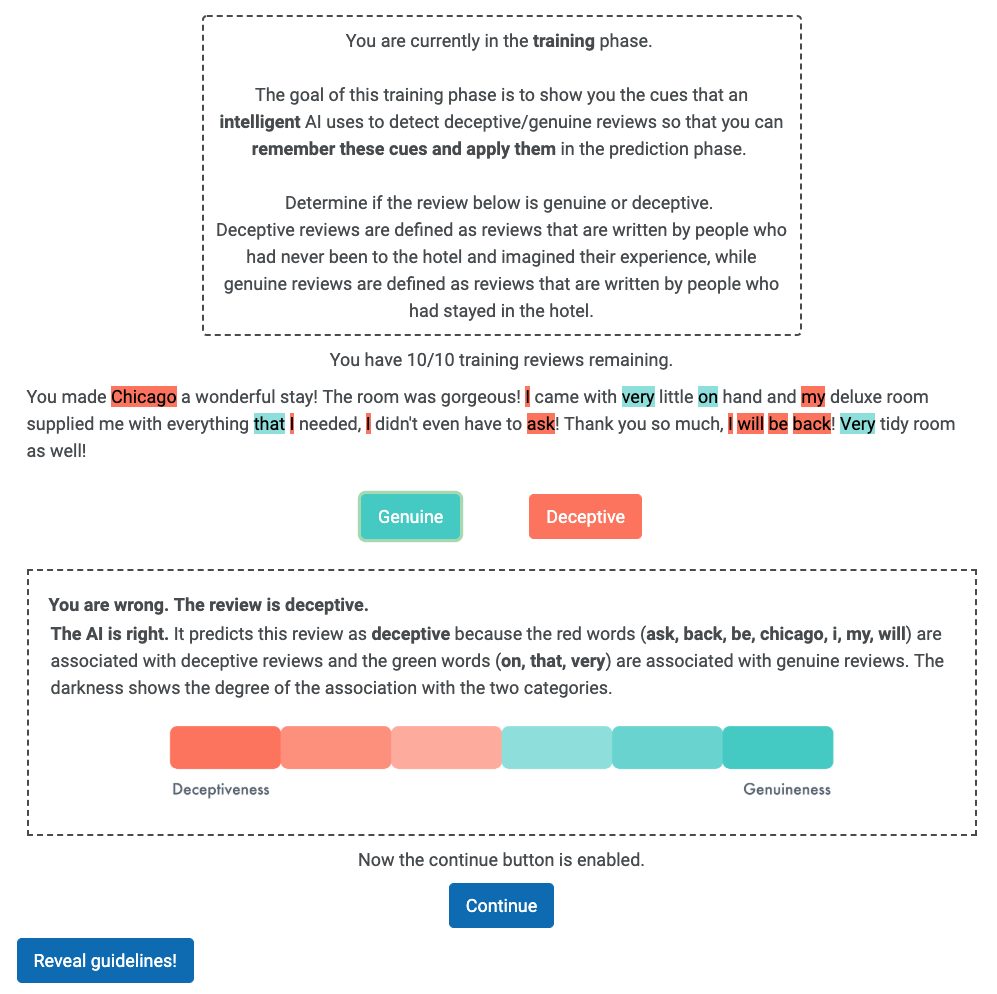}
  \caption{Experiment 1 tutorial: selected examples + guidelines. `Reveal guidelines' shows a list of guidelines as illustrated in \figref{fig:exp1-guidelines}.}
  \label{fig:exp1-selected-examples-guidelines}
\end{figure}

\figref{fig:exp1-guidelines} - \figref{fig:exp1-selected-examples-guidelines} shows tutorial interfaces for Experiment 1.

\begin{figure}[H]
  \centering
  \includegraphics[width=0.47\textwidth]{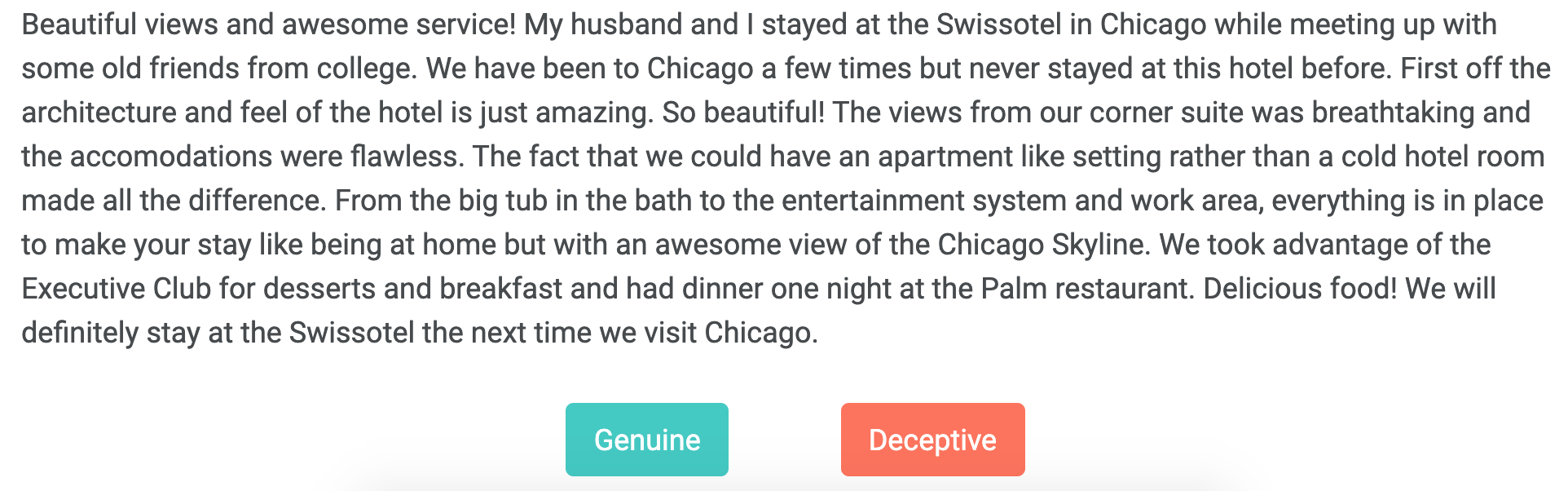}
  \caption{Experiment 2 real-time assistance: no assistance.}
  \label{fig:exp2-no-assistance}
\end{figure}

\begin{figure}[H]
  \centering
  \includegraphics[width=0.47\textwidth]{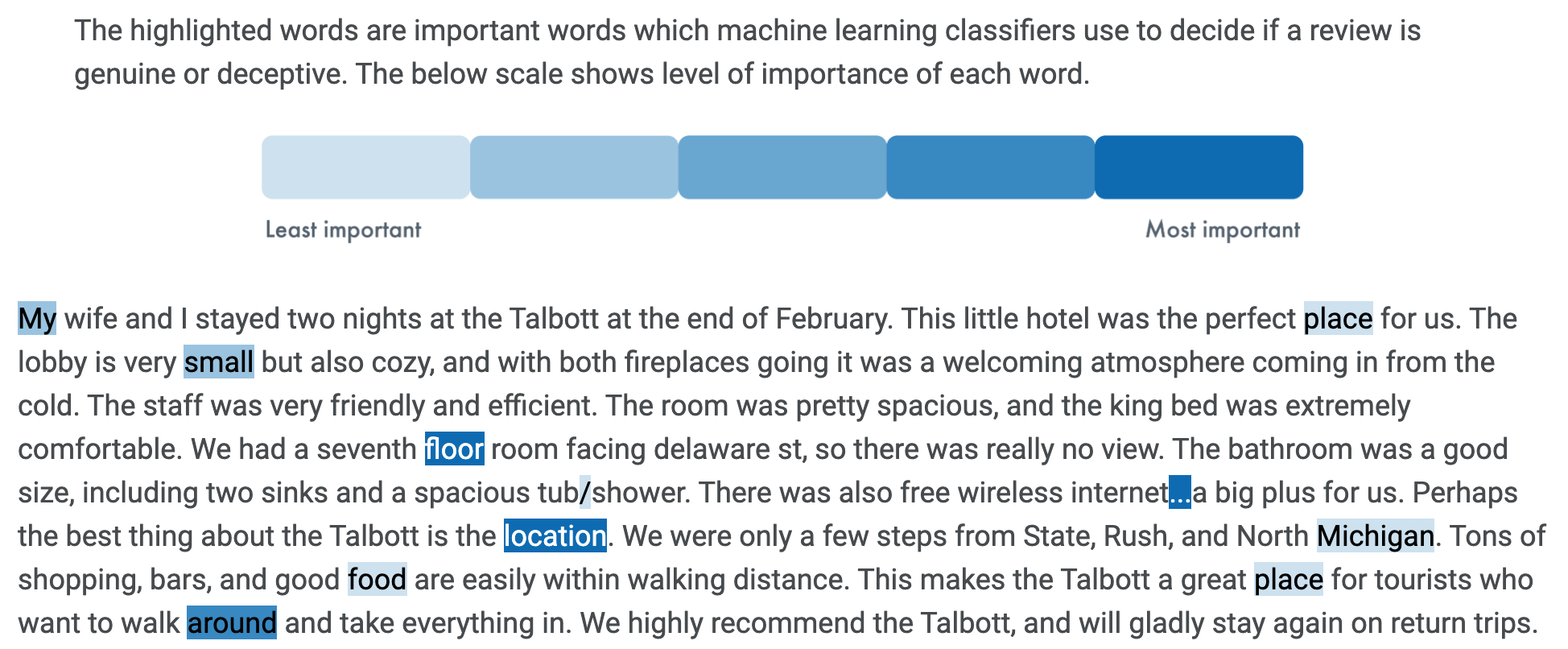}
  \caption{Experiment 2 real-time assistance: unsigned highlights.}
  \label{fig:exp2-unsigned-highlights}
\end{figure}

\begin{figure}[H]
  \centering
  \includegraphics[width=0.47\textwidth]{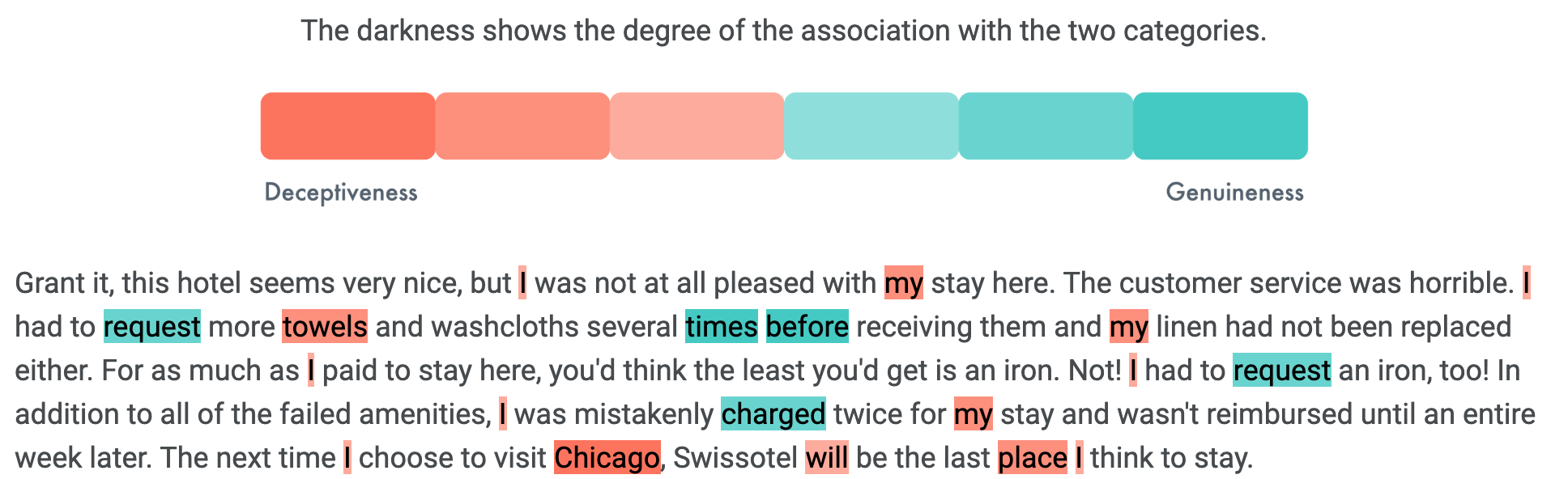}
  \caption{Experiment 2 real-time assistance: signed highlights.}
  \label{fig:exp2-signed-highlights}
\end{figure}

\begin{figure}[H]
  \centering
  \includegraphics[width=0.47\textwidth]{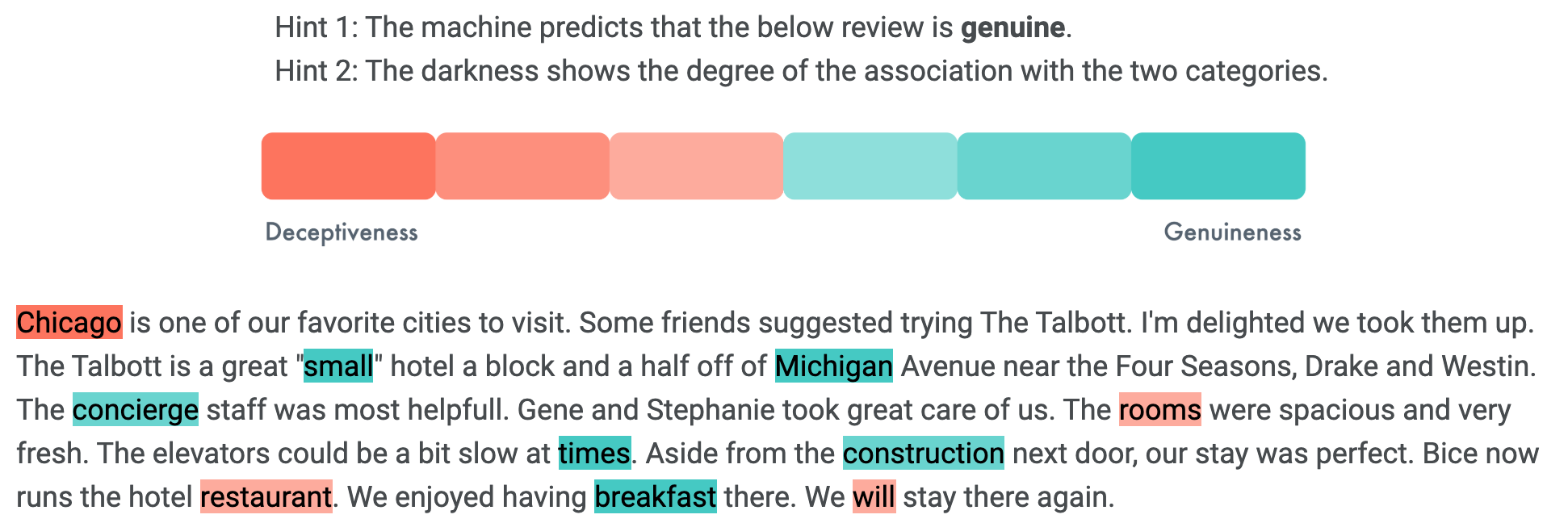}
  \caption{Experiment 2 real-time assistance: signed highlights + predicted label.}
  \label{fig:exp2-signed-highlights-pl}
\end{figure}

\begin{figure}[H]
  \centering
  \includegraphics[width=0.47\textwidth]{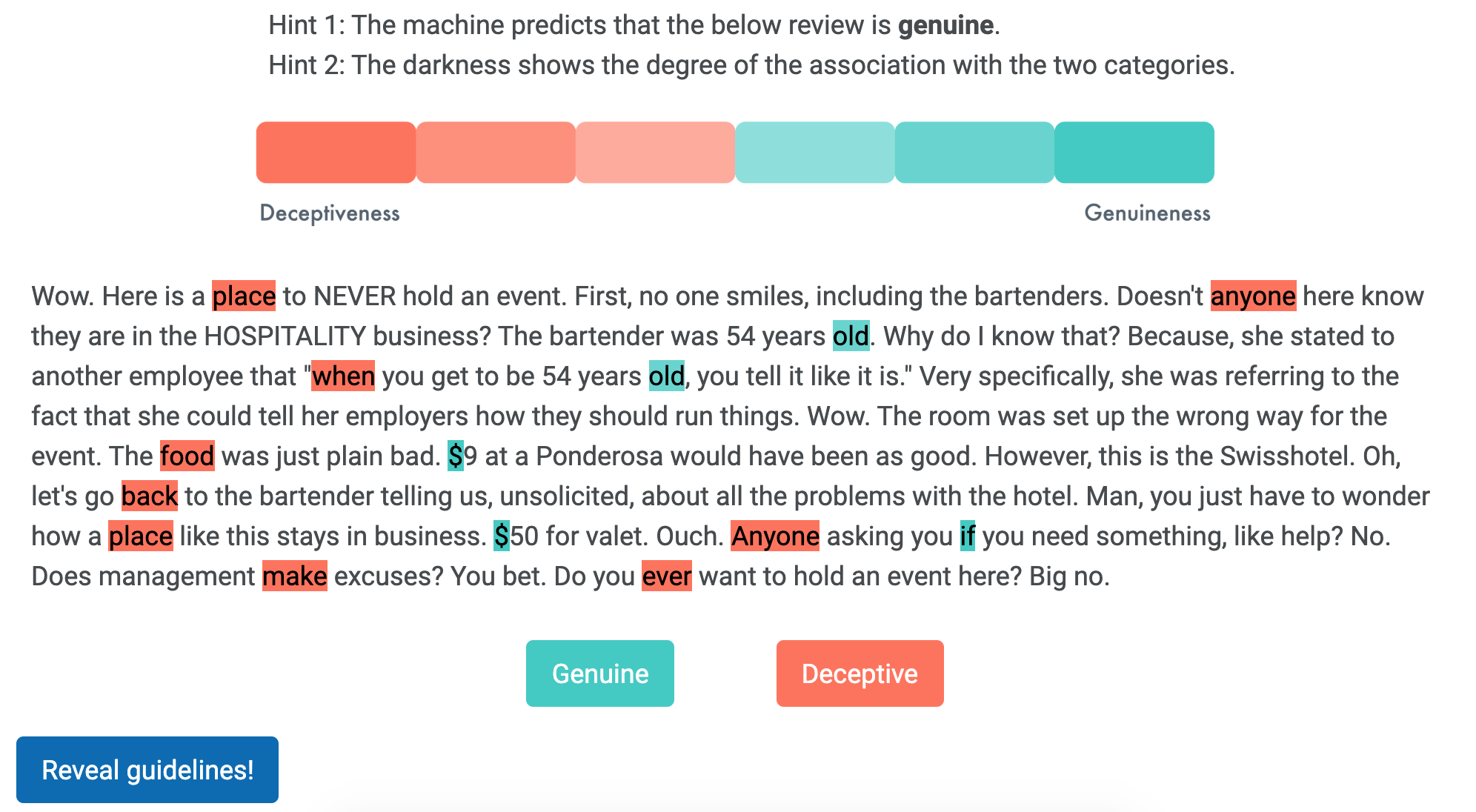}
  \caption{Experiment 2 real-time assistance: signed highlights + predicted label + guidelines. }
  \label{fig:exp2-signed-highlights-pl-guidelines-1}
\end{figure}

\begin{figure}[H]
  \centering
  \includegraphics[width=0.47\textwidth]{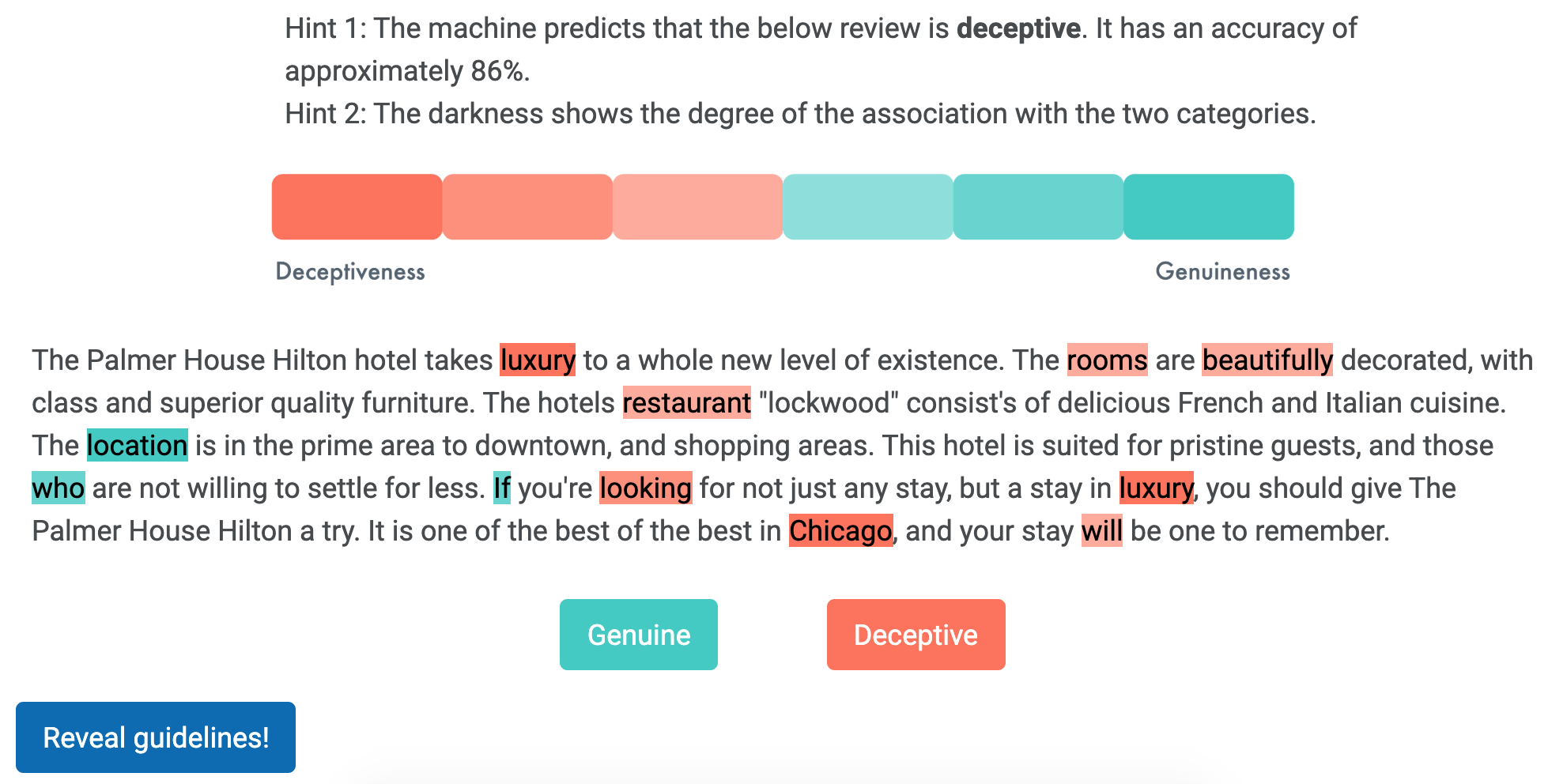}
  \caption{Experiment 2 real-time assistance: signed highlights + predicted label + guidelines + accuracy statement.}
  \label{fig:exp2-signed-highlights-pl-guidelines-acc}
\end{figure}

\figref{fig:exp2-no-assistance} - \figref{fig:exp2-signed-highlights-pl-guidelines-acc} shows the prediction phase interfaces for experiment 2.

\begin{figure}[H]
  \centering
  \includegraphics[width=0.47\textwidth]{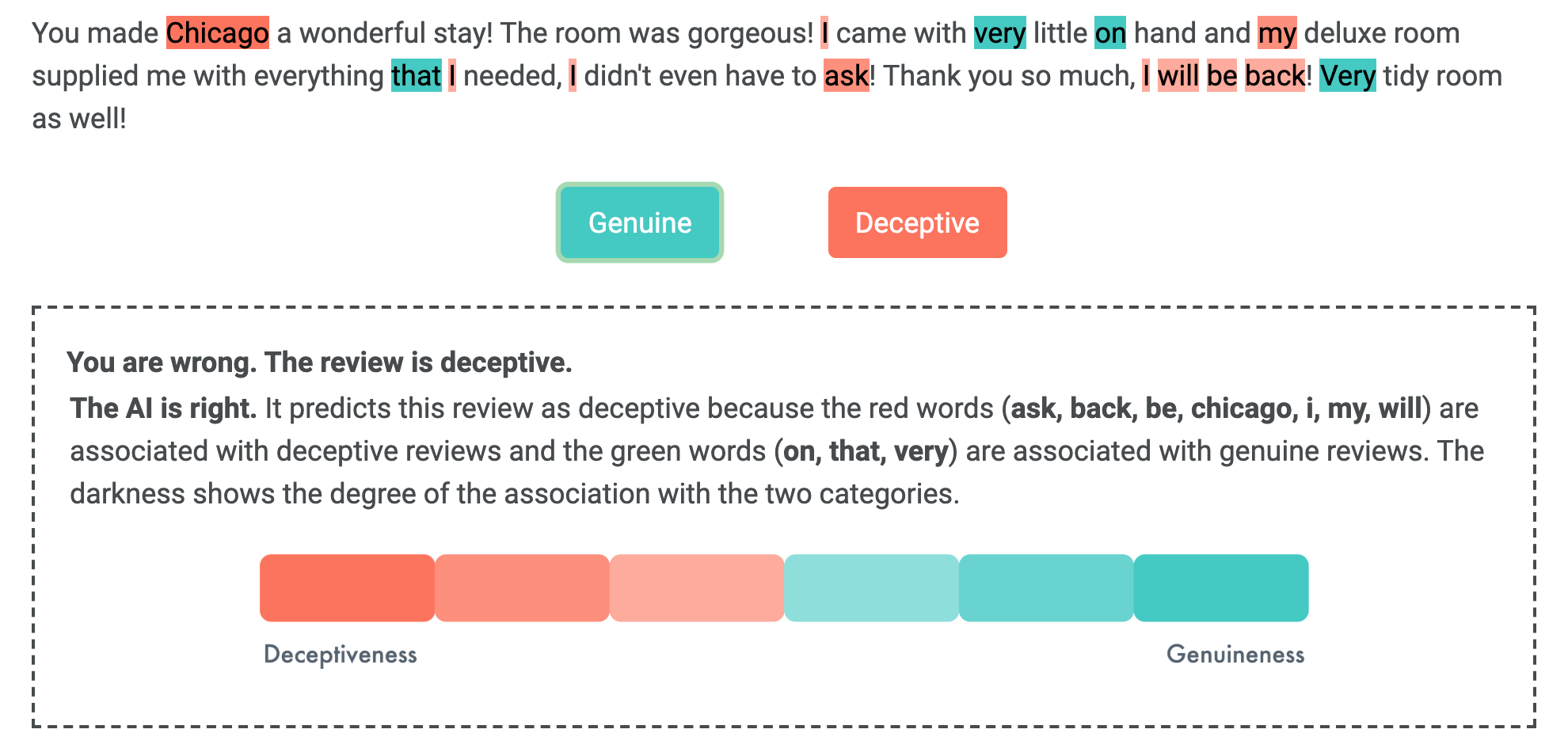}
  \caption{Experiment 3: top features from SVM are highlighted.}
  \label{fig:exp3-svm}
\end{figure}

\begin{figure}[H]
  \centering
  \includegraphics[width=0.47\textwidth]{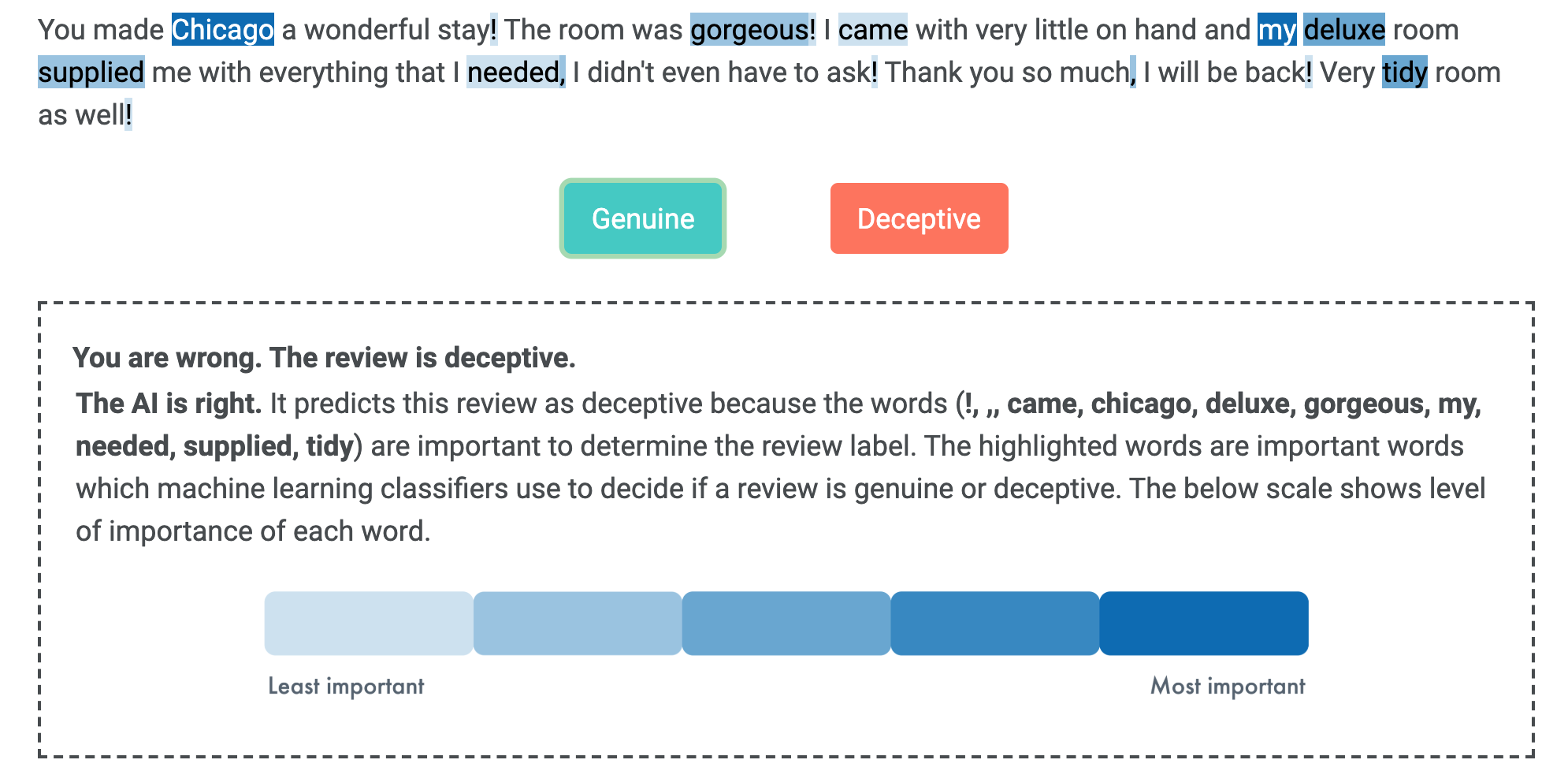}
  \caption{Experiment 3: top features from BERT attention are highlighted.}
  \label{fig:exp3-bert-att}
\end{figure}

\begin{figure}[H]
  \centering
  \includegraphics[width=0.47\textwidth]{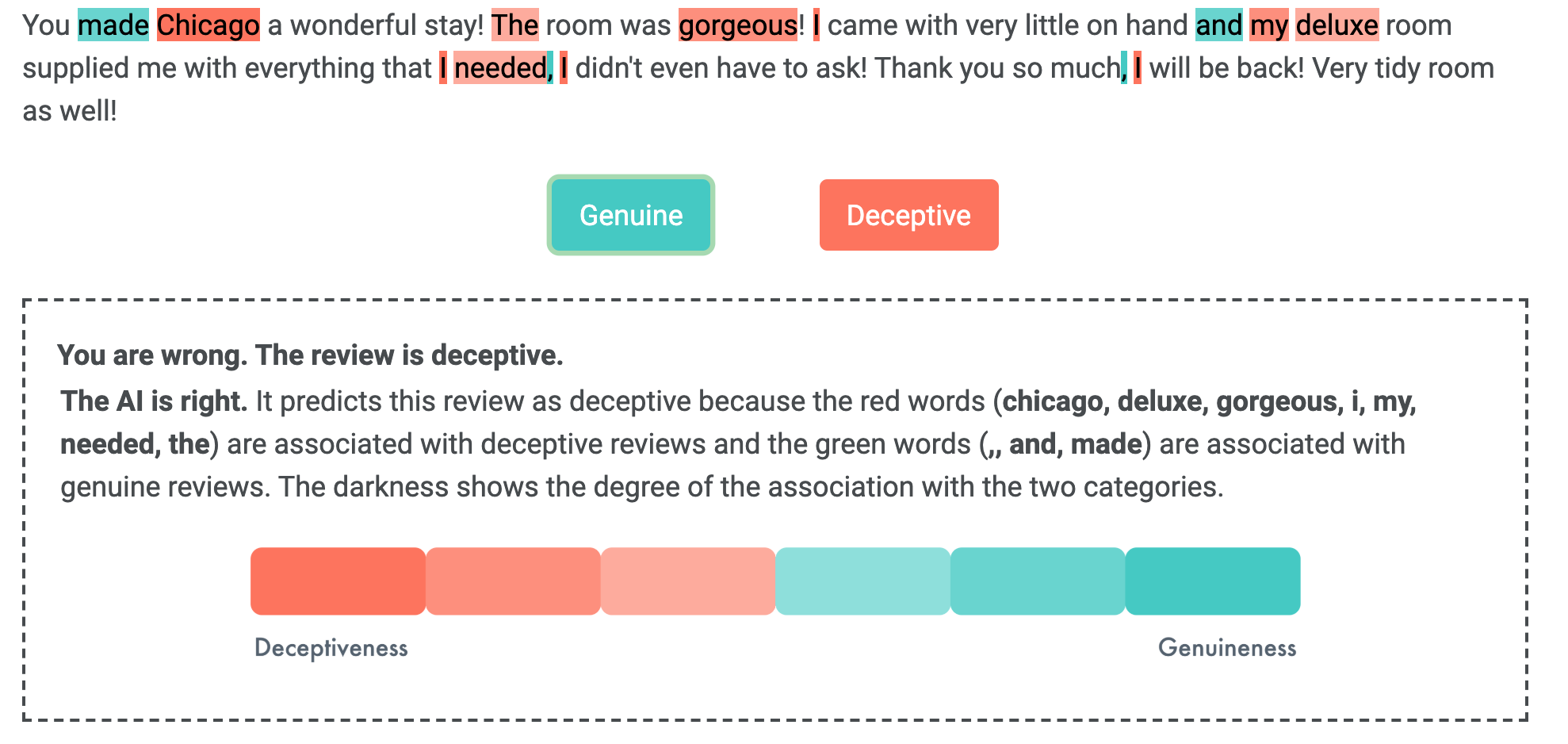}
  \caption{Experiment 3: top features from BERT LIME are highlighted.}
  \label{fig:exp3-bert-lime}
\end{figure}

\figref{fig:exp3-svm} - \figref{fig:exp3-bert-lime} shows examples in different methods deriving explanations for experiment 3.

\subsection{Experiment Details}

Among our participants in Experiment 1,
69 were between 18 and 25, 265 were between 26 and 40, 121 were between 41 and 60, 22 were 61 and above, and 3 preferred not to answer.
They had a range of education backgrounds, comprising some high school (3), high school graduate (54), some college credit (124), trade/technical/vocational training (42), Bachelor's degree and above (253), and 4 prefered not to answer.

Among our participants in Experiment 2, 64 were between 18 and 25, 270 were between 26 and 40, 116 were between 41 and 60, 26 were 61 and above, and 4 preferred not to answer.
They had a range of education backgrounds, comprising some high school (3), high school graduate (44), some college credit (120), trade/technical/vocational training (32), Bachelor's degree and above (278), and 3 prefered not to answer.

Among our participants in Experiment 3, 62 were between 18 and 25, 255 were between 26 and 40, 138 were between 41 and 60, 24 were 61 and above, and 1 preferred not to answer.
They had a range of educational attainment, comprising some high school (1), high school graduate (51), some college credit (111), trade/technical/vocational training (40), Bachelor's degree and above (274), and 3 prefered not to answer.

We only kept participants that complete the full task and submit a unique survey code.
Participants that do not comply with the criteria were not included.

\begin{figure}[H]
  \centering
  \includegraphics[width=0.4\textwidth]{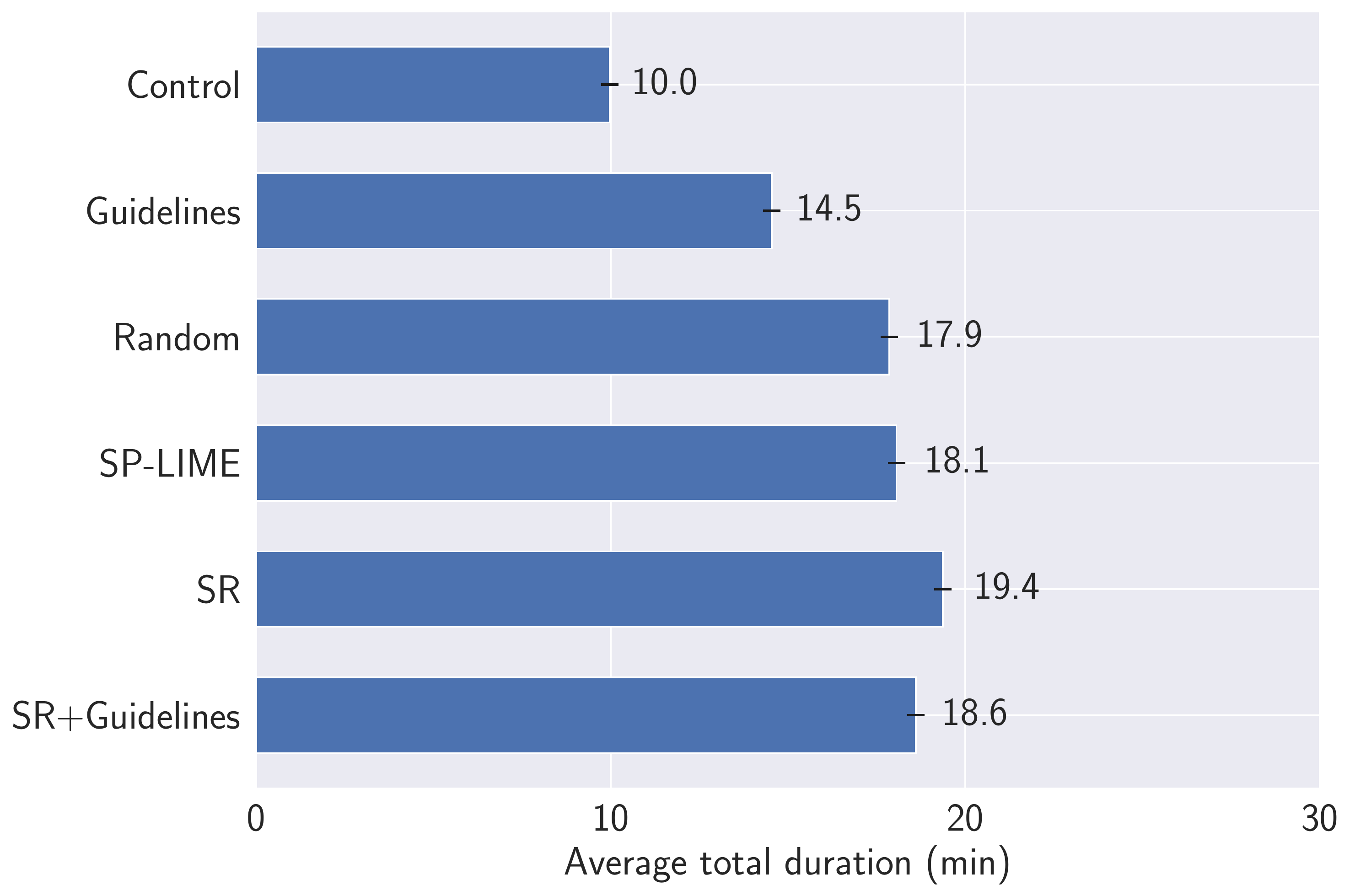}
  \caption{Average time taken for each experimental setup in experiment 1.}
  \label{fig:exp1-avg}
\end{figure}

\begin{figure}[H]
  \centering
  \includegraphics[width=0.4\textwidth]{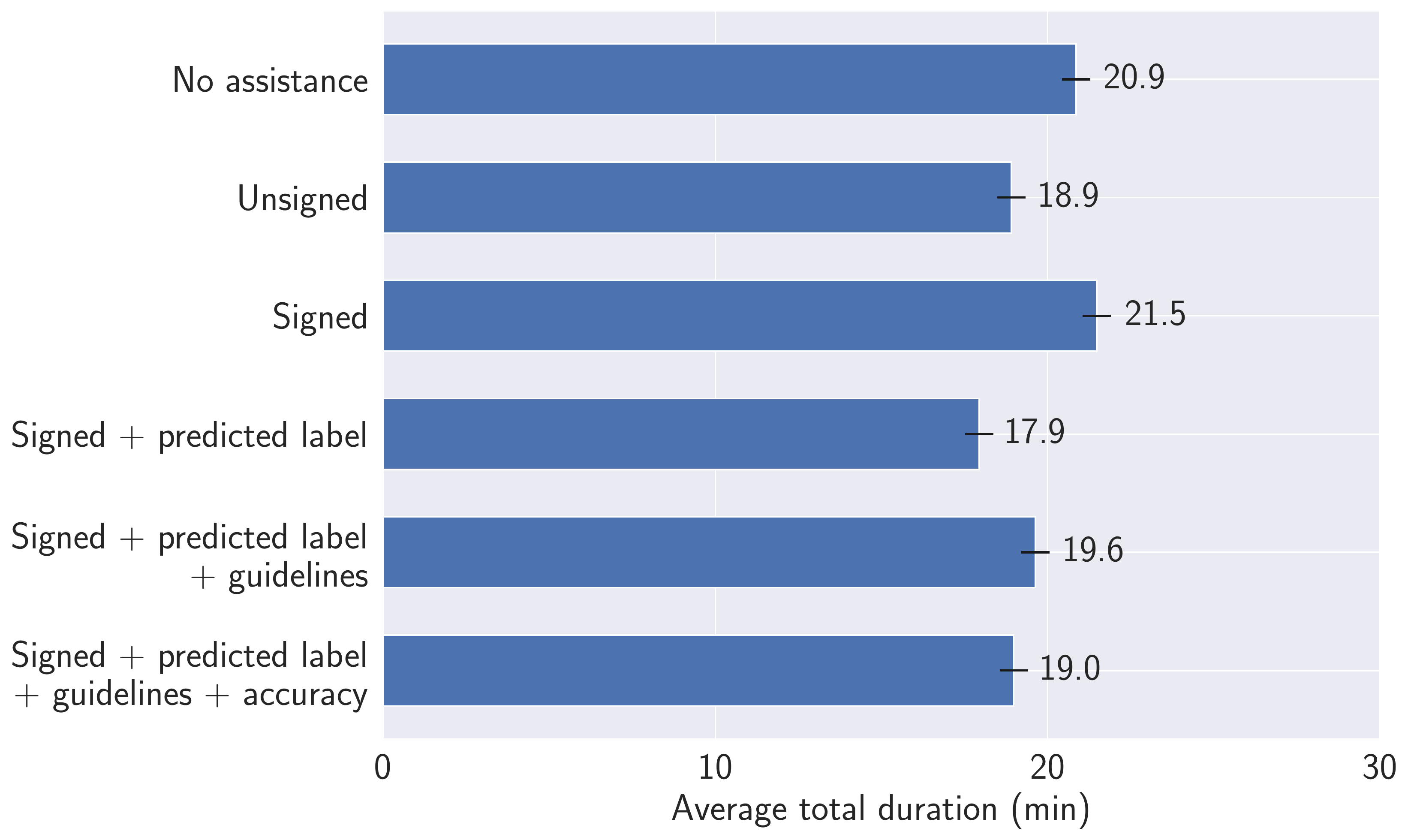}
  \caption{Average time taken for each experimental setup in experiment 2.}
  \label{fig:exp2-avg}
\end{figure}

\begin{figure}[H]
  \centering
  \includegraphics[width=0.4\textwidth]{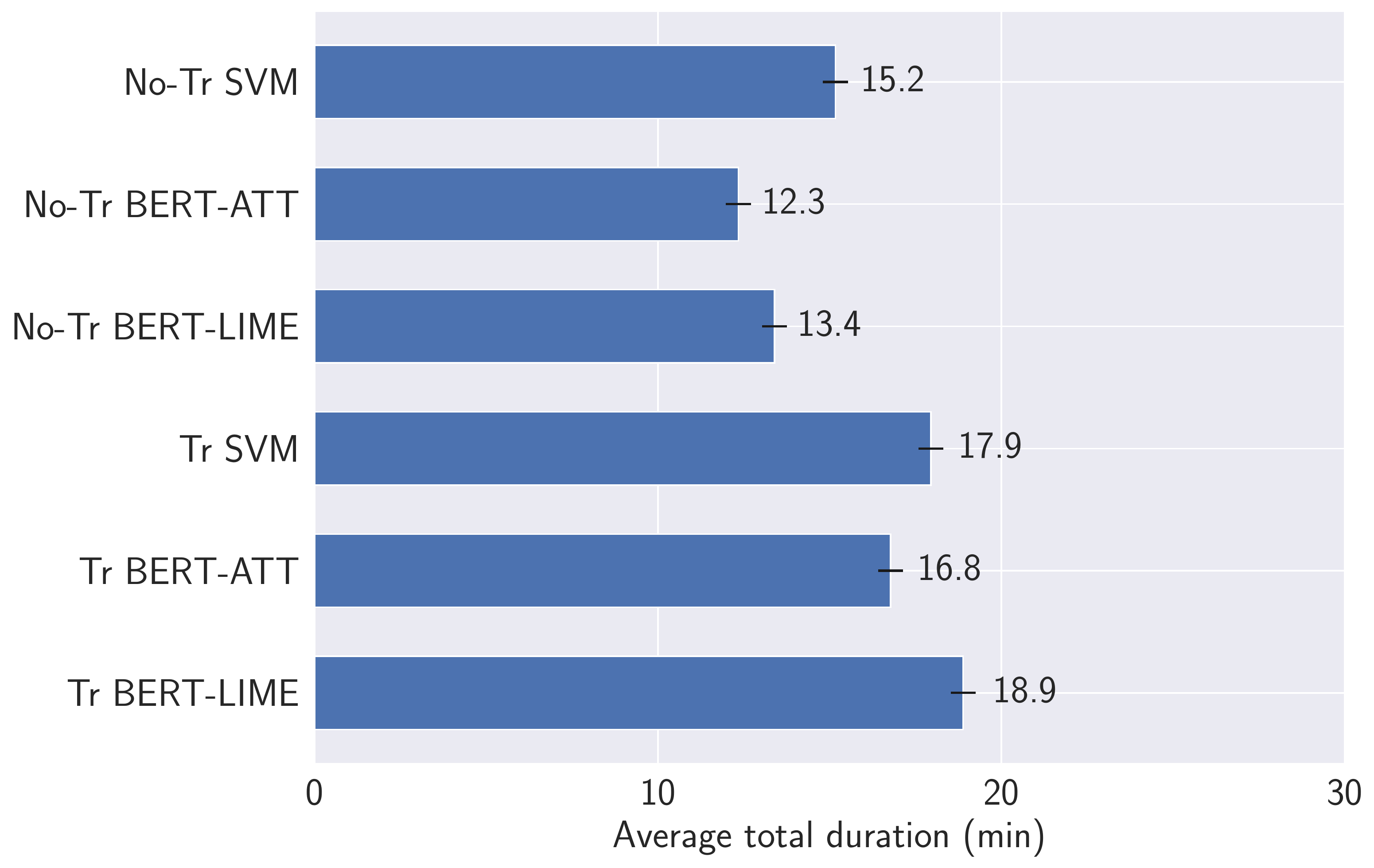}
  \caption{Average time taken for each experimental setup in experiment 3.}
  \label{fig:exp3-avg}
\end{figure}

\begin{figure}[H]
  \centering
  \includegraphics[width=0.4\textwidth]{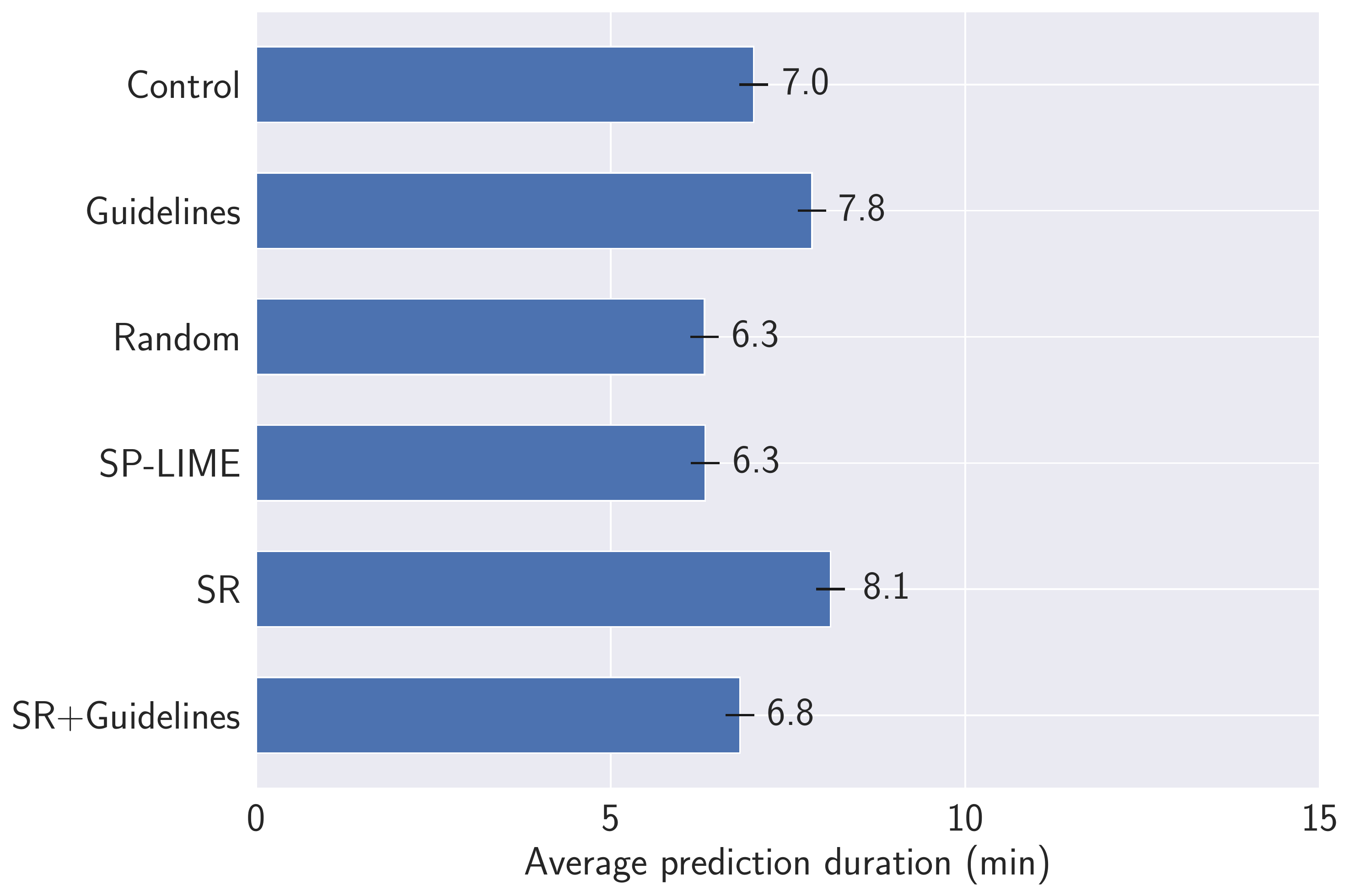}
  \caption{Average time taken for the prediction phase in each experimental setup in experiment 1.}
  \label{fig:exp1-time-distribution}
\end{figure}

\begin{figure}[H]
  \centering
  \includegraphics[width=0.4\textwidth]{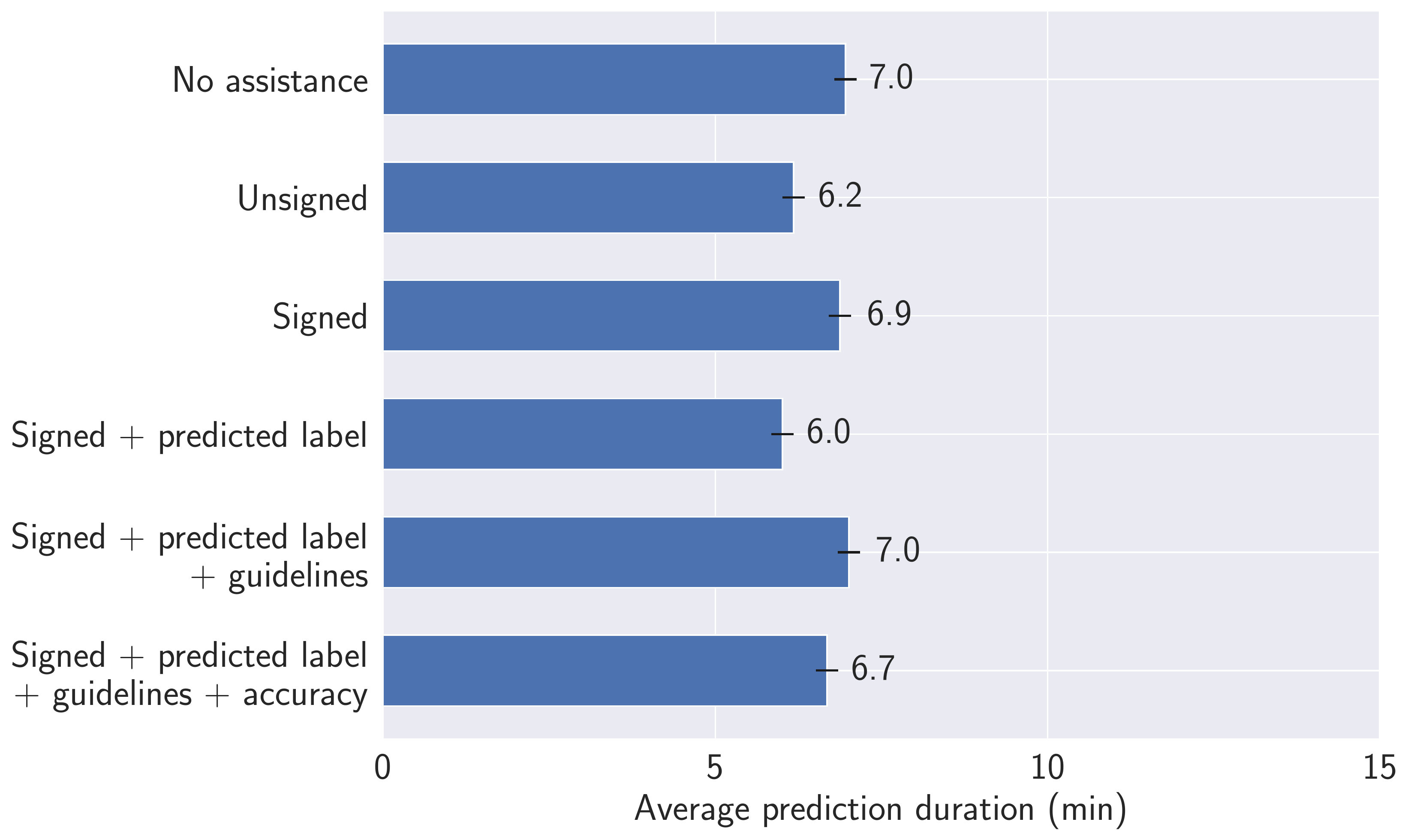}
  \caption{Average time taken for the prediction phase in each experimental setup in experiment 2.}
  \label{fig:exp2-time-distribution}
\end{figure}

\begin{figure}[H]
  \centering
  \includegraphics[width=0.4\textwidth]{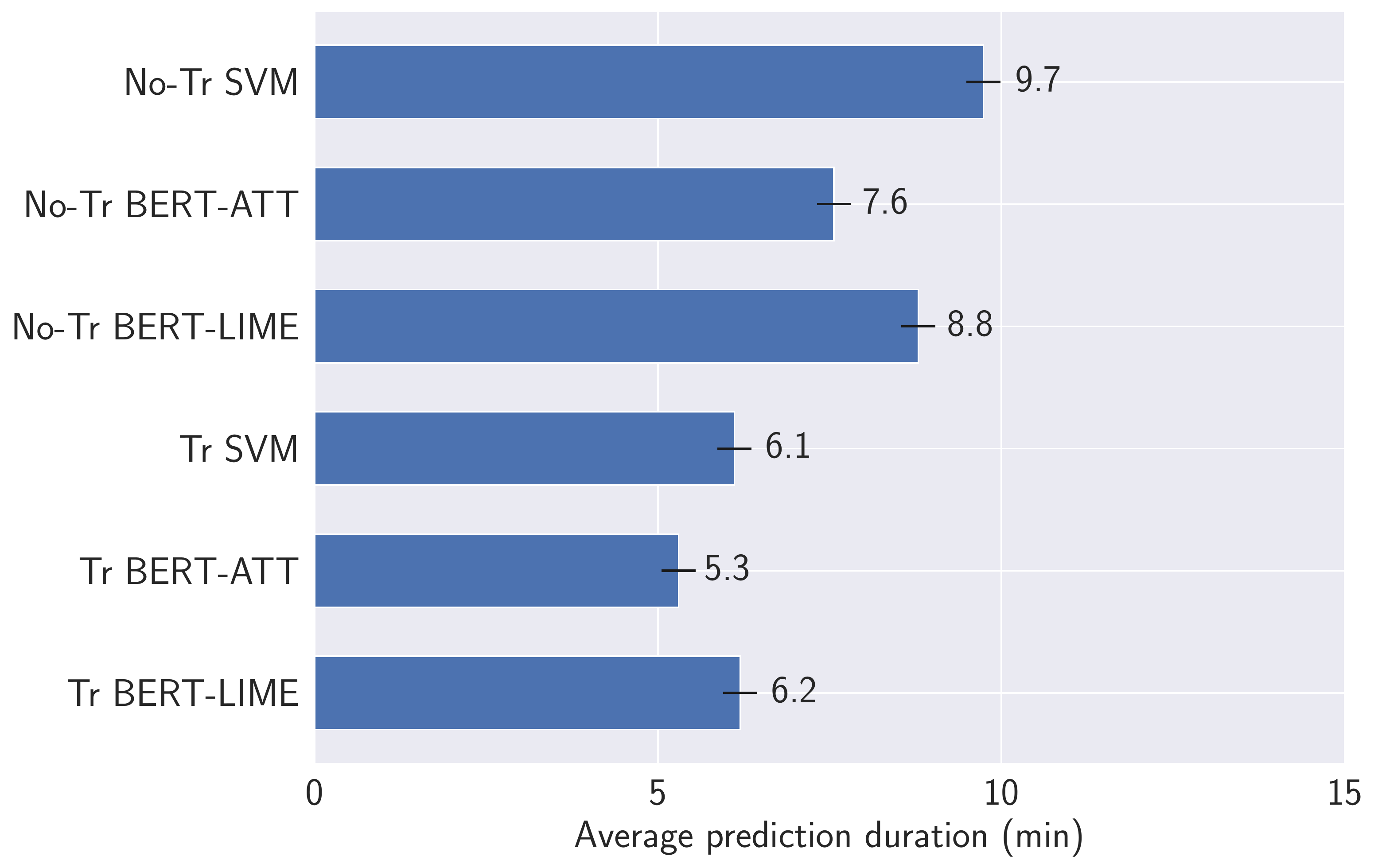}
  \caption{Average time taken for the prediction phase in each experimental setup in experiment 3.}
  \label{fig:exp3-time-distribution}
\end{figure}

\figref{fig:exp1-avg} - \figref{fig:exp3-avg} show the average time taken in each experiment. We calculated and filtered out
outliers from each experiment respectively with an interquartile range. In \figref{fig:exp1-time-distribution} - \figref{fig:exp3-time-distribution} we show the average time taken during prediction phase in each experiment.
Outliers were discarded after the same precedures.

\section{Trust Analysis}
\begin{figure}[H]
  \centering
  \includegraphics[width=0.4\textwidth]{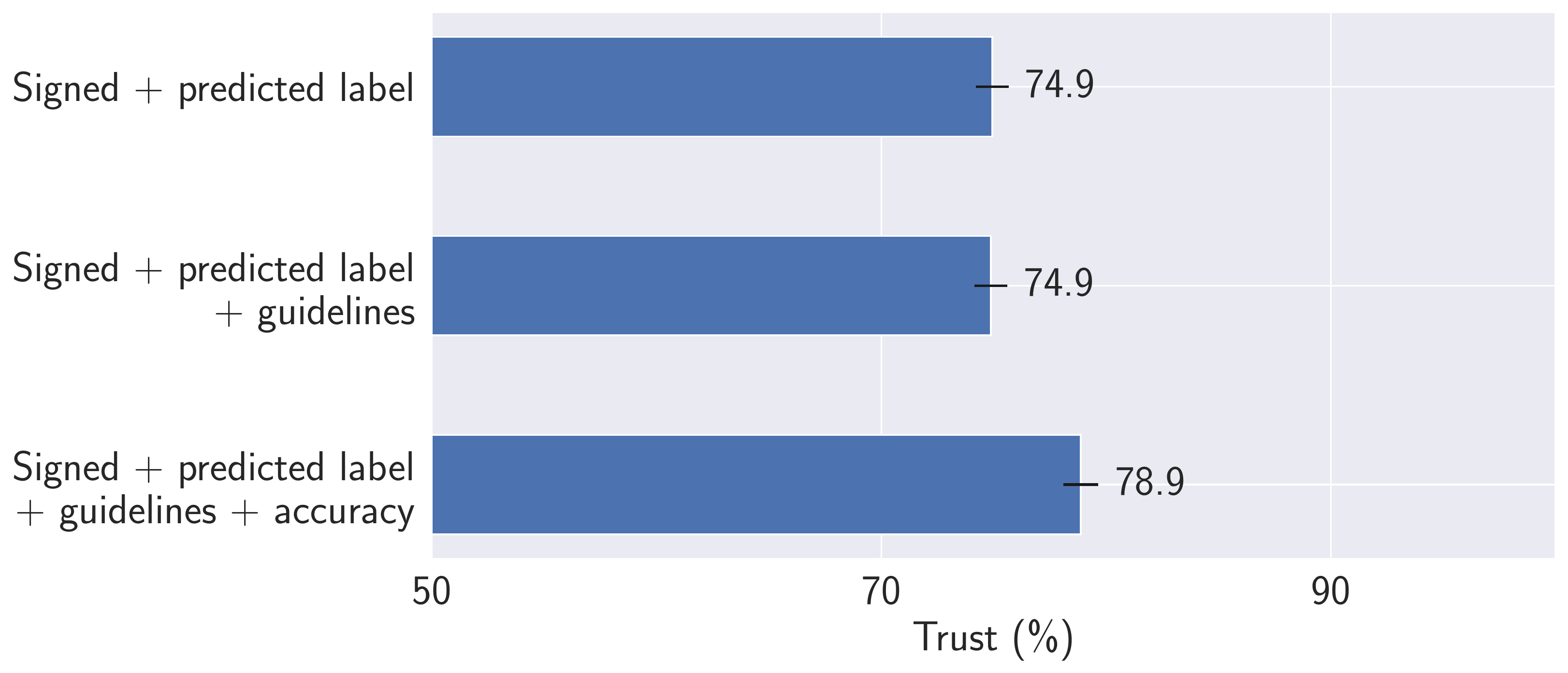}
  \caption{Human trust on machine predictions in experiment 2. Differences between all pairs are not statistically significant. These results suggest that guidelines and accuracy statement do not increase human trust in machine learning models significantly.}
  \label{fig:exp2-trust}
\end{figure}

\begin{figure}[H]
  \centering
  \includegraphics[width=0.4\textwidth]{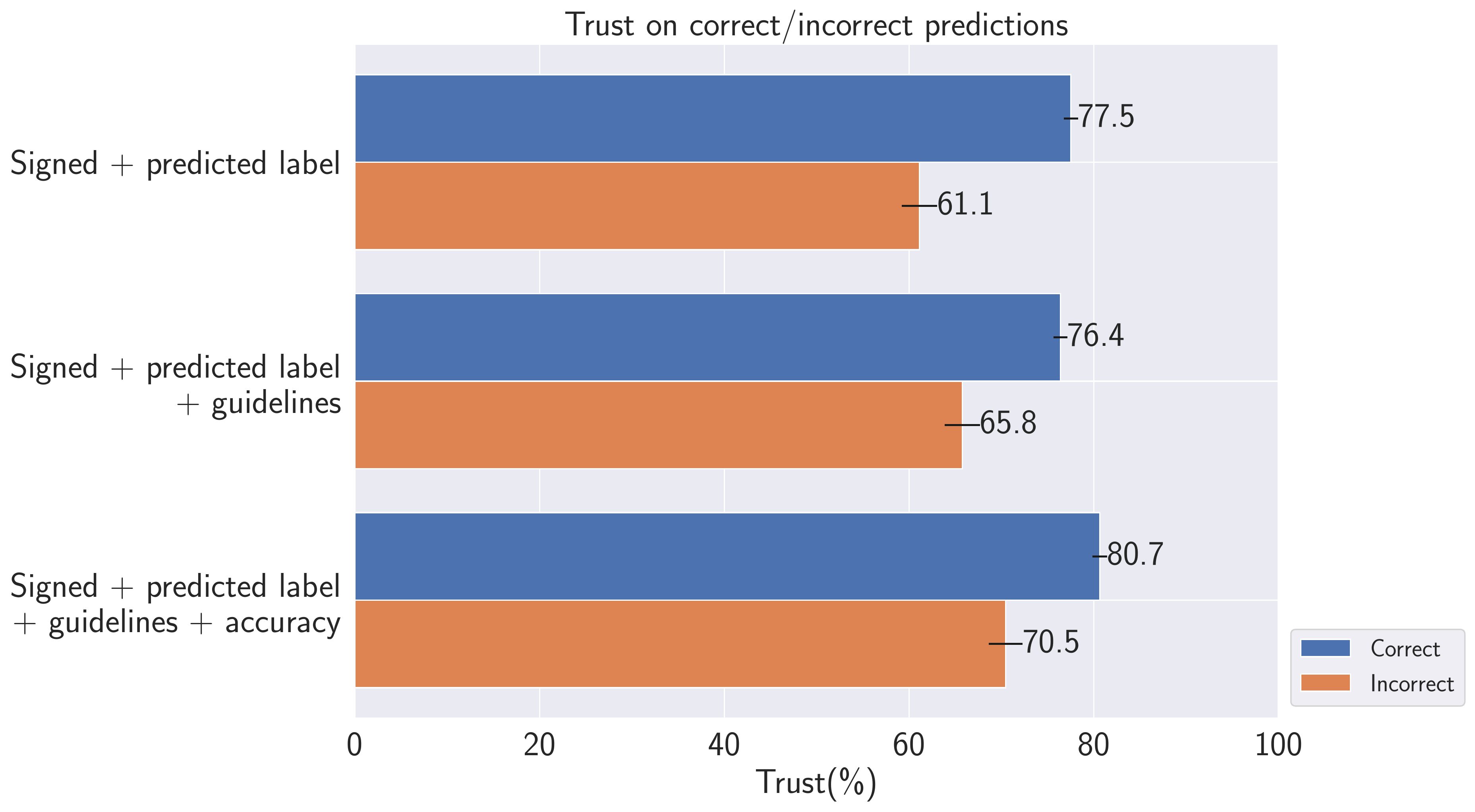}
  \caption{Human trust on correct / incorrect machine predictions in experiment 2. Differences between correct predictions and incorrect predictions are statistically significant. These results suggest that human have more trust in correct predictions than incorrect ones.}
  \label{fig:exp2-trust-correctness}
\end{figure}

\subsection{Analysis of Free Responses from Turkers}
Free responses from turkers confirmed the findings in the qualitative study.
Participants felt that tutorial was useful but could not understand why certain features are deceptive or genuine.
One participant commented, ``Although I am an English major, the training really helped me to think and consider the nuances of language. I enjoy good writing but I often overlook attempts to manipulate or deceive the reader/audience. I felt this training was very beneficial''.
Another participant remarked, ``I could not understand why words were chosen for the reason''.

\section{Human Performance Grouped by Demographics}

The is no clear trend regarding gender, education background, review writing frequency, and age among experiments.

\begin{figure}[H]
  \centering
  \includegraphics[width=0.47\textwidth]{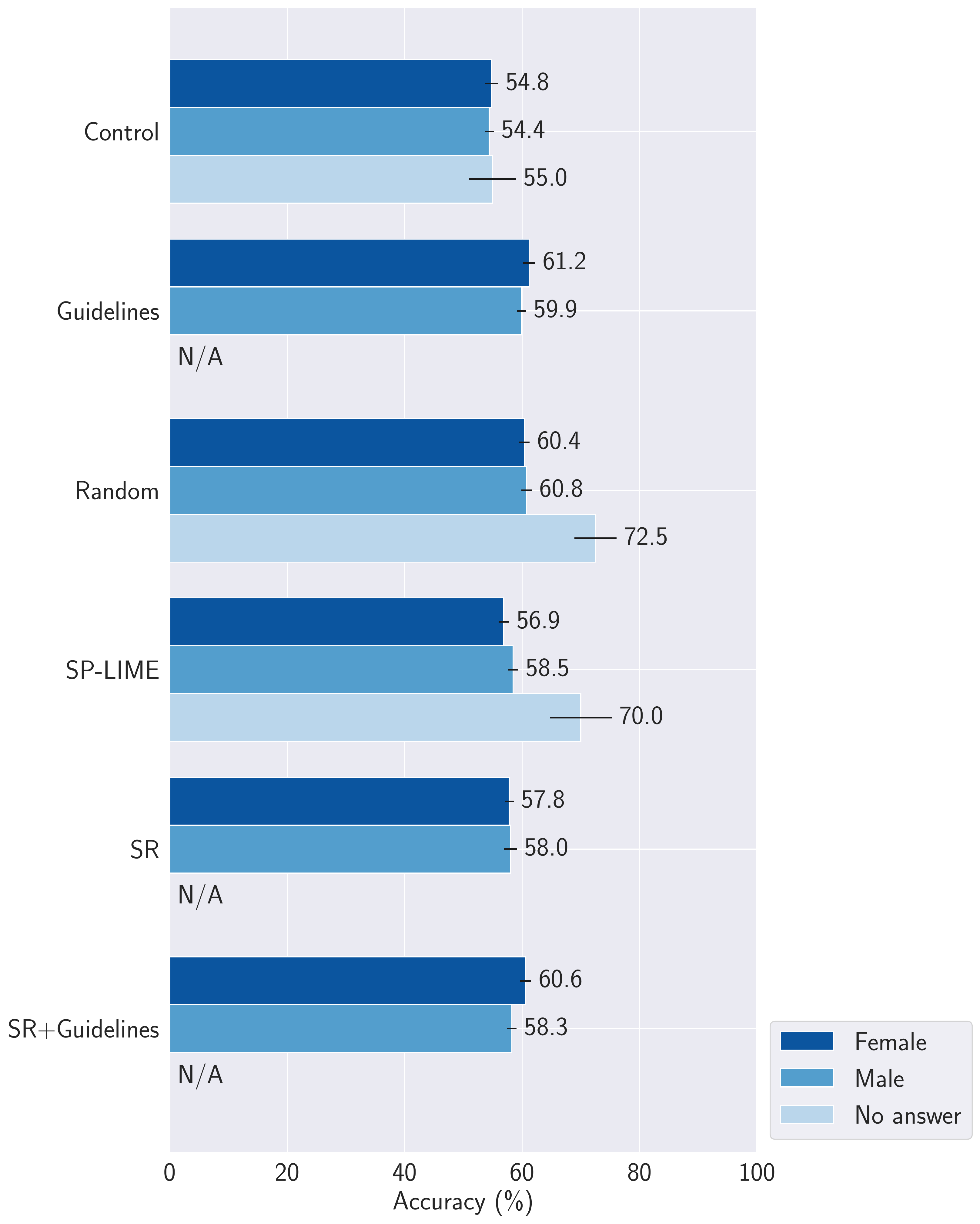}
  \caption{Experiment 1: gender. Human accuracy grouped by experimental setups and gender.}
  \label{fig:exp1-dmg-gen}
\end{figure}

\begin{figure}[H]
  \centering
  \includegraphics[width=0.47\textwidth]{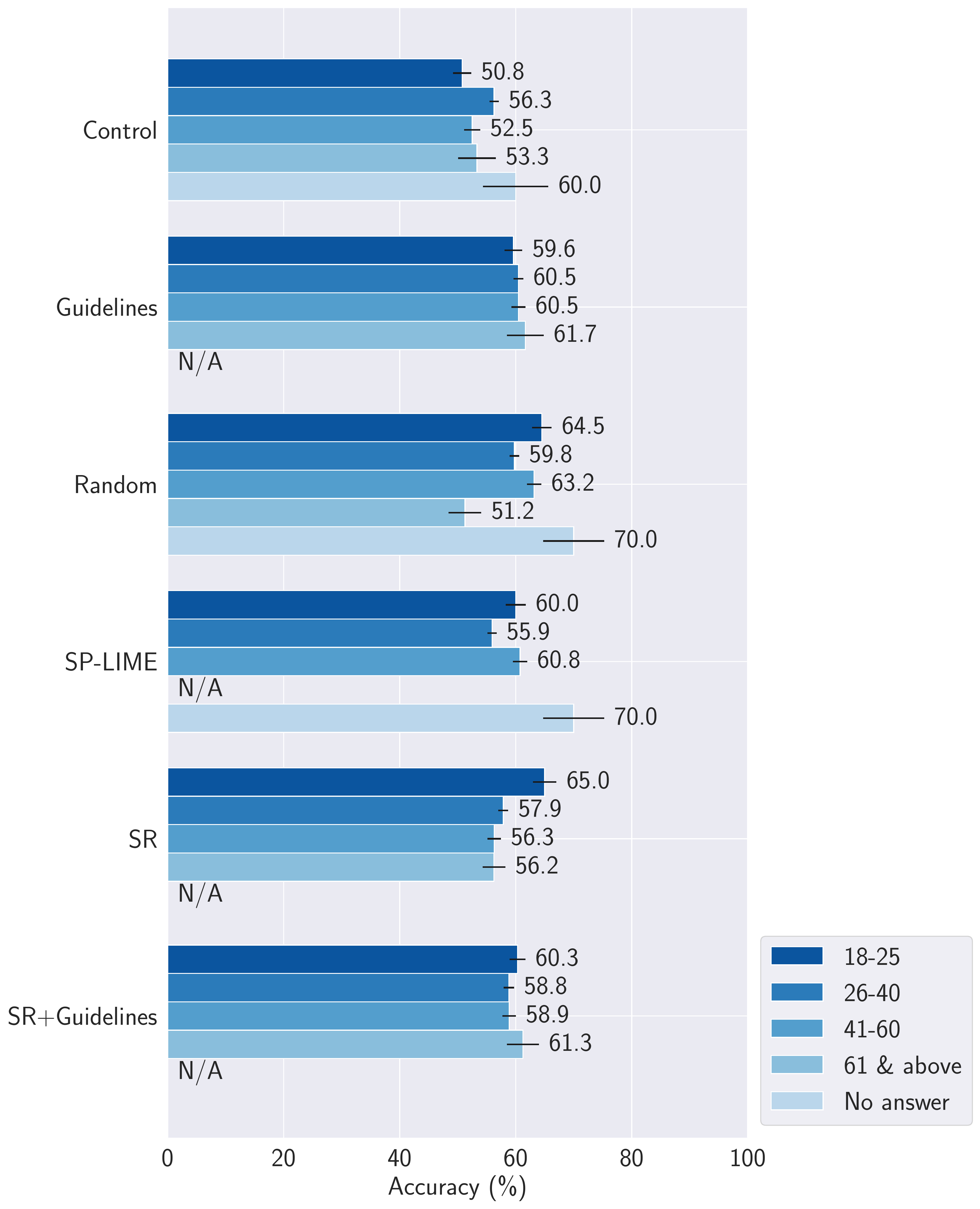}
  \caption{Experiment 1: age. Human accuracy grouped by experimental setups and age.}
  \label{fig:exp1-dmg-age}
\end{figure}

\begin{figure}[H]
  \centering
  \includegraphics[width=0.47\textwidth]{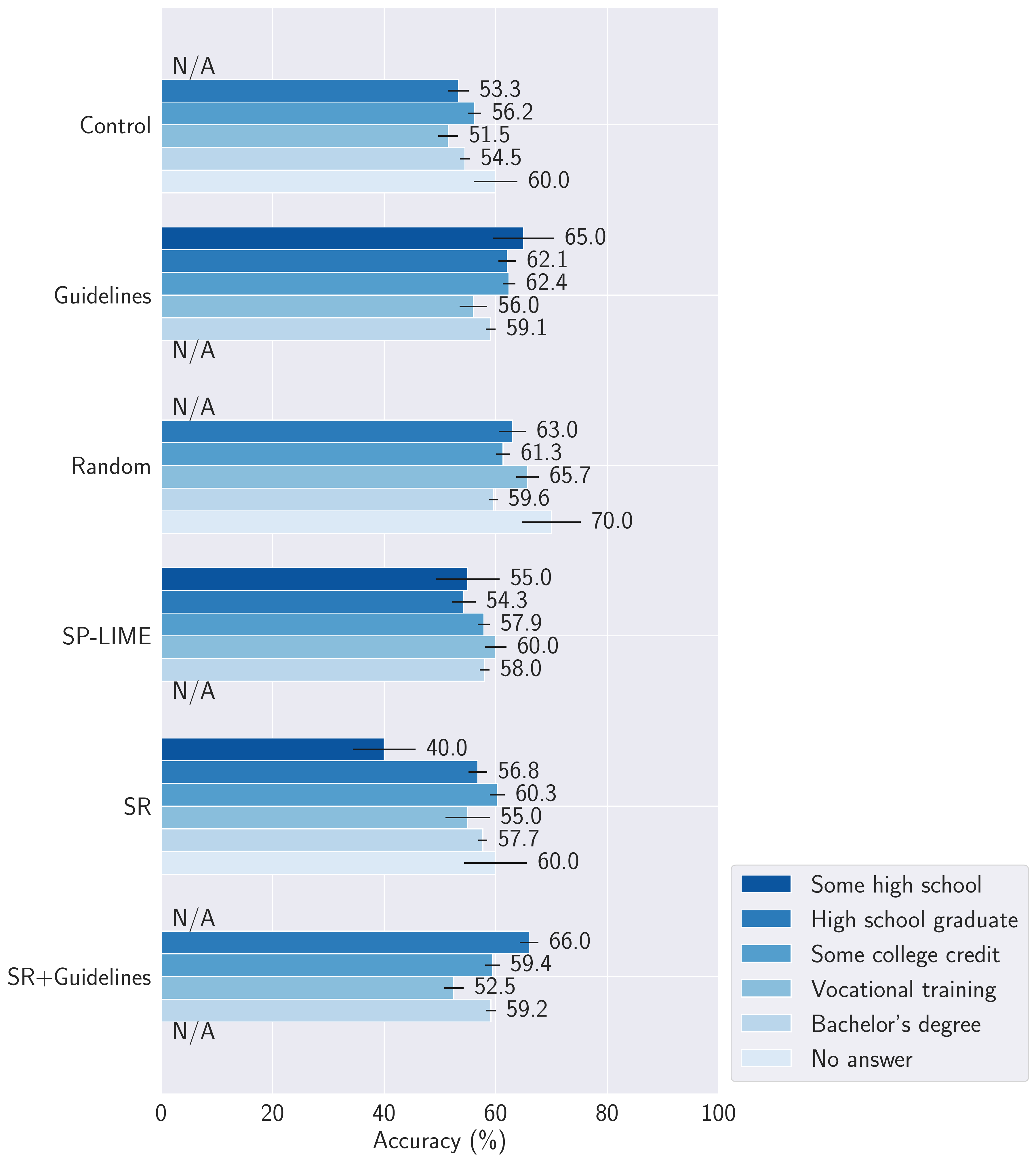}
  \caption{Experiment 1: education background. Human accuracy grouped by experimental setups and education background.}
  \label{fig:exp1-dmg-edu}
\end{figure}

\begin{figure}[H]
  \centering
  \includegraphics[width=0.47\textwidth]{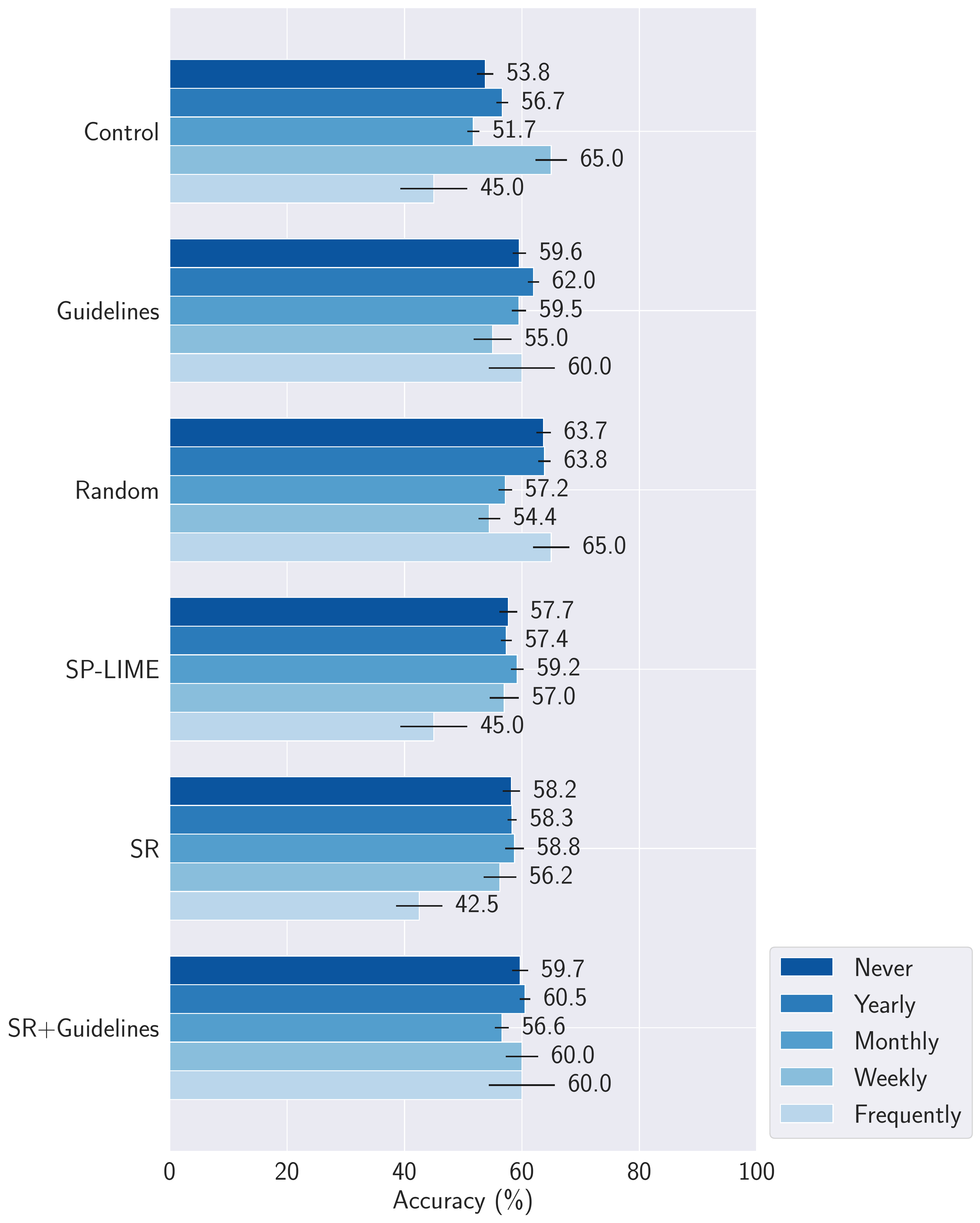}
  \caption{Experiment 1: review writing frequency. Human accuracy grouped by experimental setups and review writing frequency.}
  \label{fig:exp1-dmg-write}
\end{figure}

\begin{figure}[H]
  \centering
  \includegraphics[width=0.47\textwidth]{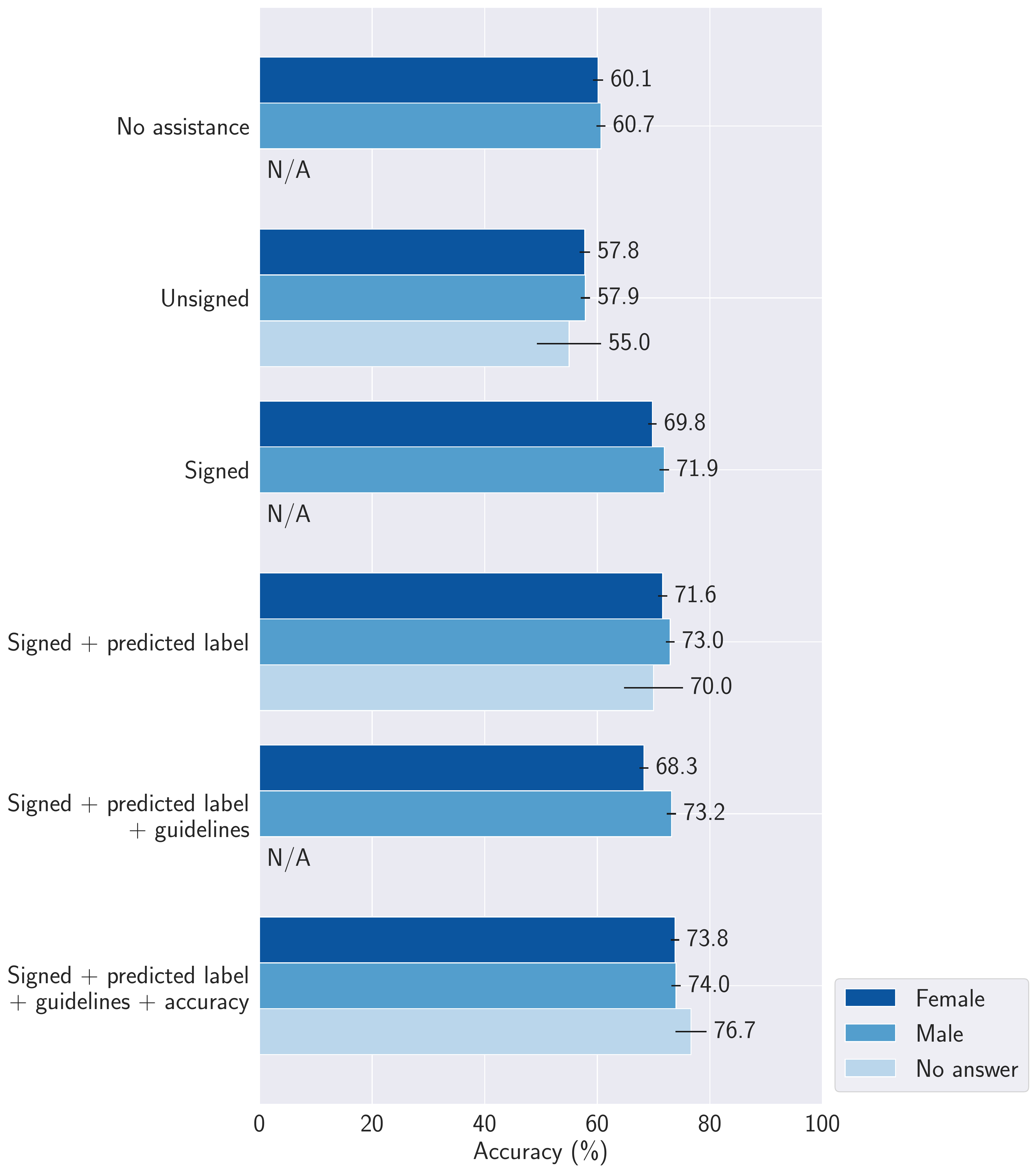}
  \caption{Experiment 2: gender. Human accuracy grouped by experimental setups and gender.}
  \label{fig:exp2-dmg-gen}
\end{figure}

\begin{figure}[H]
  \centering
  \includegraphics[width=0.47\textwidth]{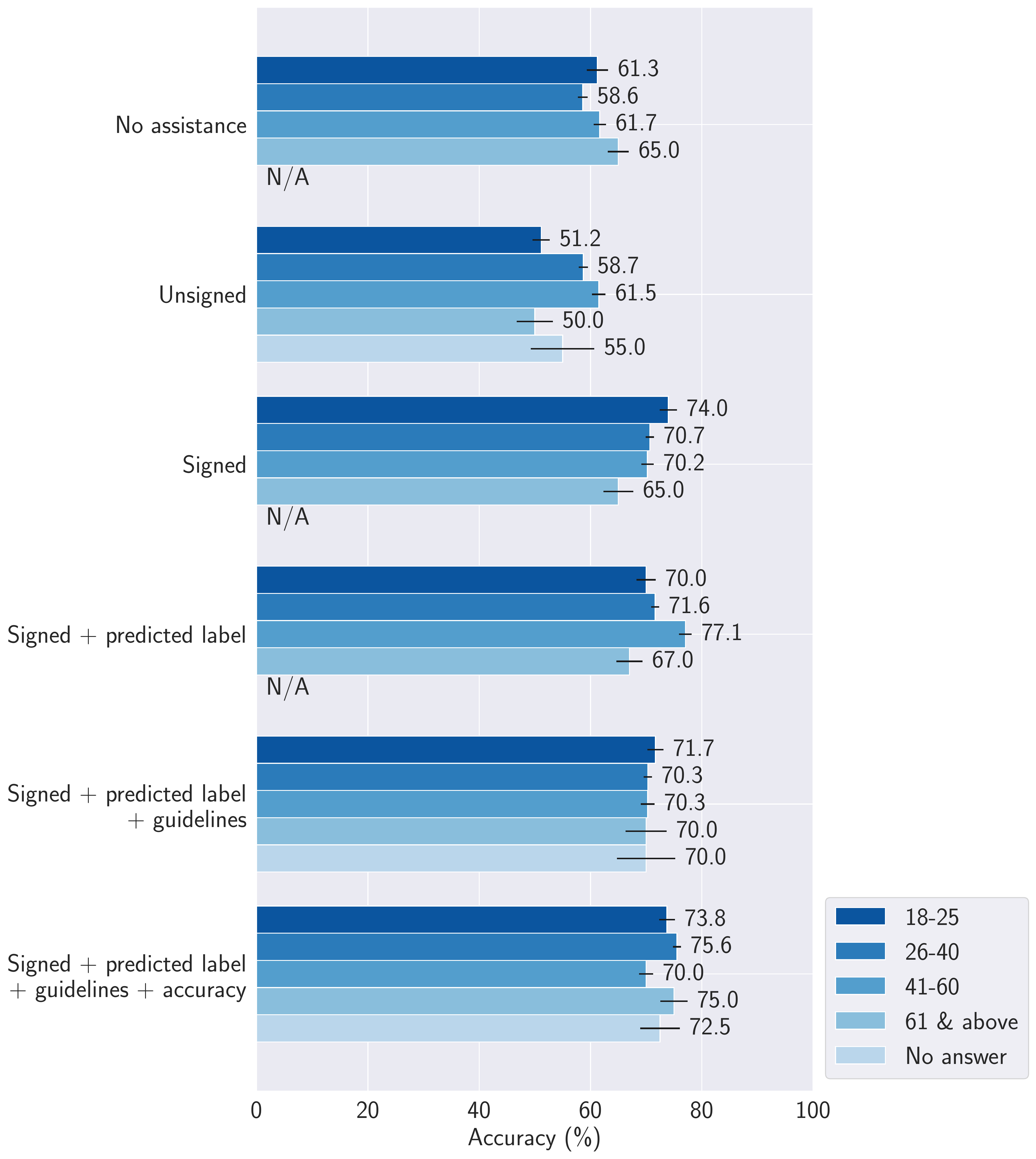}
  \caption{Experiment 2: age. Human accuracy grouped by experimental setups and age.}
  \label{fig:exp2-dmg-age}
\end{figure}

\begin{figure}[H]
  \centering
  \includegraphics[width=0.47\textwidth]{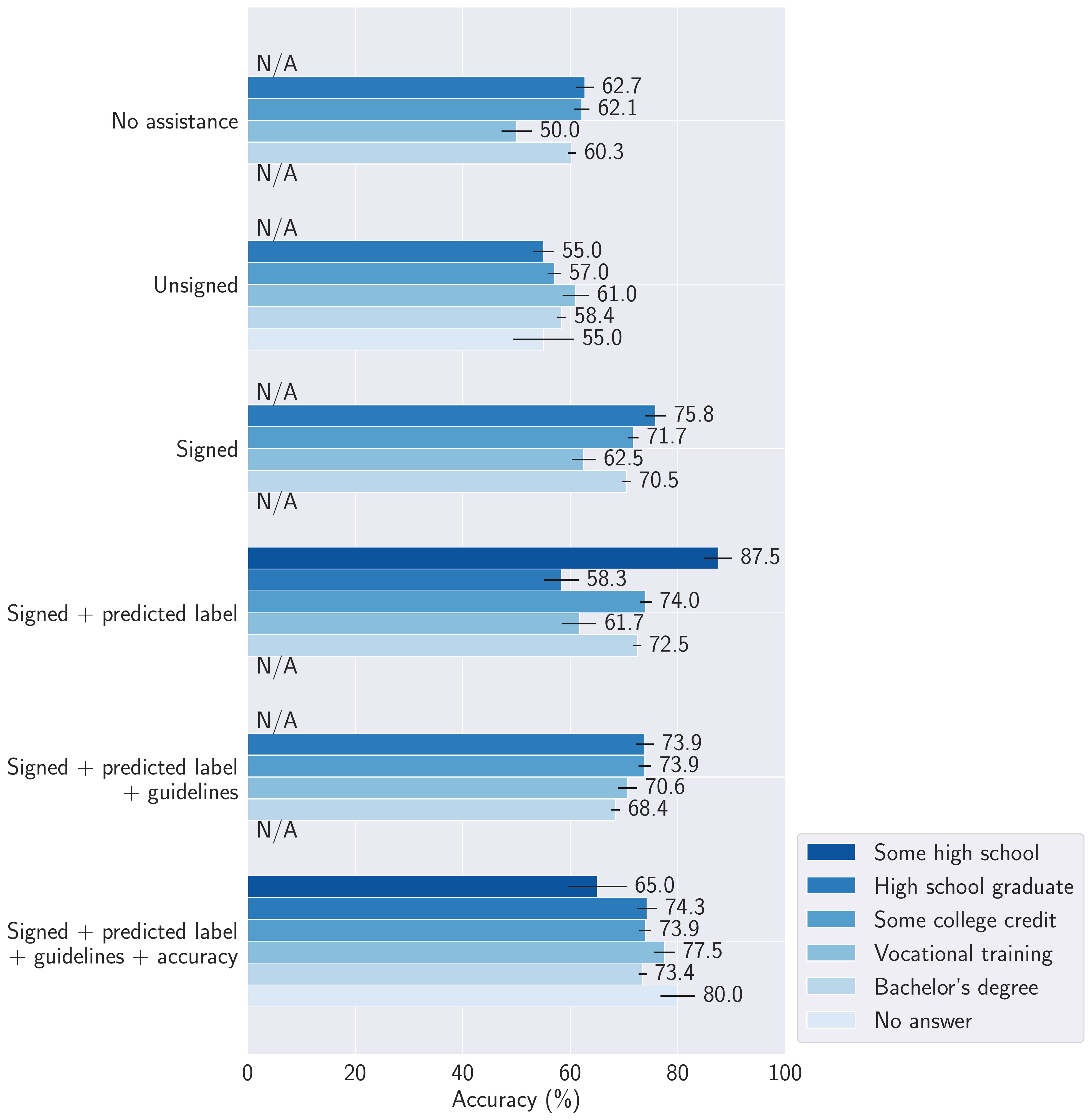}
  \caption{Experiment 2: education background. Human accuracy grouped by experimental setups and education background.}
  \label{fig:exp2-dmg-edu}
\end{figure}

\begin{figure}[H]
  \centering
  \includegraphics[width=0.47\textwidth]{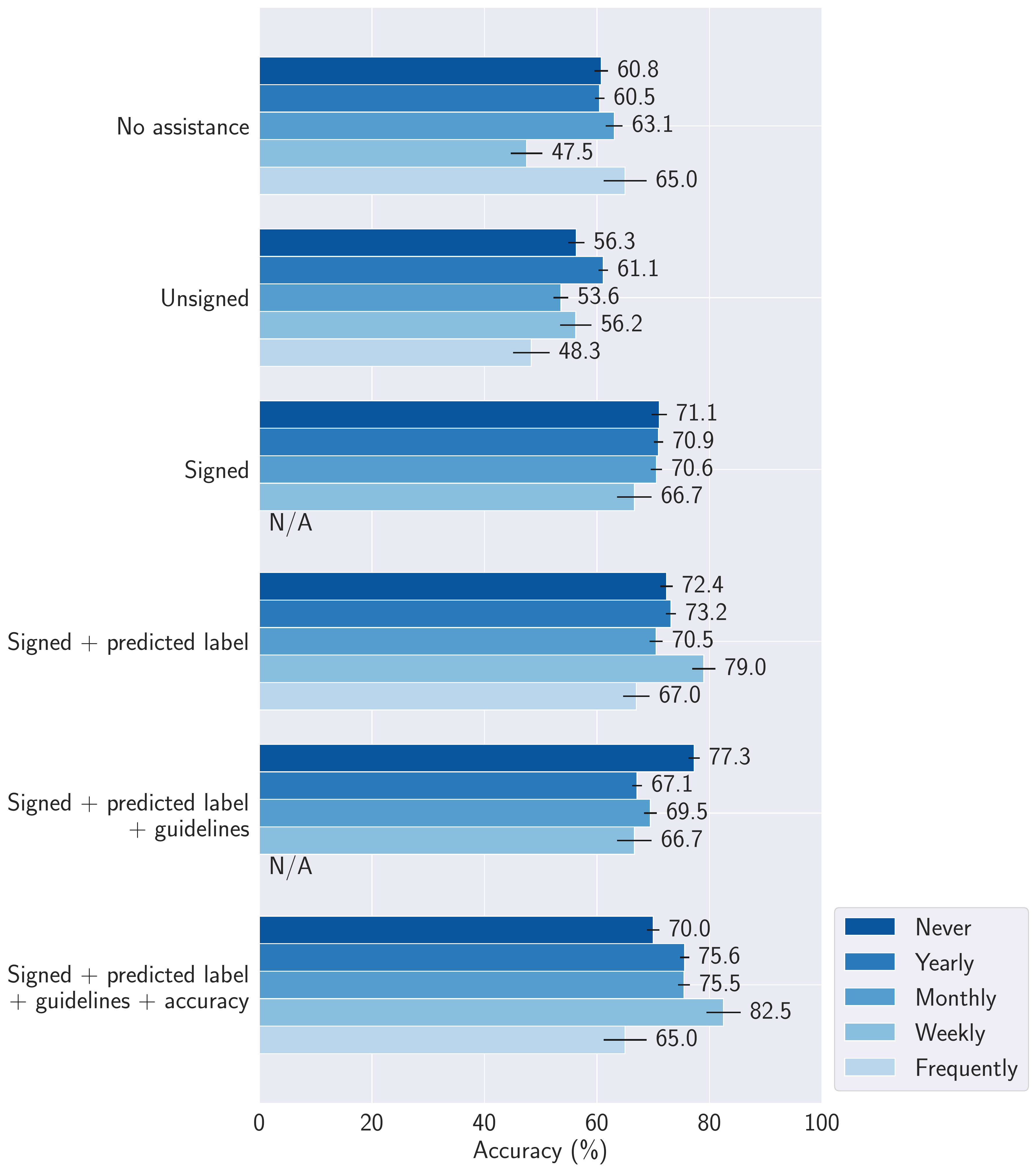}
  \caption{Experiment 2: review writing frequency. Human accuracy grouped by experimental setups and review writing frequency.}
  \label{fig:exp2-dmg-write}
\end{figure}

\begin{figure}[H]
  \centering
  \includegraphics[width=0.47\textwidth]{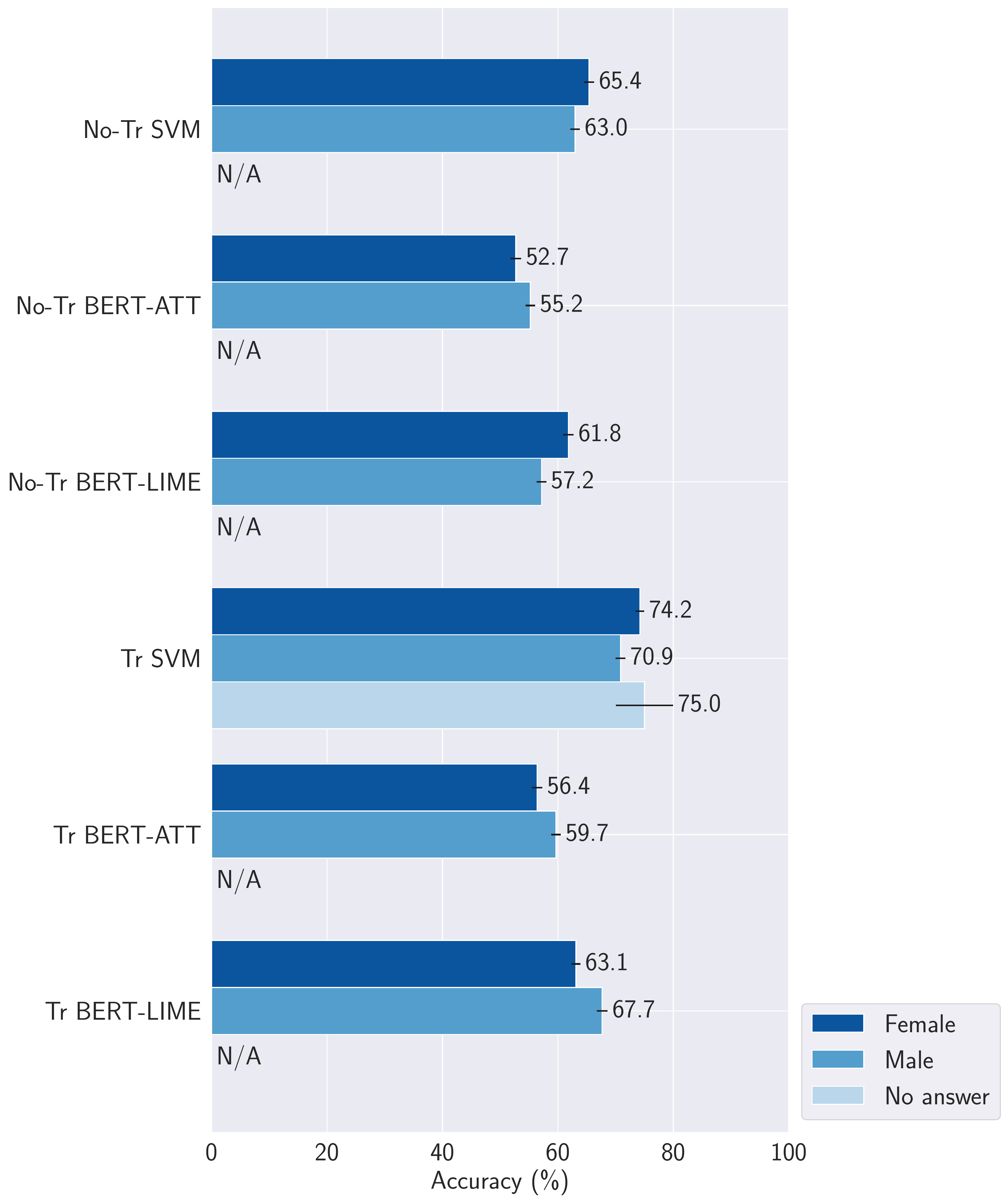}
  \caption{Experiment 3: gender. Human accuracy grouped by experimental setups and gender.}
  \label{fig:exp3-dmg-gen}
\end{figure}

\begin{figure}[H]
  \centering
  \includegraphics[width=0.47\textwidth]{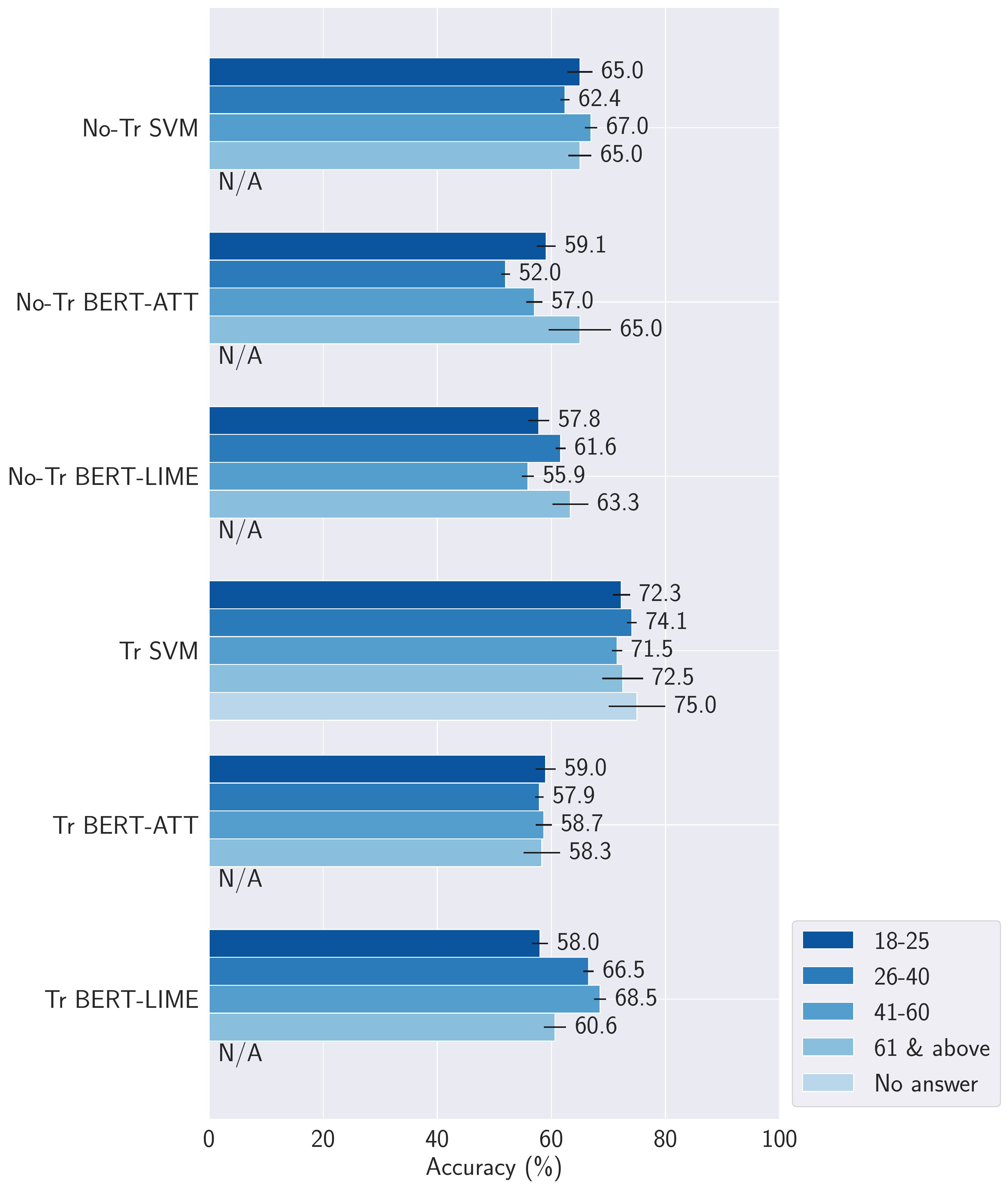}
  \caption{Experiment 3: age. Human accuracy grouped by experimental setups and age.}
  \label{fig:exp3-dmg-age}
\end{figure}

\begin{figure}[H]
  \centering
  \includegraphics[width=0.47\textwidth]{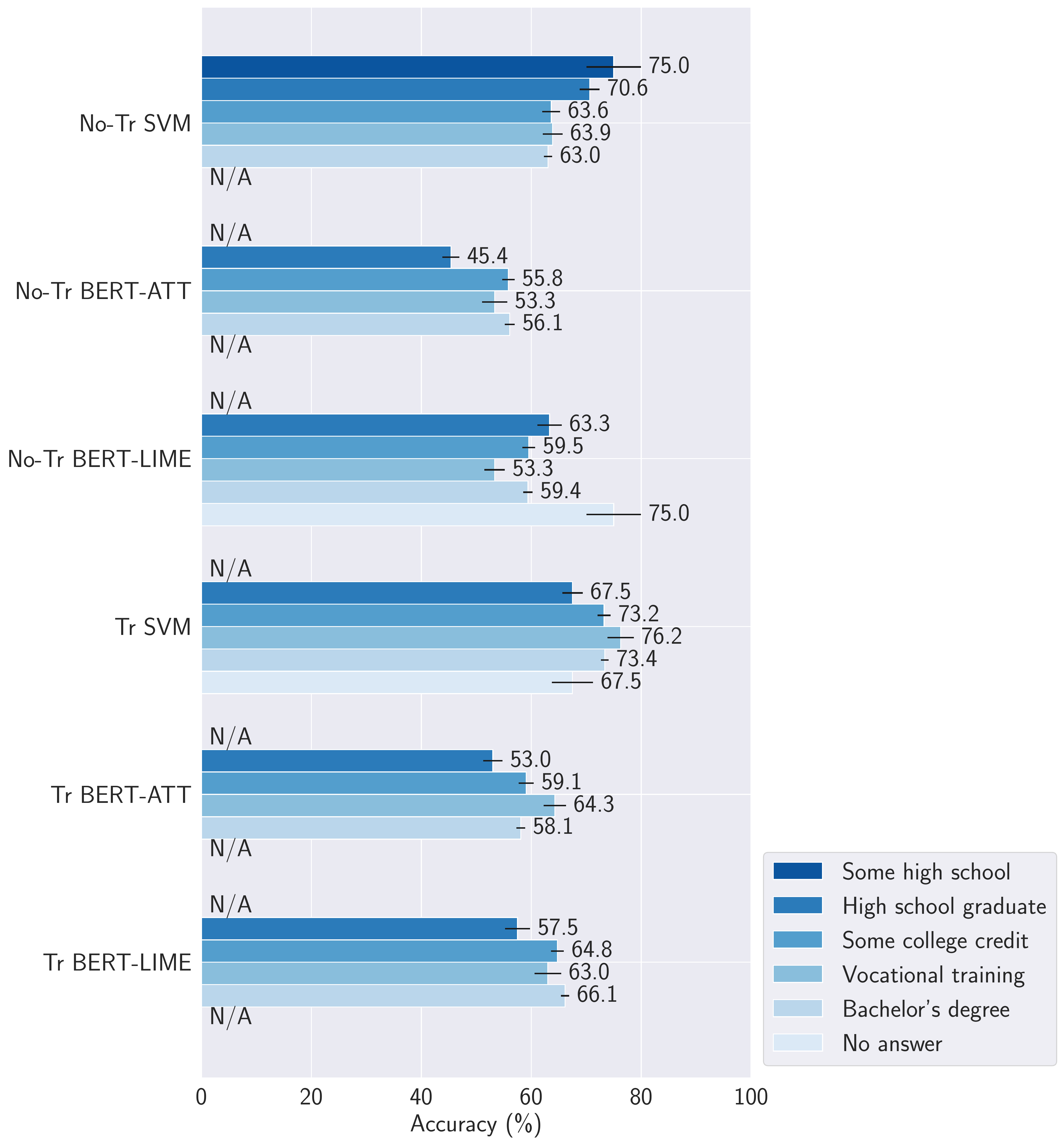}
  \caption{Experiment 3: education background. Human accuracy grouped by experimental setups and education background.}
  \label{fig:exp3-dmg-edu}
\end{figure}

\begin{figure}[H]
  \centering
  \includegraphics[width=0.47\textwidth]{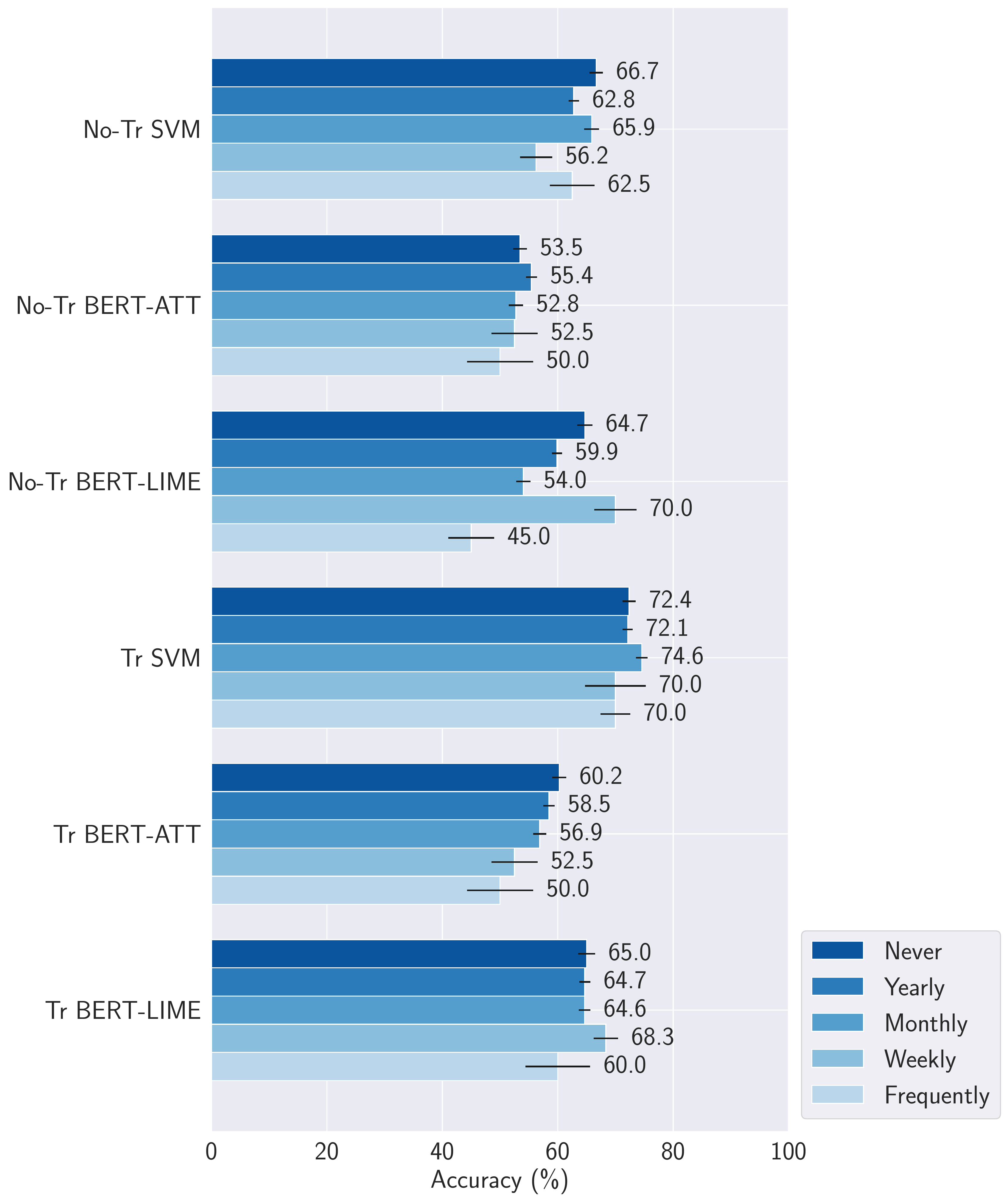}
  \caption{Experiment 3: review writing frequency. Human accuracy grouped by experimental setups and review writing frequency.}
  \label{fig:exp3-dmg-write}
\end{figure}

\section{Attention-check Design}

P11 was half way through the session and commented, ``I'm trying to think about this from a way of, like, are these reviews being generated by a computer, or are they, like, are all of these reviews from real people, and am I trying to tell if somebody's, like, lying about the review''.
The interviewer then suggested to the participant to read the instructions in the dialogue boxes.
P11 subsequently explained that he ``just didn't notice that because I was just reading the rules and skipped the box''.
Similarly, P9 asked the interviewer, ``By deceptive review do you mean users typing a review for the sake of tarnishing reputation, or uplifting reputation, or are you referring to computer-generated reviews which are trying to deceive people''.
Due to a couple of the above cases, we added additional attention-check questions to ensure that participants are aware of the definition of deceptive reviews.
Refer to the outdated and updated attention-check design below.

\begin{figure}[H]
  \centering
  \includegraphics[width=0.47\textwidth]{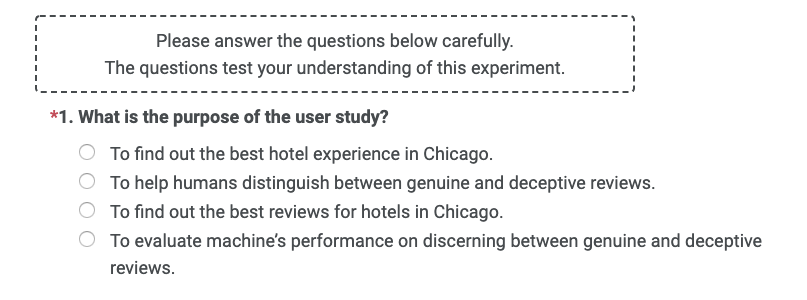}
  \caption{Outdated attention-check design. The outdated design does not allow participants to confirm on their answers. If they selected the wrong answer, they will be disqualified immediately.}
  \label{fig:survey-outdated}
\end{figure}

\begin{figure}[H]
  \centering
  \includegraphics[width=0.47\textwidth]{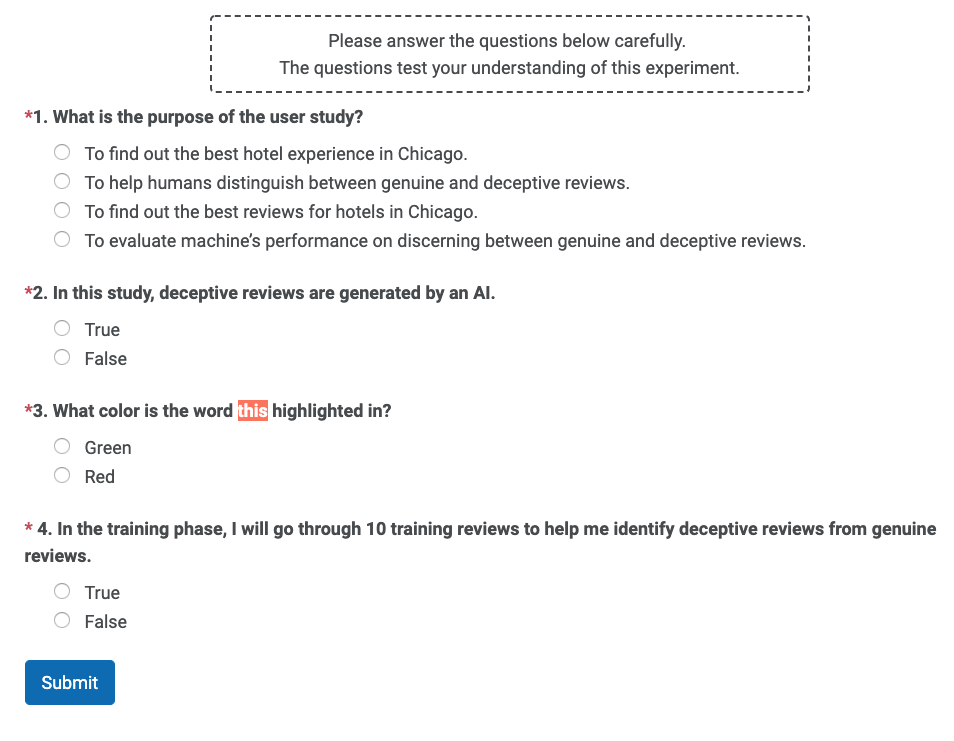}
  \caption{Updated attention-check design. The updated design allows participants to confirm on their answers.}
  \label{fig:survey-updated}
\end{figure}

\section{Exit Survey}

\begin{figure*}[t]
    \begin{center}
    \includegraphics[width=0.8\textwidth]{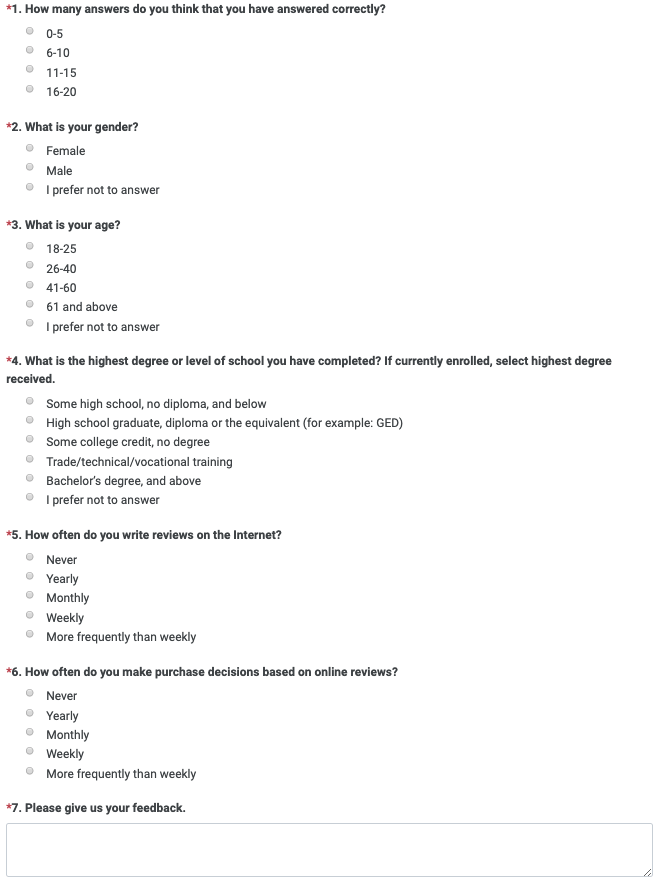}
    \end{center}
    \caption{Exit survey for control setup in Experiment 1.}
\label{fig:exp1-control-survey}
\end{figure*}

\begin{figure*}[t]
    \begin{center}
    \includegraphics[width=0.75\textwidth]{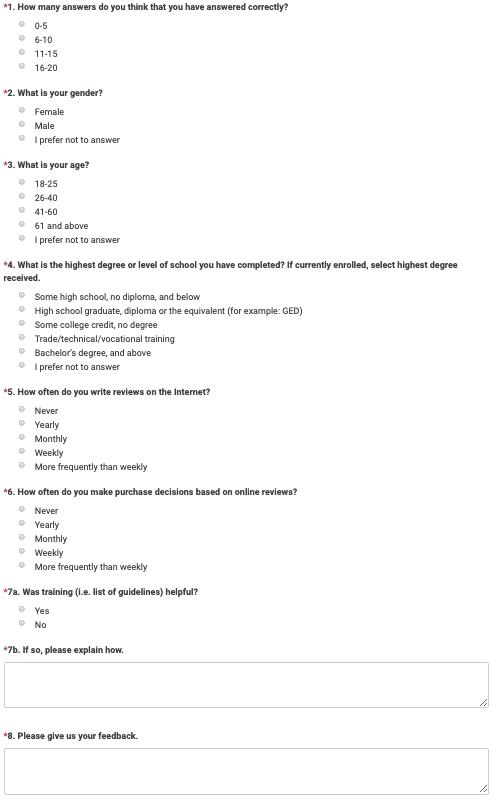}
    \end{center}
    \caption{Exit survey for guidelines setup in Experiment 1.}
\label{fig:exp1-guidelines-survey}
\end{figure*}

\begin{figure*}[t]
    \begin{center}
    \includegraphics[width=0.75\textwidth]{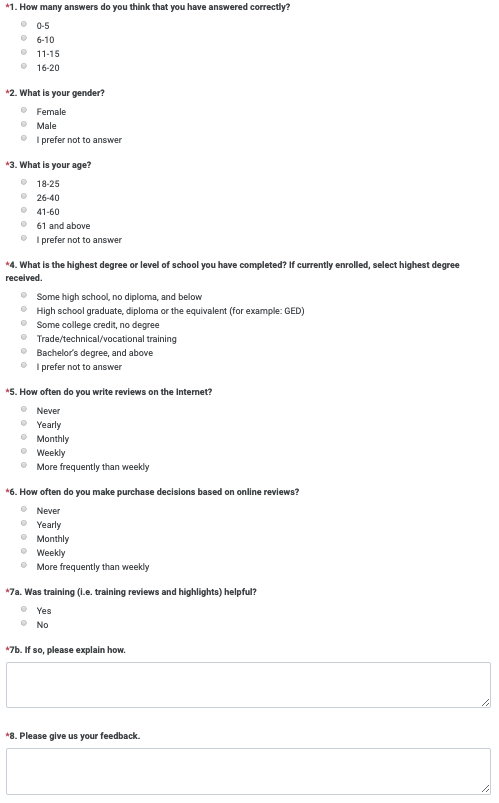}
    \end{center}
    \caption{Exit survey for examples i.e., {\em random}, {\em SP-LIME}, and {\em spaced repetition} in experiment 1. Note that question 7a changes to the following: `Was training (i.e. training reviews and list of guidelines) useful?' for {\em SR+guidelines}.}
\label{fig:exp1-treatment-survey}
\end{figure*}

\figref{fig:exp1-control-survey} - \figref{fig:exp1-treatment-survey} show exit surveys for experimental setups in Experiment 1.

\begin{figure*}[t]
    \begin{center}
    \includegraphics[width=0.75\textwidth]{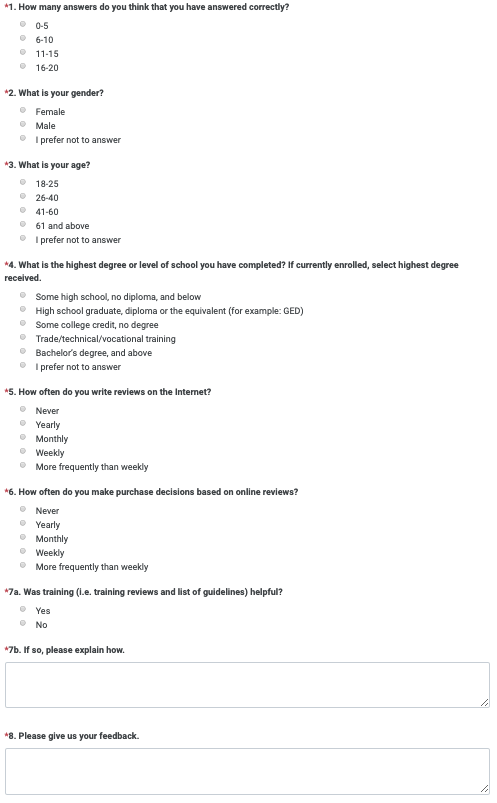}
    \end{center}
    \caption{Exit survey for experimental setup in Experiment 2.}
\label{fig:exp2-survey}
\end{figure*}

\begin{figure*}[t]
    \begin{center}
    \includegraphics[width=0.75\textwidth]{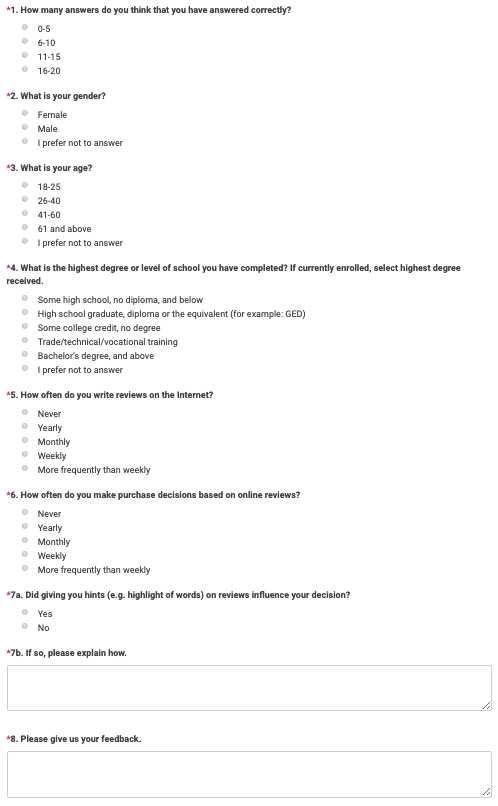}
    \end{center}
    \caption{Exit survey for non-training experimental setups in Experiment 3.}
\label{fig:exp3-no-trg}
\end{figure*}

\begin{figure*}[t]
    \begin{center}
    \includegraphics[width=0.75\textwidth]{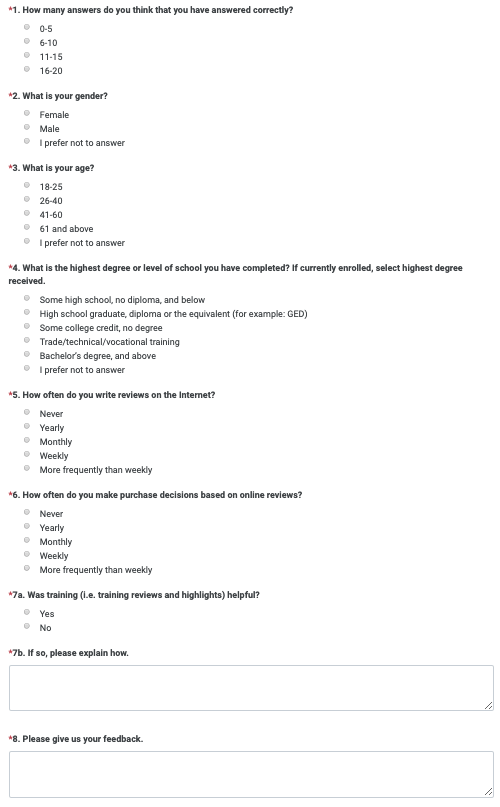}
    \end{center}
    \caption{Exit survey for training experimental setups in Experiment 3.}
\label{fig:exp3-trg}
\end{figure*}

\figref{fig:exp3-no-trg} and \figref{fig:exp3-trg} show exit surveys for experimental setups in Experiment 3.

\end{document}

